\documentclass{elsart}

 \usepackage{graphicx}

\usepackage{amssymb}

\begin{document}

\begin{frontmatter}



\title{
Electromagnetic strength of
neutron and proton single-particle halo nuclei}


\author{S. Typel\thanksref{ST}}
\address{Gesellschaft f\"{u}r Schwerionenforschung mbH, Theorie,\\
Planckstr.~1, D-64291 Darmstadt, Germany}
\thanks[ST]{Corresponding author.}
\ead{S.Typel@gsi.de}

\author{G. Baur}
\address{Institut f\"{u}r Kernphysik,
Forschungszentrum J\"{u}lich,
D-52425 J\"{u}lich, Germany}
\ead{G.Baur@fz-juelich.de}

\begin{abstract}
Electromagnetic strength functions of halo nuclei exhibit 
universal features that can be described in terms of
characteristic scale parameters.
For a nucleus with nucleon+core structure
the reduced transition probability, as determined, e.g.,
by Coulomb dissociation experiments,
shows a typical shape that depends
on the nucleon separation energy and the orbital angular
momenta in the initial and final states.
The sensitivity to the final-state interaction (FSI) between the nucleon
and the core
can be studied systematically by varying the strength of the interaction
in the continuum.
In the case of neutron+core nuclei  
analytical results for the reduced transition
probabilities are obtained by introducing 
the effective-range expansion.
The scaling with the relevant parameters is found explicitly.
General trends are observed by studying
several examples of neutron+core and proton+core nuclei
in a single-particle model assuming Woods-Saxon potentials.
Many important features of the neutron halo case can
be obtained from a square-well model. Rather
simple analytical formulas are found.
The nucleon-core interaction in the continuum
affects the determination of astrophysical S factors at zero energy
in the method of asymptotic normalisation coefficients (ANC).
It is also relevant for the extrapolation of radiative capture
cross sections to low energies.
\end{abstract}

\end{frontmatter}

\newpage 

\section{Introduction}

Light exotic nuclei are available
as secondary beams at various 
radioactive beam facilities all over the world. These
unstable nuclei 
are generally  weakly bound with few, if any, bound excited states. 
They have been studied extensively
in recent years by electromagnetic excitation with the help of the
Coulomb breakup method 
\cite{Bau86,Ber88,Lei01,Bau03,Aum04}. 
For low orbital angular momenta 
of the lowest bound valence nucleon an extended diffuse 
density distribution, a halo, develops 
resulting in a large size of the nucleus
\cite{Han87,Rii92,Rii94,Han95,Tan96}.
Simultaneously, electromagnetic transitions to the continuum 
with large strength are observed at low energies.
Properties of stable nuclei have successfully been investigated
by electromagnetic excitation
in photonuclear reactions as well as 
in heavy ion collisions for a long time.
Their electromagnetic strength functions are dominated by
the giant resonances high in the continuum.

Nuclei close to the neutron and proton drip lines
often exhibit a pronounced nucleon+core structure
that is well described by single-particle models with appropriately
chosen potentials.
In a microscopic shell-model study \cite{Sag01} 
strong low-lying dipole strength in neutron-rich ${}^{14}$Be 
and proton-rich ${}^{13}$O
was observed to originate from loosely-bound extended 
single-particle wave functions. 
These extended wave functions were obtained by 
adjusting the potential depth in order to reproduce 
the empirical binding energies. In order to study low-lying strength
theoretically, such a feature has to be added to the ab-initio 
microscopic approaches. 
A comparison of  strength functions deduced from experiment
with theoretical predictions from single-particle models
is used to extract spectroscopic factors or asymptotic normalization
coefficients (ANC). They can be compared to
more elaborated nuclear models. However, the importance of 
effects from the interaction between the nuclei in the final state
has to be assessed in order to obtain reliable
information from experimental data. 
Both the shape and the absolute magnitude of the strength function
can be affected. This will also have consequences for the application
of sum rules that relate the total excitation strength to the
properties of the ground state. 

The relevant matrix elements for the electromagnetic transition
to the continuum at low energies are esssentially determined
by the asymptotics of the bound-state and continuum wave functions
of halo nuclei. This allows to study systematically the effects of the 
final-state interaction (FSI)
without the necessity to introduce sophisticated nuclear structure
models. Magnetic contributions to the continuum transitions 
are usually much weaker than electric transitions, 
except for the deuteron breakup at low
energies \cite{Nag97}
and for the excitation of resonances. 
We will limit ourselves
to the discussion of direct electric transitions to the continuum,
with emphasis on $E1$ transitions,
but the search for low-lying $M1$ strength and 
its theoretical description \cite{Bla79} remains an interesting
challenge for future studies. 

The nuclear interaction $V_{bc}$
between the nucleon $b$ and the core $c$ in the electromagnetic breakup
of an exotic nucleus $a$ is responsible
for the binding of the nucleus $a$. However, it also affects
the structure of the continuum 
(usually in partial waves with different $l$ values)
even if there are no resonances
observed at low excitation energies. 
In the experimental analysis of neutron+core
breakup reactions it is often neglected and a plane wave is assumed
in the final state of the $b+c$ system.
The interaction between $b$ and $c$  also appears in the final state of
the photo-dissociation reaction $a(\gamma,c)b$ or in the
initial state of the radiative capture reaction $b(c,\gamma)a$.
Thus, the interaction $V_{bc}$ can affect the 
energy dependence of the astrophysical S factor of the radiative
capture reaction  that is used to extrapolate
experimental data to zero energy. Similarly, the strength of the
interaction enters into the
calculation of the zero-energy S factor from asymptotic
normalization coefficients (ANCs) that are determined experimentally
from transfer reactions in the ANC method, see, e.g., 
\cite{Xu94,Tra01,Muk01,Cre02}.

Experimentally observed excitation functions of exotic nuclei
show an approximately universal shape 
when plotted as a function of appropriately scaled variables
\cite{Ots94,Men95,Kal96,Typ01a}.
They are dominated by direct transitions
to the continuum with only few resonances
and simple scaling laws apply.
The nuclear structure in the initial and final state 
depends only on a limited number of relevant quantities
that contain all the structural information that is accessible
at low energies. Details of the nuclear
interaction are not resolved at this low energy scale.
The interaction between the fragments leads to
a change of the transition strength when compared to the
case without nuclear interaction.
The actual nuclear potential is often not 
well constrained since extrapolations of the corresponding 
systematic optical potentials, e.g.\ \cite{Per76},
from nuclei in the valley of stability 
to unstable nuclei are questionable.
The effect of the continuum interaction
was studied before only in selected cases. 
E.g., it was found that the $s$-wave ground state to $p$-wave continuum
$E1$ transition in the case of ${}^{11}$Be 
is much less affected by the potential in the final state
than the $p \to s$ transition in ${}^{13}$C \cite{Ots94,Men95}.

At low energies the effect of the nuclear potential is 
conveniently described by the effective-range expansion
\cite{Bet49,New82},
with the scattering length and the effective range as the main 
parameters.  
An effective-range approach for the FSI in electromagnetic
excitations was introduced in \cite{Typ04a} and
applied to the breakup of ${}^{11}$Be.
Recently, the same method was applied to the description of
electromagnetic dipole strength in ${}^{23}$O
\cite{Noc04}.
Here, we will discuss this approach in much more detail.
A systematic study for various transitions will shed some additional
light on the sensitivity to the interaction in the continuum.
We want to expose the dependence on the binding energy of the nucleon
and on the angular
momentum quantum numbers. Our approach extends the familiar textbook
case of the deuteron \cite{Bla79}, that can be considered as the
prime example of a halo nucleus, to arbitrary nucleon+core systems.

Our effective-range approach is closely related to effective field theories
that are nowadays used for the description of 
the nucleon-nucleon system and halo nuclei
\cite{Bir99,Kap99,Ber02}. 
(The deuteron, the only bound state in the 
nucleon-nucleon system, can be considered as
a good example of a  halo nucleus).
The characteristic low-energy
parameters are linked to QCD in systematic expansions.
Similar methods are also used in 
the study of exotic atoms ($\pi^{-} A$, $\pi^{+} \pi^{-}$, 
$\pi^{-}p$, \dots) 
in terms of effective-range parameters in Ref.\ \cite{Eri03}.
The close relation of effective field theory to the effective-range
approach for hadronic atoms was discussed in Ref.\ \cite{Hol99}.
In our approach these constants are treated as free parameters.
It is not our aim to relate them to the underlying microscopic
description. Aspects of the many-body physics are summarized, e.g., 
in terms of
spectroscopic factors or asymptotic normalization coefficients.

We also investigate in detail a specific model, the square well potential.
It has  great merits: it can be solved analytically,
it shows the main characteristic features
and it leads to rather simple and transparent formulas
where some of them seem to be new.
These explicit results can be compared to our general
results for low energies (effective-range approach) and also 
to Woods-Saxon models. 

This paper is organized as follows.
A nucleon+core potential model for halo nuclei is introduced 
in section~\ref{sec:ncm}. The relevant scaling parameters are defined 
in subsection \ref{subsec:scale} and
scaling laws for the 
the probabilities to find the nucleon inside the range of the nuclear
potential are discussed for bound and scattering states
in subsection \ref{subsec:prob}. The
scaling of the root-mean-square radius serves as another indication
for the halo nature of the bound state.
The reduced transition probabilities for the breakup of a nucleon+core
nucleus are calculated in
subsection \ref{subsec:trans} for electric transitions
with multipolarity $\lambda$. 
They only depend on the
asymptotic normalization of the bound state wave function and
on radial integrals with the asymptotic wave functions
since the radial integrals are dominated by the contribution from
outside the nuclear potential. In the square-well model explicit
expressions for the ratio of the interior to the exterior contributions
are derived.
An alternative calculation of $E1$ transition integrals with the
help of a commutator relation is presented in subsection \ref{subsec:comrel}.
The strength functions are related 
to cross sections of photo-nuclear reactions in
subsection \ref{subsec:xs} and the high-energy behaviour is discussed.
The effect of the nucleon-core potential is considered
in subsection \ref{subsec:scalen}. 
The effective-range expansion allows to 
parametrize the effects of the continuum interaction 
in a suitable way to study systematically its influence on the
transition strength.
In Section \ref{sec:rrisf} the reduced transition probabilities
are expressed in terms of characteristic shape functions that
depend on certain reduced integrals. The dependence of the
shape functions on the scaling parameters is studied in various
limits. Systematic variations of the shape functions are discussed
for neutron+core systems in subsections \ref{subsec:n+core}
and \ref{subsec:fsinc} without and with FSI, respectively,
where analytical results are obtained.
Proton+core systems are treated numerically
in subsection \ref{subsec:p+core}.
The relation of the total excitation strength to the 
properties of the bound state
with the help of sum rules
is considered in section \ref{sec:totsum}.
The findings of the model with asymptotic wave functions
are corroborated in more realistic
calculations using wave function generated from Woods-Saxon potentials
in section \ref{sec:examples}.
The implication
of the nucleon-core interaction in the continuum state
on the ANC method is discussed
in section \ref{sec:anc}.
We close with a summary and conclusions.
The appendix contains detailed derivations and explicit expressions
of our analytic calculations.

\section{Nucleon+core model}
\label{sec:ncm}

Exotic nuclei close to the driplines often exhibit a pronounced
structure with a nucleon $b$ (proton or neutron) 
weakly bound to a core $c$. 
They are often well
described by simple single-particle models
that are able to explain 
the basic features of low-energy excitations.
For small separation energies $S_{b}$ of the nucleon and
low orbital angular momenta $l$ the exotic nucleus $a$ develops
a proton or neutron halo where the nucleon wave function extends
to large radii and there is a large probability of finding
the nucleon outside the classically allowed region of the
potential $V_{bc}$. Matrix elements for electromagnetic transitions
only depend on a small number of characteristic scaling parameters.
 
\subsection{Scaling parameters of halo nuclei}
\label{subsec:scale}

The main scale is set by the nucleon separation energy $S_{b}$
($b=n,p$)
that is related to an inverse decay length 
\begin{equation}
 q = \frac{\sqrt{2\mu S_{b}}}{\hbar} 
\end{equation}
of the bound state
with the reduced mass $\mu = m_{b}m_{c}/(m_{b}+m_{c})$.
It becomes very small for typical halo nuclei with small
binding energies. The scattering state is characterized by 
the momentum $\hbar k$ that is related to the relative energy
in the continuum $E$ by
\begin{equation}
 k = \frac{\sqrt{2\mu E}}{\hbar}  \: .
\end{equation}
A third parameter is the size of the nuclear core given
by the range of the core-nucleon potential $R$.
These three parameters allow to define the dimensionless
quantities
\begin{equation}
 \gamma = q R \qquad \mbox{and} \qquad \kappa = k R
\end{equation}
and the ratio
\begin{equation}
 x = \frac{\kappa}{\gamma} = \frac{k}{q} = \sqrt{\frac{E}{S_{b}}} 
\end{equation}
independent of $R$.
It can be considered as a definition of halo nuclei that
the  parameter $\gamma$ is small. This means that the extension of the 
wave function, characterized by $1/q$, is much larger than R. 
The parameter $\gamma$ can be used as
a convenient expansion parameter in a systematic approach to calculate
matrix elements. On the other hand, the
relevant range of the parameter $x$ 
extends from zero to a value of the order of one.
For larger values of $x$ other degrees of freedom, like
core excitations, will come into play and 
tend to invalidate the simple model.

For proton+core systems an additional scale enters the problem set by
the Gamov energy
\begin{equation}
 E_{G} = \left(\frac{Z_{b}Z_{c}e^{2}}{\hbar}\right)^{2}
 \frac{\mu}{2} 
\end{equation}
or the nuclear Bohr radius 
\begin{equation} \label{eq:nbr}
 a_{N} = \frac{\hbar}{\sqrt{2\mu E_{G}}}
 = \frac{\hbar^{2}}{Z_{b}Z_{c}e^{2}\mu}
\end{equation}
with charge numbers $Z_{b}$ and $Z_{c}$ of the nucleon
$b$ and the core $c$.
They serve to define the Sommerfeld parameters
\begin{equation}
 \eta_{i} = \frac{1}{qa_{N}} 
 = \sqrt{\frac{E_{G}}{S_{b}}}
 \qquad \mbox{and} \qquad 
 \eta_{f} = \frac{1}{ka_{N}} 
 = \sqrt{\frac{E_{G}}{E}} 
\end{equation}
of the bound ($i$) and the scattering ($f$) state
with the relation
\begin{equation}
 x \eta_{f} = \eta_{i} \: .
\end{equation}
The nuclear Bohr radius (\ref{eq:nbr}) is related to
another important parameter: 
half the distance of closest approach in a head on collision
with energy $E$. It is given by $a_{0}=1/(k^{2} a_{N})$.
Typical features of a halo system essentially depend on the three
independent parameters
$\gamma$, $x$, and $\eta_{i}$. 
The characteristic scaling parameters are summarized in Table~\ref{tab:1}.
For a neutron+core system one obviously has
$\eta_{i}=0$. In case of a proton+core system with
a large charge number
of the core and/or small binding energy $S_{b}$
the parameter $\eta_{i}$ can become quite large.

\begin{table}
\caption{\label{tab:1}Characteristic scaling parameters for electromagnetic
strength in single-particle halo nuclei.}
\begin{tabular}{ccc}
 \hline 
 origin & energy scale & dimensionless parameter \\ 
 \hline 
 bound state & 
 $\displaystyle S_{b}= \frac{\hbar^{2}q^{2}}{2\mu}$ & 
 $\gamma = q R$ \\
 & one-nucleon separation energy & \\
 \hline
 scattering state & 
 $\displaystyle E= \frac{\hbar^{2}k^{2}}{2\mu}$ &  
 $\displaystyle x =
 \frac{\kappa}{\gamma} = \frac{k}{q} = \sqrt{\frac{E}{S_{b}}}$  \\ 
 & nucleon-core relative energy & with $\kappa = kR$\\
 \hline
 Coulomb field & 
 $\displaystyle 
 E_{G}= \left( \frac{Z_{b}Z_{c}e^{2}}{\hbar}\right)^{2} \frac{\mu}{2}
 = \frac{\hbar^{2}}{2\mu a_{N}^{2}}$ &
 $\displaystyle
 \eta_{i} = \sqrt{\frac{E_{G}}{S_{b}}} = \frac{1}{qa_{N}} = x \eta_{f}$ \\
 & Gamov energy & 
 \\
 \hline 
\end{tabular}
\end{table}

\subsection{Probabilities in nucleon+core systems}
\label{subsec:prob}

Common to all nucleon+core nuclei with small binding energy
is the large probability in the bound state
of finding the nucleon outside the range of the nuclear potential.
A similar observation is made for the scattering state. For halo
nuclei the nucleon does not penetrate strongly into the range
of the nuclear potential that describes the halo nature of
the bound state.

\begin{figure}
\begin{center}
\includegraphics[width=135mm]{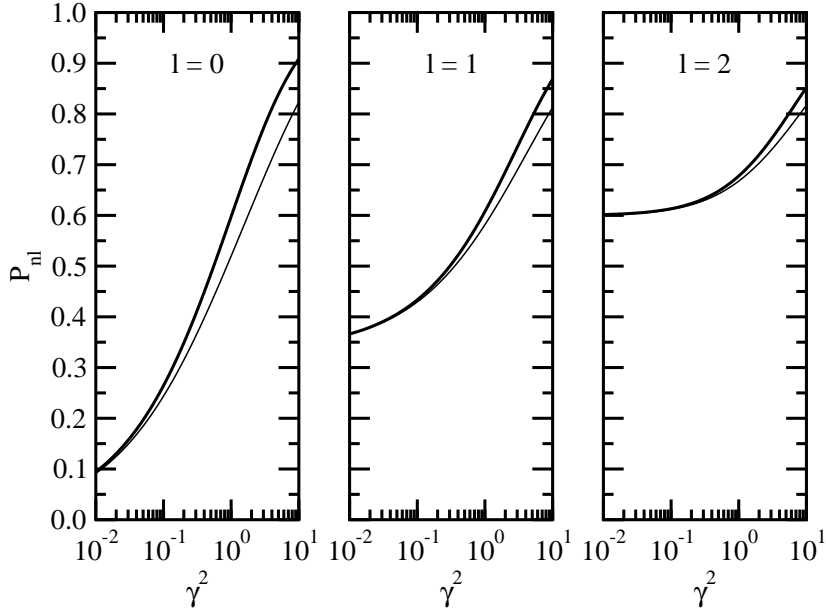}
\end{center}
\caption{\label{fig:0a} 
Probability $P_{nl}$ of finding a neutron with separation energy
$S_{n}$ inside the radius $R$ of a square-well potential
as a function of the parameter $\gamma^{2} = 2\mu S_{n}R^{2}/\hbar^{2}$
for orbital angular momenta $l=0,1,2$ and principal quantum numbers
$n=1$ (thick line) and $n=2$ (thin line).
}
\end{figure}

In a simple neutron+core model assuming a square-well potential
of radius $R$
the probability $P_{nl}$ of finding the neutron with 
principal quantum number $n$ ($=$ number of nodes of the radial
wave function including the node at $r=0$) and orbital angular
momentum $l$ inside the range of the potential can be
calculated analytically (see Appendix \ref{app:A}).
It essentially depends on the parameter $\gamma$.
One finds that the typical halo structure appears only for
low $l$, i.e.\ $s$, $p$  waves, and small neutron 
separation energies $S_{n}$, i.e.\ small $\gamma$.
Figure~\ref{fig:0a} clearly shows the
increase of the probability to find the neutron outside the
range of the potential with decreasing $\gamma$.
For larger values of $l$ the
centrifugal barrier hinders the occurence of a halo structure
and the penetration into the classically forbidden region
is reduced.
A larger number of nodes in the wave function 
increases the halo effect again. This effect is most pronounced for 
s waves. In the extreme halo limit $\gamma \to 0$, however, there
is no dependence on $n$ any more and the probability 
approaches finite values of 
$0$, $1/3$, and $3/5$
for $s$, $p$, and $d$ waves, respectively.
In case of a more realistic Woods-Saxon shape of the nuclear potential
the probability is even smaller as in a square-well 
potential with the same radius \cite{Liu04}. 
This is easily understood since the depth of the potential
is reduced inside the radius and increased outside the radius. As a
consequence the probability is shifted to larger radii.

\begin{figure}
\begin{center}
\includegraphics[width=135mm]{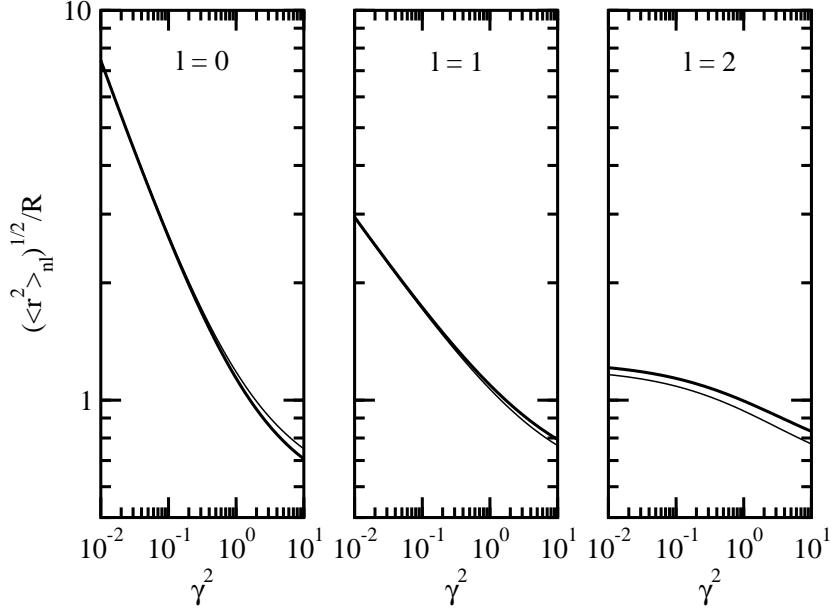}
\end{center}
\caption{\label{fig:0c} 
Root-mean-square radius $\langle r^{2} \rangle_{nl}^{\frac{1}{2}}$ 
of a neutron with separation energy
$S_{n}$ in a square-well potential of radius $R$
as a function of the parameter $\gamma^{2} = 2\mu S_{n}R^{2}/\hbar^{2}$
for orbital angular momenta $l=0,1,2$ and principal quantum numbers
$n=1$ (thick line) and $n=2$ (thin line).}
\end{figure}

\begin{figure}
\begin{center}
\includegraphics[width=135mm]{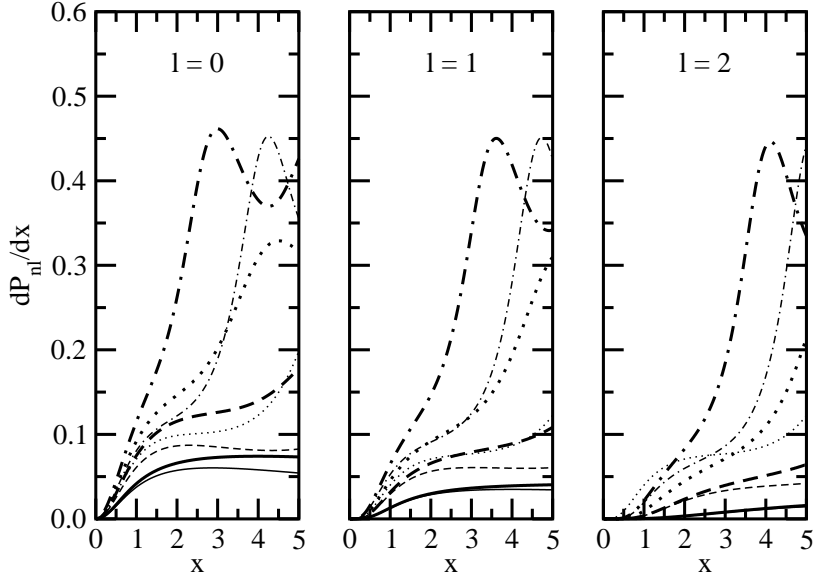}
\end{center}
\caption{\label{fig:0b} 
Differential probability $dP_{nl}/dx$ 
as a function of $x=\kappa/\gamma$ 
of finding a neutron with orbital angular momenta $l=0,1,2$ 
inside the radius $R$ of a square-well potential
that binds a neutron with
separation energy $S_{n}$, orbital angular momentum $l$
and principal quantum numbers
$n=1$ (thick lines) and $n=2$ (thin lines)
for various values of the parameter
$\gamma^{2} = 2\mu S_{n}R^{2}/\hbar^{2}=0.1$ (solid line),
$0.5$ (dashed line), $1.0$ (dotted line), $2.0$ (dot-dashed line).
}
\end{figure}

A different way to characterize the halo effect for small orbital angular
momenta is the dependence of the root-mean-square (rms) radius 
$\sqrt{\langle r^{2} \rangle_{l}}$ on the parameter $\gamma$. Explicit 
expressions of the rms radius for $l=0$, $1$, and $2$ in the square-well
model were given
in Ref.\ \cite{Ham98}, see also \cite{Liu04}. However, simpler
expressions for arbitrary values of $l$ can be obtained
(see Appendix \ref{app:A}). The scaling behaviour of the
rms radius is well known, see, e.g.,
\cite{Han87,Rii92,Rii94,Han95,Fed93,Rii00,Jen04} 
and Figure \ref{fig:0c}. For $l=0$ and $l=1$ one finds the scaling laws
\begin{eqnarray} \label{eq:rms}
 \langle r^{2} \rangle_{l} \to
 \left\{ \begin{array}{lll}
 \frac{R^{2}}{\gamma^{2}} & \mbox{if} & l = 0 \\ 
 \frac{5R^{2}}{6\gamma} & \mbox{if} & l = 1
 \end{array} \right.
\end{eqnarray}
for $\gamma \to 0$. For $l\geq 2$ the rms radius approaches
a finite value. The divergence of $\langle r^{2} \rangle_{l}$
for $l=0$ and $1$ is the typical sign of the halo nature.

Large values for $l$ and small parameters $\gamma$ 
prevent the neutron in a continuum state to enter into
the nuclear interior
as long as there is no resonance. In figure \ref{fig:0b}
the corresponding differential probability $dP_{nl}/dx$ 
(see Appendix \ref{app:A}) as a function
of  the ratio $x = \kappa/\gamma$ is shown for various parameters $\gamma$
and $l$. The explicit expression (\ref{eq:pnlcont})
shows a scaling proportional to $\gamma$ and inversely
proportional to the penetrability. For larger $x$ where resonances
in the scattering occur the probability exhibits clear maxima.
However, the neutron does not 
penetrate deeply into the nuclear interior for small $x$,
small $\gamma$ and large $l$.
Again, a larger number of nodes in the bound-state wave function
that determines the potential depth also reduces the
probability to find the neutron in the potential well
for continuum states.

In the case of proton+core systems the probabilities of finding
the nucleon inside and outside the range of the potential, respectively,
cannot be calculated analytically in the model with a
square-well potential. Obviously, some qualitative changes are expected.
Due to the additional Coulomb barrier the proton penetrates less
deeply into the classically forbidden region and the
probability for the proton to be inside the nuclear radius is enlarged
for bound states. On the other hand, a proton in a continuum state
will be found less probable inside the range of the nuclear potential
as compared to a corresponding neutron with the same energy.
This is described by the Coulomb penetration factor.

Since the nucleon in halo systems can be found predominantly
outside the range of the nuclear potential one can expect that
the relevant transition matrix elements mainly depend on the asymptotics
of the bound and scattering wave functions. This is especially true
for electric multipole transitions that contain an additional
$r^{\lambda}$ dependence enhancing contributions from large radii.

\subsection{Reduced transition probabilities and radial integrals}
\label{subsec:trans}

The reduced transition probability 
$dB(\pi \lambda) / dE$ of multipolarity $\pi \lambda$ ($\pi=E,M$;
$\lambda=1,2,\dots$)
for the electromagnetic breakup  of the nucleus $a$ 
into $b+c$ with relative energy $E$
is the basic quantity of our study. It 
contains the information
on the nuclear structure in the initial ground state and the 
interaction in
the final continuum state. This strength function 
determines the response of the system to a photon and 
enters the expressions for the corresponding cross sections.
It can be extracted
from experimental data in order to compare
directly to predictions of nuclear models.

Assuming the nucleon+core picture for the halo nucleus the
reduced transition probabilities are easily expressed in terms of
certain radial integrals
with the wave functions in the 
initial (bound) and final (scattering) states denoted
by $i$ and $f$, respectively, in the following.
In general, the spin $s=1/2$ of the nucleon $b$
couples with the orbital angular momentum $l_{i/f}$ 
to the total angular momentum $j_{i/f}$.
The total angular momentum $J_{i/f}$
of the system $a=b+c$ is obtained by coupling 
$j_{i/f}$ with the spin of the core $j_{c}$.

Usually there are various combinations
of $j_{i/f}$ and $j_{c}$ that are possible to contribute to
a given  value of $J_{i/f}$.
The reduced transition probability for a specific 
electromagnetic transition
$\pi \lambda$ 
to a final state with momentum $\hbar k$ in the
continuum is given by
\begin{eqnarray} \label{eq:dbde}
  \lefteqn{\frac{dB}{dE} (\pi\lambda, J_{i} s  \to k J_{f} s)
  =}
\\ \nonumber & &  
\frac{2J_{f}+1}{2J_{i}+1}  \sum_{j_{f} l_{f}}
 \left| \sum_{j_{i} l_{i} j_{c}} 
 \langle k J_{f}j_{f}l_{f}s j_{c} || {\mathcal M}(\pi\lambda) 
  || J_{i} j_{i} l_{i} s j_{c} \rangle \right|^{2}
 \frac{\mu k}{(2\pi)^{3}\hbar^{2}}
\end{eqnarray}
depending on reduced multipole matrix elements.
In the following we will only consider electric excitations ($\pi = E$)
with multipole operator
\begin{equation}
 {\mathcal M}(E\lambda \mu) = 
 Z_{\rm eff}^{(\lambda)}e  r^{\lambda} Y_{\lambda \mu}(\hat{r})
\end{equation}
that dominate the continuum breakup of exotic nuclei.
The effective charge number is given by
\begin{equation}
 Z_{\rm eff}^{(\lambda)} = 
 Z_{b}\left(\frac{m_{c}}{m_{b}+m_{c}}\right)^{\lambda}
 +Z_{c}\left(-\frac{m_{b}}{m_{b}+m_{c}}\right)^{\lambda}  \: .
\end{equation}
For proton+core systems the effective charge numbers for
$E1$ and $E2$ transitions are of comparable magnitude and, generally,
one has to consider both contributions 
in the cross sections for Coulomb breakup,
photo dissociation or radiative capture. In the case of a neutron+core
system, the $E2$ effective charge number is suppressed by a factor
$1/A$ as compared to $E1$ since $Z_{b}=0$. $E1$ transitions
dominate the low-lying electromagnetic strength and the $E2$ contribution
can be neglected.
Neverthess we include the $E2$ case here for completeness.
The strong reduction of contributions from higher multipolarities in the
neutron+core case was noticed before, e.g.\ in
Refs.\ \cite{Typ01a,Ber92}.

In the single-particle model
the wave functions of the initial and final state are given by
\begin{eqnarray}
 \Phi_{i}(\vec{r}) 
 & = & \langle \vec{r} | J_{i} j_{i} l_{i} s j_{c} \rangle 
 \\ \nonumber 
 & = & \frac{1}{r}
 \sum_{m_{i}m_{c}} (j_{i} \: m_{i} \: j_{c} \: m_{c} | J_{i} \: M_{i} )
 f_{J_{i}j_{i}l_{i}}^{j_{c}}(r) {\mathcal Y}_{j_{i}m_{i}}^{l_{i}s}
(\hat{r}) \phi_{j_{c}m_{c}}
\end{eqnarray}
and
\begin{eqnarray}
 \Phi_{f}(\vec{r})
 & = & \langle \vec{r} | \vec{k} J_{f} j_{f} l_{f} s j_{c} \rangle 
 \\ \nonumber 
 & = & \frac{4\pi}{kr}
 \sum_{m_{f}m_{c}} (j_{f} \: m_{f} \: j_{c} \: m_{c} | J_{f} \: M_{f} )
 g_{J_{f}j_{f}l_{f}}^{j_{c}}(r) i^{l_{f}} 
 Y_{l_{f}m_{f}}^{\ast}(\hat{k})
 {\mathcal Y}_{j_{f}m_{f}}^{l_{f}s}
(\hat{r}) \phi_{j_{c}m_{c}} \: ,
\end{eqnarray}
respectively, 
with the radial wave functions $f^{j_{c}}_{J_{i}j_{i}l_{i}}(r)$
and $g_{J_{f}j_{f}l_{f}}^{j_{c}}(r)$
and with the spinor spherical harmonics
\begin{equation}
{\mathcal Y}_{jm}^{ls} = \sum_{m_{l} m_{s}}
 ( l \: m_{l} \: s \: m_{s} | j \: m) Y_{lm}(\hat{r}) \chi_{s m_{s}} \: .
\end{equation}
The wave function of the core is denoted by $\phi_{j_{c}m_{c}}$.
The reduced matrix element in (\ref{eq:dbde}) can be expressed as
\begin{eqnarray} \label{eq:rme}
 \lefteqn{\langle k J_{f} j_{f} l_{f} s j_{c} || {\mathcal M}(E\lambda) 
 || J_{i} j_{i} l_{i} s j_{c}\rangle =}
 \\ \nonumber & & 
 \frac{4\pi Z_{\rm eff}^{(\lambda)}e}{k} 
 D_{J_{i}j_{i}l_{i}}^{J_{f}j_{f}l_{f}}(\lambda s j_{c}) \:
 (-i)^{l_{f}} I_{J_{i}j_{i}l_{i}}^{J_{f}j_{f}l_{f}}(\lambda j_{c})
\end{eqnarray}
with 
the angular momentum coupling coefficient
\begin{eqnarray}
D_{J_{i}j_{i}l_{i}}^{J_{f}j_{f}l_{f}}(\lambda s j_{c})  & = & 
(-1)^{s+j_{i}+l_{f}+\lambda} 
(-1)^{j_{c}+J_{i}+j_{f}+\lambda}
 (l_{i} \: 0 \: \lambda \: 0 | l_{f} \: 0 ) 
  \\ \nonumber & & 
\sqrt{2j_{i}+1} \sqrt{2l_{i}+1}
 \sqrt{2J_{i}+1} \sqrt{2j_{f}+1}
\\ \nonumber & & 
 \sqrt{\frac{2\lambda+1}{4\pi}} 
\left\{ \begin{array}{ccc}
 l_{i} & s & j_{i} \\ j_{f} & \lambda & l_{f}
 \end{array} \right\}
 \left\{ \begin{array}{ccc}
 j_{i} & j_{c} & J_{i} \\ J_{f} & \lambda & j_{f}
 \end{array} \right\} \: ,
\end{eqnarray}
and the radial integral
\begin{equation} \label{eq:radint}
 I_{J_{i}j_{i}l_{i}}^{J_{f}j_{f}l_{f}}(\lambda j_{c})
 = \int_{0}^{\infty} dr \: 
 g^{j_{c}\ast}_{J_{f}j_{f}l_{f}}(r)
 r^{\lambda} f^{j_{c}}_{J_{i}j_{i}l_{i}}(r)
\end{equation}
that contains the radial wave functions 
of the bound state $f^{j_{c}}_{J_{i}j_{i}l_{i}}(r)$
and the  continuum state $g^{j_{c}}_{J_{f}j_{f}l_{f}}(r)$,
respectively.

The asymptotic of the radial wave functions for the ground state
\begin{equation} \label{eq:asymb}
 f^{j_{c}}_{J_{i}j_{i}l_{i}}(r) 
 \to C^{j_{c}}_{J_{i}j_{i}l_{i}} W_{-\eta_{i}, l_{i}+1/2}
 (2qr)
\end{equation}
is determined by the asymptotic normalization coefficient 
$C^{j_{c}}_{J_{i}j_{i}l_{i}}$ of the true many-body wave function and
a Whittaker function $W_{-\eta_{i}, l_{i}+1/2}$ \cite{Abr65}.
The bound state is characterized by the parameters
$q$, $\eta_{i}$, and the orbital angular momentum $l_{i}$.
In the pure single-particle model for the nucleon+core system
the corresponding 
ANC $C^{j_{c}}_{J_{i}j_{i}l_{i}}(\rm sp)$ is determined by the
normalization of the wave function $f^{j_{c}}_{J_{i}j_{i}l_{i}}(r)$.
Because the radial wave function for small $r$ depends on the
nuclear potential of the single-particle model there is a model-dependence
of $C^{j_{c}}_{J_{i}j_{i}l_{i}}(\rm sp)$. Similarly, the 
spectroscopic factor $S^{j_{c}}_{J_{i}j_{i}l_{i}}$ for the
nucleon+core configuration 
depends on the single-particle model that is used for the
calculation. In contrast, the
actual ANC of the true many-body wave function
\begin{equation}
 C^{j_{c}}_{J_{i}j_{i}l_{i}}
 = C^{j_{c}}_{J_{i}j_{i}l_{i}}({\rm sp}) 
 \left[ S^{j_{c}}_{J_{i}j_{i}l_{i}} \right]^{\frac{1}{2}}
\end{equation}
is a model-independent quantity. 
It is directly inferred from transfer reactions for example.
In the following we always assume
a spectroscopic factor of one, i.e.\ the single-particle ANC
and the true ANC are identical.

For the scattering state we have the asymptotic form
\begin{eqnarray} \label{eq:asyms}
 g^{j_{c}}_{J_{f}j_{f}l_{f}}(r) & \to &
  \exp\left[i(\sigma_{l_{f}}+\delta^{j_{c}}_{J_{f}j_{f}l_{f}})\right]
\\  \nonumber & & \times
 \left[ \cos (\delta^{j_{c}}_{J_{f}j_{f}l_{f}}) \: F_{l_{f}}(\eta_{f};kr) 
 + \sin ( \delta^{j_{c}}_{J_{f}j_{f}l_{f}}) \:  G_{l_{f}}(\eta_{f};kr) \right]
\end{eqnarray}
with regular and irregular Coulomb wave functions
$F_{l_{f}}$ and $G_{l_{f}}$ \cite{Abr65}, 
respectively, and Coulomb phase shifts
$\sigma_{l_{f}}$ that depend on 
the Sommerfeld parameter $\eta_{f}=\eta_{i}/x$.
Effects of the  nuclear interaction in the final-state
are contained in the nuclear phase shifts $\delta^{j_{c}}_{J_{f}j_{f}l_{f}}$.

In general the radial integral (\ref{eq:radint}) has to be calculated
with the relevant bound and scattering wave functions
with the correct asymptotics (\ref{eq:asymb}) and (\ref{eq:asyms}), 
respectively. 
They are obtained by
solving the Schr\"{o}dinger equation for a given potential, e.g.\
of Woods-Saxon form $V(r) = -V_{0}/(1+\exp[(r-R_{0})/a])$
with depth $V_{0}$, radius $R_{0}$ and diffuseness parameter $a$.
In the following calculations 
we will only consider nuclear central potentials and neglect 
contributions from the spin-orbit interaction to simplify
the discussion. 
In case of proton+core systems 
the contribution of the Coulomb potential
can be assumed to be that of a homogeneously charged sphere
of the same radius $R_{0}$ as the nuclear potential. The potential
depth $V_{0}$ has to be adjusted to give the correct separation
energy $S_{b}$ of the nucleon in the ground state of $a$.
However, in the continuum state the parameter $V_{0}$ can be
varied freely to investigate the dependence of the transition
strength on the final-state interaction. 

\begin{figure}
\begin{center}
\includegraphics[width=135mm]{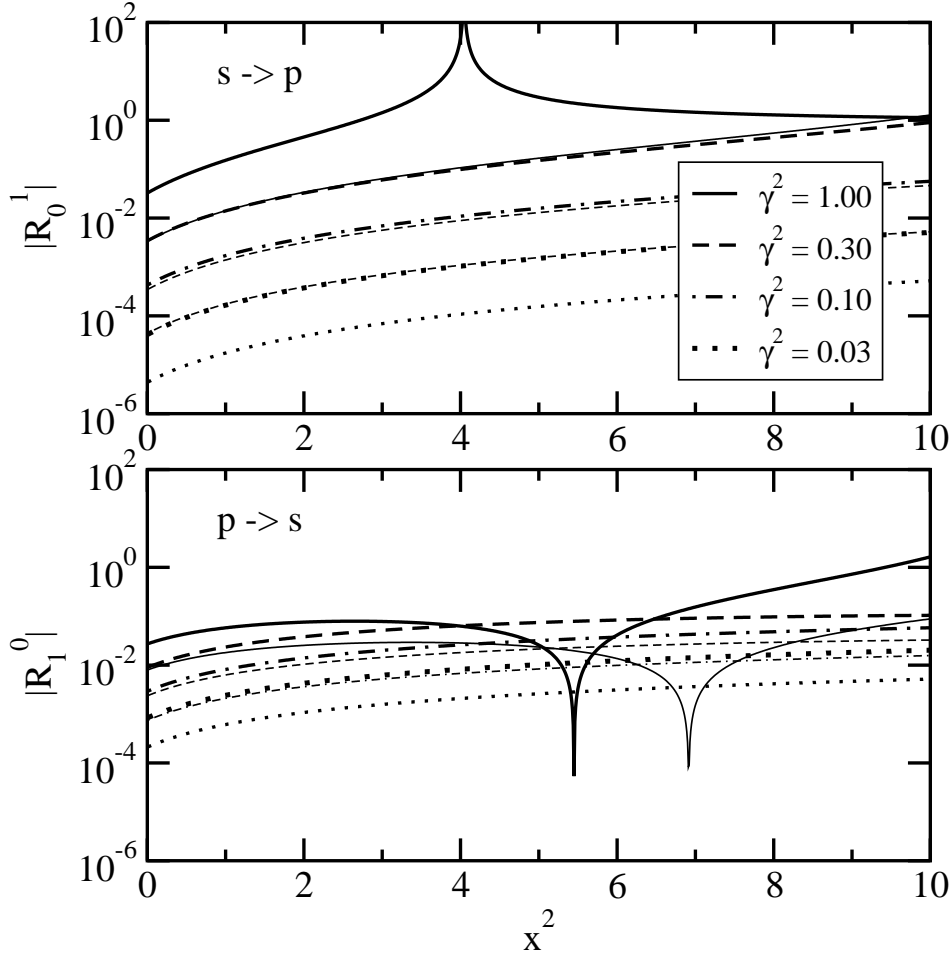}
\end{center}
\caption{\label{fig:0d}
Modulus of the ratio $R_{l_{i}}^{l_{f}}(1)$ of the interior and
exterior radial integrals for dipole transitions $s \to p$ (top) and
$p \to s$ (bottom) as a functions of $x^{2}=E/S_{n}$ for various values of
the parameter $\gamma^{2}= 2\mu S_{n}R^{2}/\hbar^{2}$. 
Thick (thin) lines correspond to the pricipal quantum number $n=1$ ($n=2$).}
\end{figure}

A remarkably good approximation 
for the nuclear interaction is the
square-well potential and many results can be derived 
analytically. They already show the main 
features for the scaling laws of the matrix elements. 
In this case the contributions to the integral
(\ref{eq:radint}) from the interior and exterior part can be calculated
independently to estimate their relevance in the transition matrix element.
In Appendix \ref{app:A4} explicit expressions for the dipole integrals
(suppressing irrelevant quantum numbers)
\begin{equation} 
 I_{l_{i}}^{l_{f}}(1 <)
 = \int_{0}^{R} dr \: 
 g^{\ast}_{l_{f}}(r) r f_{l_{i}}(r)
\quad \mbox{and} \quad
 I_{l_{i}}^{l_{f}}(1 >)
 = \int_{R}^{\infty} dr \: 
 g^{\ast}_{l_{f}}(r) r f_{l_{i}}(r)
\end{equation}
in the neutron+core case for $s \to p$ and $p \to s$ transitions
are derived. They only depend on the value of the radial wave functions
and their logarithmic derivative at the radius $R$ of the square well.
The ratio
\begin{equation}
 R_{l_{i}}^{l_{f}}(1) = \frac{I_{l_{i}}^{l_{f}}(1 <)}{I_{l_{i}}^{l_{f}}(1 >)}
\end{equation}
as a function of $x^{2}$ for $s \to p$ and $p \to s$ transitions,
respectively, is displayed in figure \ref{fig:0d} for various values of
$\gamma^{2}$
assuming the same depth of the square-well potential for the bound
and scattering state. For halo nuclei with small $\gamma^{2}$ the 
total radial integral is clearly dominated by the exterior integral
even at values $x^{2}=E/S_{n} \gg 1$. In the case of a resonance in the
scattering wave (e.g. for the $s \to p$ transition with $\gamma^{2} = 1$)
the ratio $R_{l_{i}}^{l_{f}}(1)$ shows a distinct peak since the
scattering wave function penetrates into the core of the nucleus.
In the limit $x \to 0$ one finds the scaling laws 
\begin{equation}
 R_{0}^{1}(1)  \to  \frac{\gamma^{4}}{2(2n-1)^{2}\pi^{2}}
 \quad \mbox{and} \quad
 R_{1}^{0}(1)  \to  -\frac{\gamma^{2}}{4n^{2}\pi^{2}}
\end{equation}
depending on the principal quantum number $n=1,2,\dots$ 
(see appendix \ref{app:A4}).
For larger values of $n$ the ratios
become smaller but they remain finite for a given $\gamma$ in the limit
$x \to 0$. As can be seen from figure \ref{fig:0d} these
scaling laws are also well satisfied for larger values of $x^{2}$
as long as there are no resonances or accidental zeros in the 
interior integral. For halo nuclei one 
clearly sees that the total radial integral is well
approximated by the exterior part with the asymptotic form of the
wave functions that is independent of the details of the potential.
This will be even more true for higher multipolarities due to the $r^{\lambda}$
factor in the radial integral.
Only the parameters $\gamma$, $\kappa$, the ANC $C_{l_{i}}$ 
and the phase shift $\delta_{l_{f}}$ are really relevant.

\subsection{Dipole integrals and commutator relations}
\label{subsec:comrel}

In the case of electric dipole transitions the relevant reduced matrix element
(\ref{eq:rme}) can also be calculated with the help of the
commutator relation \cite{Ehr27,Gor29}
\begin{equation} \label{eq:commrel}
 \langle \Phi_{f} | \left[ H , \left[ H, \vec{r} \right] \right] 
| \Phi_{i} \rangle
 =  (E_{f}-E_{i})^{2} \langle \Phi_{f} | \vec{r} | \Phi_{i} \rangle \: .
\end{equation}
Introducing the scaling parameters $\gamma$ and $x$, we
find
\begin{equation} \label{eq:dme}
\langle \Phi_{f} | \vec{r} | \Phi_{i} \rangle 
 = \frac{4\mu^{2}R^{4}}{\hbar^{4}\gamma^{4}(1+x^{2})^{2}}
 \langle \Phi_{f} | \left[ H , \left[ H, \vec{r} \right] \right] 
| \Phi_{i} \rangle
\end{equation} 
for the dipole matrix element. This equation
directly shows the occurence
of a pole at $x^{2} = -1$, 
i.e., $E=-S_{b}$. This derivation is much more transparent than
the corresponding discussion in \cite{Jen98} for $E1$ radiative capture
reactions.
For halo systems with small nucleon
separation energies $S_{b}$ one immediately sees that a series expansion
of the dipole matrix element (\ref{eq:dme})
in the parameter $x$ (or the energy $E$) 
is only of limited value
because of the small radius of convergence 
$x_{\rm conv} = 1$
($E_{\rm conv} = S_{b}$). It would be more advantageous to expand the
matrix element with the double commutator 
$\left[ H , \left[ H, \vec{r} \right]\right]$ in a power series in $x$.

The double commutator can be calculated assuming different forms
of the potential $V$ in the Hamiltonian $H=p^{2}/(2\mu)+V$ of the system.
The simplest case is a central potential $V(r)$
that commutes with $\vec{r}$, i.e.\ $\left[ V, \vec{r} \right] = 0$. 
With $\left[ H , \left[ H, \vec{r} \right] \right]
= \hbar^{2} \vec{\nabla} V(r)/\mu$
the radial integral (\ref{eq:radint}) can be expressed as
\begin{equation} \label{eq:ricom}
 I_{J_{i}j_{i}l_{i}}^{J_{f}j_{f}l_{f}}(1 j_{c})
 = \frac{4\mu R^{4}}{\hbar^{2}\gamma^{4}(1+x^{2})^{2}} 
 \int_{0}^{\infty} dr \: 
 g^{j_{c}\ast}_{J_{f}j_{f}l_{f}}(r)
 \left( \frac{d}{dr} V(r) \right) f^{j_{c}}_{J_{i}j_{i}l_{i}}(r) \: .
\end{equation}

In the neutron+core case with a square-well potential 
\begin{equation}
 V(r) = - V_{0} \theta(R-r)
\end{equation}
we find that the radial integral is given by
\begin{equation} \label{eq:radintgf}
 I_{J_{i}j_{i}l_{i}}^{J_{f}j_{f}l_{f}}(1 j_{c})
 = \frac{2R^{2}v}{(\gamma^{2}+\kappa^{2})^{2}} 
 g^{j_{c}\ast}_{J_{f}j_{f}l_{f}}(R)
 f^{j_{c}}_{J_{i}j_{i}l_{i}}(R)
\end{equation}
It depends only on the values of the initial and final radial wave function
at the radius $R$
and the depth of the potential in the dimensionless parameter
\begin{equation}
 v = \frac{2 \mu V_{0} R^{2}}{\hbar^{2}} 
\end{equation}
that characterizes the strength of the interaction.
It is not surprising that this result can also be derived
by evaluating the radial integral (\ref{eq:radint}) directly
(see appendix \ref{app:A4}).
For more realistic potentials the derivative $dV/(dr)$
also peaks close to the nuclear radius and the radial integral is
mainly sensitive to the bound and scattering wave functions near the
nuclear surface.

For finite values of $R$ the radial integral (\ref{eq:radintgf})
shows a $x^{-4}$ dependence for large energies independent of the
orbital angular momenta in the initial and final state. 
(The wave function $g_{l_{i}}$ is independent of $x$ and $f_{l_{f}}$
shows an oscillatory behaviour for large $x$.)
This $E^{-2}$ dependence is sufficient for the convergence of
the non energy-weighted and energy-weighted sum rules as discussed in
section \ref{sec:totsum}.
%

\begin{figure}
\begin{center}
\includegraphics[width=135mm]{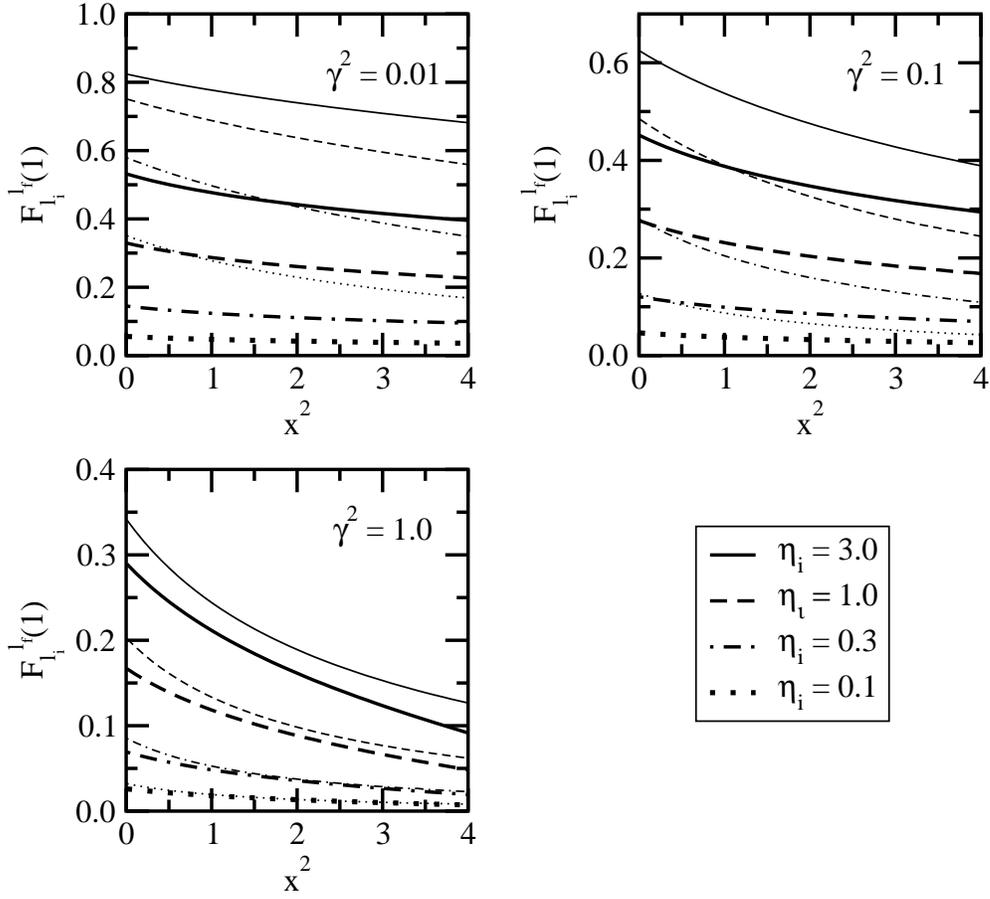}
\end{center}
\caption{\label{fig:0e}
Ratio $F_{l_{i}}^{l_{f}}(1)$ of Eq.\ (\ref{eq:coulc}) 
as a function of $x^{2}=E/S_{p}$
for various values of $\gamma^{2}= 2\mu S_{p}R^{2}/\hbar^{2}$ and $\eta_{i}$
for the bound state without a node (principal quantum number $n=1$).
Thick (thin) lines correspond to transitions $s \to p$ ($p \to s$).}
\end{figure}

In the proton+core case the potential is well approximated by
\begin{equation}
 V(r) = -V_{0} \: \theta(R-r)
 + \frac{Z_{c}e^{2}}{r} \: \theta(r-R)
\end{equation}
and the radial integral becomes
\begin{eqnarray} \label{eq:icoul}
 I_{J_{i}j_{i}l_{i}}^{J_{f}j_{f}l_{f}}(1 j_{c})
 & = & \frac{2R^{2}}{(\gamma^{2}+\kappa^{2})^{2}}
 \\ \nonumber & &
 \times
 \left[ (v+2\gamma \eta_{i}) g^{j_{c}\ast}_{J_{f}j_{f}l_{f}}(R)
 f^{j_{c}}_{J_{i}j_{i}l_{i}}(R)
 -  2 \gamma \eta_{i}K_{J_{i}j_{i}l_{i}}^{J_{f}j_{f}l_{f}}(1 j_{c}, R) \right]
\end{eqnarray}
with the 
integral
\begin{equation}
 K_{J_{i}j_{i}l_{i}}^{J_{f}j_{f}l_{f}}(1 j_{c}, R) 
 = R  \int_{R}^{\infty} \frac{dr}{r^{2}} \: 
 g^{j_{c}\ast}_{J_{f}j_{f}l_{f}}(r)
 f^{j_{c}}_{J_{i}j_{i}l_{i}}(r) 
\end{equation} 
that converges much more rapidly than the usual $E1$ transition
integral.
This contribution only depends on the parameters $\gamma$, $\eta_{i}$,
and $x$.
The ratio 
\begin{equation} \label{eq:coulc}
  F_{l_{i}}^{l_{f}}(1) =
 \frac{2 \gamma \eta_{i}}{(v+2\gamma\eta_{i})}
 \frac{K_{l_{i}}^{l_{f}}(1,R)}{g^{\ast}_{l_{f}}(R)f_{l_{i}}(R)} 
\end{equation}
of the two contributions in (\ref{eq:icoul}) is depicted in Fig.\
\ref{fig:0e} as a function of $x^{2}$ for typical values of $\gamma^{2}$
and $\eta_{i}$ for both $s \to p$ and $p \to s$ transitions.
It is clearly seen that the ratio $F_{l_{i}}^{l_{f}}(1)$ of the two
contributions to the total radial integral increases with increasing 
$\eta_{i}$ and decreasing $\gamma$. This is expected since Coulomb
effects become stronger for larger $\eta_{i}$ and the bound state
wave function at larger radii is less steep for smaller $\gamma$. 
The ratio (\ref{eq:coulc}) decreases for larger values of $x^{2}$ with
a larger effect for large values of $\gamma$. Generally, transitions $p \to s$
are more strongly 
affected by the correction than transitions $s \to p$, however,
for large $\gamma$ (less halo effect) the difference becomes smaller.

Ususally, one can expect that the Hamiltonian 
$H=p^{2}(2\mu)+V$ contains more general potentials that do not commute
with $\vec{r}$ \cite{Lev50}, e.g.,
\begin{equation} \label{eq:vls}
 V=V_{c}(r)
+V_{ls}(r)\vec{\ell} \cdot \vec{s}
\end{equation}
with a central potential $V_{c}(r)$ and
a spin-orbit contribution $V_{ls}(r)$. In this case the double commutator
in (\ref{eq:commrel}) can be calculated explictly and a rather complicated
radial integral is obtained. Alternatively, we can
introduce $H_{f} = P_{f}HP_{f}$
and $H_{i}=P_{i}HP_{i}$ with projection operators $P_{f}$ and $P_{f}$ on the
final and initial state, respectively. Then the commutator 
relation (\ref{eq:commrel}) can be generalized to
\begin{eqnarray} \label{eq:commrel2}
  (E_{f}-E_{i})^{2} \langle \Phi_{f} | \vec{r} | \Phi_{i} \rangle
 & = & 
\langle \Phi_{f} | \left(
 \left[ \Sigma , \left[ \Sigma, \vec{r} \right] \right] 
 + \left[ \Sigma, \{ \Delta, \vec{r} \} \right]
\right.  \\ \nonumber & &  \left.
 + \{ \Delta, \left[ \Sigma, \vec{r} \right] \}
 + \{ \Delta, \{ \Delta, \vec{r} \} \} \right)
| \Phi_{i} \rangle
\end{eqnarray}
where
$\Sigma=(H_{f}+H_{i})/2$, $\Delta = (H_{f}-H_{i})/2$ and $\{ . ,. \}$ denotes
the anticommutator. For the potential (\ref{eq:vls}) we find 
$H_{k} = p^{2}/(2\mu)+V_{k}(r)$ with $k=i,f$ and
\begin{equation}
 V_{k}(r)  =   V_{c}(r) + \frac{V_{ls}(r)}{2}
 \left[ j_{k}(j_{k}+1)-l_{k}(l_{k}+1)-s(s+1)\right]
\end{equation}
where $j_{k}$, $l_{k}$, and $s$ are the total angular momentum, the orbital
angular momentum, and the spin of the nucleon, respectively. The central
potentials $V_{k}(r)$ with $\left[ V_{k}(r), \vec{r} \right]=0$ are
now different in the bound and scattering states, however, the various
contributions in (\ref{eq:commrel2}) can be calculated easily. Assuming 
a square-well potential for both $V_{c}(R)$ and $V_{ls}(r)$ with the same
radius but different depths the radial integral corresponding
to (\ref{eq:radintgf}) becomes a more complicated 
expression that contains also derivatives of the wave functions at the radius
$R$. Explicit expresssions are obtained from adding the
interior and exterior integrals as derived in appendix \ref{app:A4}.
Also other forms of the potentials, e.g., a surface-peaked spin-orbit
potential $V_{ls} = \mbox{const.} \times \delta(r-R)$, 
can be considered. However, we omit the details of the calculation.

\subsection{Cross sections}
\label{subsec:xs}

The reduced transition probability determines
the cross sections for photo-nuclear reactions.
The photo dissociation cross section
\begin{equation} \label{eq:sigabs}
 \sigma_{\pi \lambda}(a+\gamma \to b+c) = 
 \frac{\lambda+1}{\lambda} \frac{(2\pi)^{3}}{[(2\lambda+1)!!]^{2}}
 \left(\frac{E_{\gamma}}{\hbar c}\right)^{2\lambda-1}
 \frac{dB(\pi \lambda)}{dE}
\end{equation}
is proportional to $dB(\pi \lambda)/dE$ where
$E_{\gamma}= E+S_{b}$ is the
sum of the binding energy $S_{b}>0$ of the nucleus $a$ with respect
to the breakup into $b+c$ and the relative energy $E$
in the final state. The photo absorption cross section (\ref{eq:sigabs}) also
enters the cross section for
the electromagnetic dissociation reaction $a+X \to b+c+X$
during the scattering of an exotic nucleus $a$ on a target nucleus $X$.
A first-order calculation  gives
\begin{equation} \label{eq:cdxs}
 \frac{d^{2}\sigma}{d\Omega_{aX}dE} =
 \frac{1}{E_{\gamma}} \sum_{\pi\lambda} 
 \sigma_{\pi \lambda}(a+\gamma \to b+c) \frac{dn_{\pi \lambda}}{d\Omega_{aX}}
\end{equation}
with virtual photon numbers $dn_{\pi \lambda} / d\Omega_{aX}$
that can be calculated in the semiclassical theory or in prior-form
DWBA in the quantal approach.
The factorization of the cross section (\ref{eq:cdxs})
into contributions that are related to the  nuclear structure 
of the exotic nucleus $\sigma_{\pi \lambda}$ and to the excitation mechanism
$dn_{\pi \lambda} / d\Omega_{aX}$ is 
no longer valid if higher-order effects from the target-fragment
interactions are significant.
From the photo dissociation reaction it is also possible to
extract information on the
radiative capture reaction $b+c \to a +\gamma$
that for low energies is relevant for nuclear astrophysics.
The capture cross section
\begin{equation} \label{eq:detbal}
 \sigma_{\pi \lambda}(b+c \to a + \gamma)  = 
 \frac{2(2J_{a}+1)}{(2J_{b}+1)(2J_{c}+1)}
 \frac{k_{\gamma}^{2}}{k^{2}}
 \sigma_{\pi \lambda}(a + \gamma \to b + c)
\end{equation}
is obtained by applying the theorem of detailed balance.
The photo absorption cross section (\ref{eq:sigabs})
is multiplied with a factor that contains the spins $J_{i}$ of the particles
$i=a,b,c$, the photon momentum $\hbar k_{\gamma} = E_{\gamma}/c$,
and the final state momentum $\hbar k$.

For electric dipole transitions it is easy to find
the high-energy behaviour of the various cross sections.
From equation (\ref{eq:ricom}) the dependence 
$I_{l_{i}}^{l_{f}}(1) \propto x^{-4} \propto E^{-2}$ of the radial transition
integral is extracted. With (\ref{eq:dbde}) and (\ref{eq:rme})
one finds $dB(E1)/dE \propto x^{-1} |I_{l_{i}}^{l_{f}}(1)|^{2}  
\propto x^{-9}$ or $dB(E1)/dE \propto E^{-9/2}$. This corresponds to
a dependence $\sigma_{E1}(a+\gamma \to b +c) \propto E^{-7/2}$
and $\sigma_{E1}(b +c \to a + \gamma) \propto E^{-5/2}$
for the photo dissociation and radiative capture cross sections
at high energiess $E$, respectively.
We note that in the case of the photodissociation of atoms,
with the $1/r$ shape of the Coulomb potential,
one finds the same $E_{\gamma}^{-7/2}$ law, see 
\cite{Bet77}.

\subsection{Effective-range expansion}
\label{subsec:scalen}

The nucleon-core interaction that is responsible for the
binding of the system also generates structures in the continuum.
At low energies in the continuum the phase shifts are insensitive
to the details of the nuclear potential as long as no resonance
appears. They
are determined by only a few parameters that are given by
the effective-range approximation. 
E.g., in the case of charged-particle scattering
the low-energy phase shift $\delta_{l}$ in the partial wave 
with orbital angular momentum $l$ is determined
by the scattering length $a_{l}$ and the effective range
$r_{l}$ in the expansion \cite{Bru96}
\begin{equation} \label{eq:fre}
 C_{l}^{2}(\eta) k^{2l+1} \left[ \cot \delta_{l} 
 + \frac{2\eta h(\eta)}{C_{0}^{2}(\eta)}\right]
 = -\frac{1}{a_{l}} + \frac{r_{l}}{2} k^{2} + \dots
\end{equation}
with the function
\begin{eqnarray}
 h(\eta) & = & 
 \frac{1}{2} \left[ \psi (1+i\eta) + \psi(1-i\eta)\right] - \ln \eta
 \\ \nonumber 
 & = & \eta^{2} \sum_{n=1}^{\infty} \frac{1}{n(n^{2}+\eta^{2})} 
 - \gamma -\ln \eta 
\end{eqnarray}
that depends on the Sommerfeld parameter $\eta$ in the argument
of the Digamma function $\psi$ and
$\gamma=0.5772156649\dots$ is Euler's constant.
The constants $C_{l}^{2}$
for increasing $l$
are obtained by means of a recursion relation
\begin{equation} \label{eq:coulfac}
 C_{l}^{2}(\eta) = C_{l-1}^{2}(\eta) \left( 1 + \frac{\eta^{2}}{l^{2}}\right)
 \qquad \mbox{with} \qquad
 C_{0}^{2}(\eta) = \frac{2\pi\eta}{\exp(2\pi\eta) -1}  \: .
\end{equation}
Note that $a_{l}$ and $r_{l}$ for $l>0$ do not have the dimension
of a length.
The effective-range expansion can be used to express the
phase shifts directly as functions of the momentum $\hbar k$.

Taking only the scattering length $a_{l}$ and the effective range
$r_{l}$ into account one does not necessarily find a good approximation
of the phase shift $\delta_{l}$ over a wide range of 
energies in the continuum. 
However, the effective-range expansion motivates to introduce 
an energy-dependent function $b_{l}$ via the equation
\begin{equation}
 C_{l}^{2} x^{2l+1} \left[ \cot \delta_{l} 
 + \frac{2\eta h(\eta)}{C_{0}^{2}}\right]
 = -\frac{1}{b_{l}^{2l+1}} 
\end{equation}
with the dimensionless parameter 
$x=k/q$ where $q$ is constant for a given nucleus.
In general, the function $b_{l}$ 
depends on the momentum $\hbar k$
in the final state. 
It fully describes the effects of the
interaction in the scattering state at all energies.
The quantity $b_{l}$ varies slowly with the parameter $x=k/q$
as long as no resonance is approached.
For small $k$ one can identify $b_{l}$ with a reduced scattering length
since the usual scattering length is obtained in the limit
\begin{equation} \label{eq:abrel}
 a_{l} = \lim_{x\to 0} \left(\frac{b_{l}}{q}\right)^{2l+1}  
\end{equation}
and for $\delta_{l} = 0$ we obviously have $b_{l} = 0$.

\section{Reduced radial integrals and shape functions}
\label{sec:rrisf}

Considering the asymptotics of the radial wave functions
one can define dimensionless reduced radial integrals 
${\mathcal I}_{J_{i}j_{i}l_{i}}^{J_{f}j_{f}l_{f}}(\lambda j_{c})$
by the relation
\begin{equation}
 I_{J_{i}j_{i}l_{i}}^{J_{f}j_{f}l_{f}}(\lambda j_{c})
 = \frac{C^{j_{c}}_{J_{i}j_{i}l_{i}}}{q^{\lambda+1}} 
 \exp\left[i(\sigma_{l_{f}}+\delta^{j_{c}}_{J_{f}j_{f}l_{f}})\right]
 {\mathcal I}_{J_{i}j_{i}l_{i}}^{J_{f}j_{f}l_{f}}(\lambda j_{c}) \: .
\end{equation}
At low energies the main contribution to the radial integral
arises from radii $r$ larger than the radius of the nucleus.
This is especially true for halo nuclei where the probability
of finding the nucleon inside the range of the nuclear potential
is small (see subsection \ref{subsec:prob}).
Neglecting contributions from radii smaller than a cutoff radius $R$
the reduced radial integrals can be approximated by
\begin{eqnarray} \label{eq:idef}
 {\mathcal I}_{l_{i}}^{l_{f}}(\lambda) & = & 
  \left[ \cos (\delta_{l_{f}}) \:  {\mathcal F}_{l_{i}}^{l_{f}}(\lambda)
 + \sin ( \delta_{l_{f}}) \:  {\mathcal G}_{l_{i}}^{l_{f}}(\lambda)\right]
\end{eqnarray}
where
$ {\mathcal F}_{l_{i}}^{l_{f}}(\lambda)$
and
$  {\mathcal G}_{l_{i}}^{l_{f}}(\lambda)$
denote the real and the imaginary part of the function
\begin{eqnarray} \label{eq:hdef}
 {\mathcal H}_{l_{i}}^{l_{f}}(\lambda) 
 & = & q^{\lambda+1} 
 \int_{R}^{\infty} dr \: r^{\lambda} 
  \left[ F_{l_{f}}(\eta_{f};kr)+ i G_{l_{f}}(\eta_{f};kr)   \right]
 W_{-\eta_{i}, l_{i}+\frac{1}{2}} (2qr)  
 \\ \nonumber 
 & = & \gamma^{\lambda+1} 
 \int_{1}^{\infty} dt \: t^{\lambda} 
  \left[ F_{l_{f}}(\eta_{i}/x;x\gamma t)
     + i G_{l_{f}}(\eta_{i}/x;x\gamma t)   \right]
 W_{-\eta_{i}, l_{i}+\frac{1}{2}} (2\gamma t)  
 \: .
\end{eqnarray}
Here, as in the following, the quantum
numbers $J_{i}$, $j_{i}$, $J_{f}$, $j_{f}$ and $j_{c}$
have been suppressed if no confusion arises.
The nuclear phase shifts $\delta_{l_{f}}$ encode the effects of
the final-state interaction. For $\delta_{l_{f}}=0$
one obtains the results without the nuclear interaction between
the nucleon and the core. Note that the function
${\mathcal H}_{l_{i}}^{l_{f}}(\lambda)$ depends only on the
parameters $\gamma$, $\eta_{i}$ and $x$.

If there is only one fixed pair $(j_{i},l_{i})$ in the initial
state and similar $(j_{f},l_{f})$ in the final state the expression 
for the reduced transition probability reduces to
\begin{eqnarray} \label{eq:dbelde}
 \lefteqn{\frac{dB}{dE} (E\lambda, J_{i} s j_{c}  \to k J_{f} s j_{c})=}
 \\ \nonumber & & 
  \left[Z_{\rm eff}^{(\lambda)}e\right]^{2} 
 \frac{2\mu}{\pi\hbar^{2}} 
 \frac{2J_{f}+1}{2J_{i}+1}  
 \left[  D_{J_{i}j_{i}l_{i}}^{J_{f}j_{f}l_{f}}(\lambda s j_{c}) \right]^{2}
  \frac{\left|C^{j_{c}}_{J_{i}j_{i}l_{i}}\right|^{2}}{q^{2\lambda+3}}
 {\mathcal S}_{J_{i}j_{i}l_{i}}^{J_{f}j_{f}l_{f}}(\lambda j_{c})
\end{eqnarray}
with the dimensionless shape function of the transition strength
\begin{equation} \label{eq:sdef}
 {\mathcal S}_{J_{i}j_{i}l_{i}}^{J_{f}j_{f}l_{f}}(\lambda j_{c})
 =  \frac{1}{x} \left| 
 {\mathcal I}_{J_{i}j_{i}l_{i}}^{J_{f}j_{f}l_{f}}(\lambda j_{c})
\right|^{2}
\end{equation} 
that completely contains the dependence on the
momentum in the continuum. Similarly, one can define the 
characteristic shape functions
for the photo absorption
\begin{equation} \label{eq:sabs}
  {\mathcal S}_{J_{i}j_{i}l_{i}}^{J_{f}j_{f}l_{f}}({\rm abs},\lambda j_{c})
 = \left(1+x^{2}\right)^{2\lambda-1}
 {\mathcal S}_{J_{i}j_{i}l_{i}}^{J_{f}j_{f}l_{f}}(\lambda j_{c})
\end{equation}
and for the capture cross section
\begin{equation} \label{eq:scapt}
  {\mathcal S}_{J_{f}j_{f}l_{f}}^{J_{i}j_{i}l_{i}}({\rm capt},\lambda j_{c})
 = \frac{(1+x^{2})^{2\lambda+1}}{x^{2}}
 {\mathcal S}_{J_{i}j_{i}l_{i}}^{J_{f}j_{f}l_{f}}(\lambda j_{c}) \: .
\end{equation}
The expression (\ref{eq:dbelde})
directly shows
the dependence of the transition strength on the characteristic parameters
$C^{j_{c}}_{J_{i}j_{i}l_{i}}$ and $q$ of the ground state. For small
separation energies $S_{b}$ of the nucleon the strength becomes very large
since $q$ is a small number. Note that the ANC in general depends on $q$, too.
In a model for neutrons in a square-well potential 
one finds a dependence $C^{j_{c}}_{J_{i}j_{i}l_{i}} \propto \sqrt{q}$ 
for  $l_{i}=0$ 
and $C^{j_{c}}_{J_{i}j_{i}l_{i}} \propto q^{l_{i}}$ for $l_{i}>0$,
respectively,  for small $q$ (see Appendix \ref{app:A}).

When the nucleon $b$ is a neutron the reduced radial integrals can be
calculated analytically (see Appendix \ref{app:B}). Then the 
functions (\ref{eq:idef}) only
depend on the dimensionless variables $\gamma = qR$ and 
$\kappa = kR = x \gamma$ 
and the phase shift $\delta_{l_{f}}$ in the final state. 
It is found that the reduced radial integrals have the general form
\begin{eqnarray} \label{eq:iredgen}
  {\mathcal I}_{l_{i}}^{l_{f}}(\lambda) & = &
 \frac{\gamma \exp(-\gamma)}{(\gamma^{2}+\kappa^{2})^{\lambda+1}} 
 \left(\frac{\gamma}{\kappa}\right)^{l_{f}}
 \\ \nonumber & & \times
 \left[ {\mathcal R}_{l_{i}}^{(+)l_{f}}(\lambda) \cos (\kappa+\delta_{l_{f}}) 
 + {\mathcal R}_{l_{i}}^{(-)l_{f}}(\lambda) \sin(\kappa+\delta_{l_{f}})\right]
\end{eqnarray}
with polynomials/rational functions 
${\mathcal R}_{l_{i}}^{(\pm)l_{f}}(\lambda)$.
Explicit expressions are given in Appendix \ref{app:B}. In general they are 
complicated functions of $\gamma$ and $\kappa$. 

From the discussion in subsection \ref{subsec:comrel} a $x^{-4}$
dependence of the reduced radial integrals 
${\mathcal I}_{l_{i}}^{l_{f}}(1)$ for dipole transitions is expected
for large $x$. 
However, the integrals (\ref{eq:iredgen}) for $\lambda=1$
show a different high-$x$
behaviour and the convergence of the sum rules (see section
\ref{sec:totsum}) is not guaranteed. This is a consequence of neglecting
the interior contribution to the radial integral that becomes relevant
at high relative energies since the nucleon penetrates into the core.
Thus, the reduced radial integrals (\ref{eq:iredgen}) are only a
good approximation for not too high relative energies in the continuum.
In comparison, equation (\ref{eq:radintgf}) is  valid 
irrespective of the halo nature of the system and for all energies
but only 
for $\lambda =1$ and the square-well case; it contains all contributions 
to the radial integral from zero to infinity. Equation (\ref{eq:iredgen}) 
is a good approximation for halo nuclei at small relative energies, 
since the interior contributions are small 
in this case (cf.\ Fig.\ \ref{fig:0d}). 
It can also be applied easily to cases where
the potential is different in the bound and scattering states.
It is worthwhile to study some limiting cases.

\begin{table}
\caption{\label{tab:2}Characteristic shape functions 
${\mathcal S}_{l_{i}}^{l_{f}}(\lambda)$ for $\gamma=0$ and no final-state
interaction.}
\begin{tabular}{lll}
 \hline 
 $\lambda=0$ & $\lambda = 1$ & $\lambda=2$ \\
 \hline 
 $\displaystyle {\mathcal S}_{0}^{0}(0) = \frac{x}{(1+x^{2})^{2}}$ &
 $\displaystyle {\mathcal S}_{1}^{0}(1) = 
 \frac{x(3+x^{2})^{2}}{(1+x^{2})^{4}}$ &
 $\displaystyle {\mathcal S}_{2}^{0}(2) = 
 \frac{x(15+10x^{2}+3x^{4})^{2}}{(1+x^{2})^{6}}$ \\
 $\displaystyle {\mathcal S}_{1}^{1}(0) = \frac{x^{3}}{(1+x^{2})^{2}}$ &
 $\displaystyle {\mathcal S}_{0}^{1}(1) = \frac{4x^{3}}{(1+x^{2})^{4}}$ &
 $\displaystyle {\mathcal S}_{1}^{1}(2) = 
 \frac{4x^{3}(5+x^{2})^{2}}{(1+x^{2})^{6}}$ \\
  &
 $\displaystyle {\mathcal S}_{2}^{1}(1) = 
 \frac{x^{3}(5+3x^{2})^{2}}{(1+x^{2})^{4}}$ &
 \\
 &
 $\displaystyle {\mathcal S}_{1}^{2}(1) = \frac{4x^{5}}{(1+x^{2})^{4}}$ &
 $\displaystyle {\mathcal S}_{0}^{2}(2) = \frac{64x^{5}}{(1+x^{2})^{6}}$ \\
 \hline 
\end{tabular}
\end{table}

\subsection{Shape functions in n+core systems without final-state interaction}
\label{subsec:n+core}

\begin{figure}
\begin{center}
\includegraphics[width=135mm]{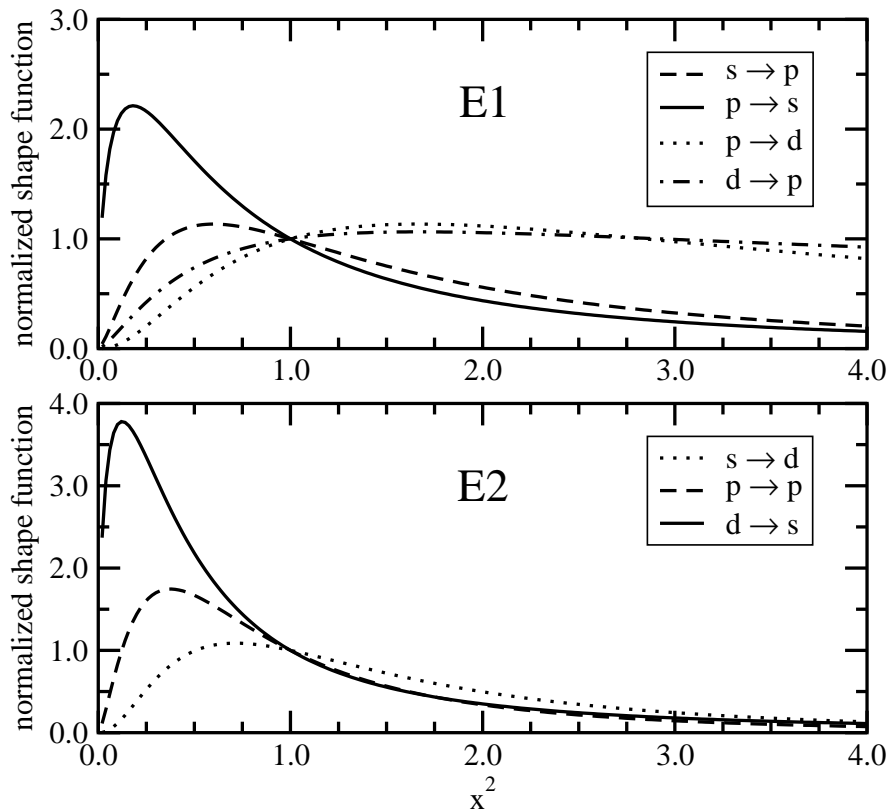}
\end{center}
\caption{\label{fig:1} 
Dependence of the generic 
shape functions ${\mathcal S}_{l_{i}}^{l_{f}}(\lambda)$
with cutoff radius $R=0$
on $x^{2}=E/S_{n}$ for various 
transitions $l_{i} \to l_{f}$ of multipolarity
$E1/E2$ (top/bottom). The functions are normalized to 1
at $x^{2}=1$.}
\end{figure}

Without nuclear interaction between the neutron and the core 
in the final state the phase shift $\delta_{l_{f}}$ is zero 
and there is no contribution from the integral with the
irregular wave function in  (\ref{eq:idef}). The contribution
for radii $r < R$  in the radial integral is small and it is
possible to take the limit $R \to 0$ keeping the neutron separation
energy $S_{n}$ or equivalently the inverse bound-state decay 
length $q$ constant. In this limit both $\gamma$
and $\kappa$ approach zero but the ratio $x=\kappa/\gamma=
\sqrt{E/S_{n}}$, i.e.\ the relevant variable for the shape
of the strength distribution, is independent of $R$.
In this case the reduced radial integrals (\ref{eq:iredgen})
assume a particular simple form. 
They only depend on this dimensionless variable $x$.
From the reduced radial integrals the shape functions (\ref{eq:sdef})
of the reduced transition probability are easily derived.
The functions ${\mathcal S}_{l_{i}}^{l_{f}}(\lambda)$ in the limit
$R \to 0$ with $\delta_{l_{f}}=0$ are given in table \ref{tab:2}.
Especially the expression ${\mathcal S}_{0}^{1}(1)$ is well known,
see, e.g., \cite{Bla79,Nak94}. It can be found in a different notation
also in Refs.\ \cite{Ber88,Ber92}.

In Fig.~\ref{fig:1} the generic form of the shape functions
${\mathcal S}_{l_{i}}^{l_{f}}(\lambda)$ is shown as a 
function of $x^{2} = E/S_{n}$ for $E1$ and $E2$ transitions
from a bound state with orbital angular momentum $l_{i}$ to
a scattering state with orbital angular momentum $l_{f}$.
The dependence of the transition shape on the centrifugal barrier
is clearly seen. The peak at small energies is more pronounced
for low orbital angular momenta $l_{f}$. Only for $s$ and $p$ waves
in the continuum a large transition strength is found close to the threshold.
At low $x$ we have the typical dependence
\begin{equation}
 {\mathcal S}_{l_{i}}^{l_{f}}(\lambda) \propto x^{2l_{f}+1} \: .
\end{equation}
The shape function for photo absorption 
${\mathcal S}_{l_{i}}^{l_{f}}(\mbox{abs},\lambda)$
has the same $x$ dependence as ${\mathcal S}_{l_{i}}^{l_{f}}(\lambda)$
for small final state momenta. 
The shape function
for radiative capture
${\mathcal S}_{l_{f}}^{l_{i}}(\mbox{capt},\lambda)$,
on the other hand,
shows a $x^{2l_{f}-1}$ dependence for small  momenta in the continuum
due to the additional momentum dependent factor from the theorem
of detailed balance (\ref{eq:detbal}), see equation (\ref{eq:sabs}).

\begin{figure}
\begin{center}
\includegraphics[width=135mm]{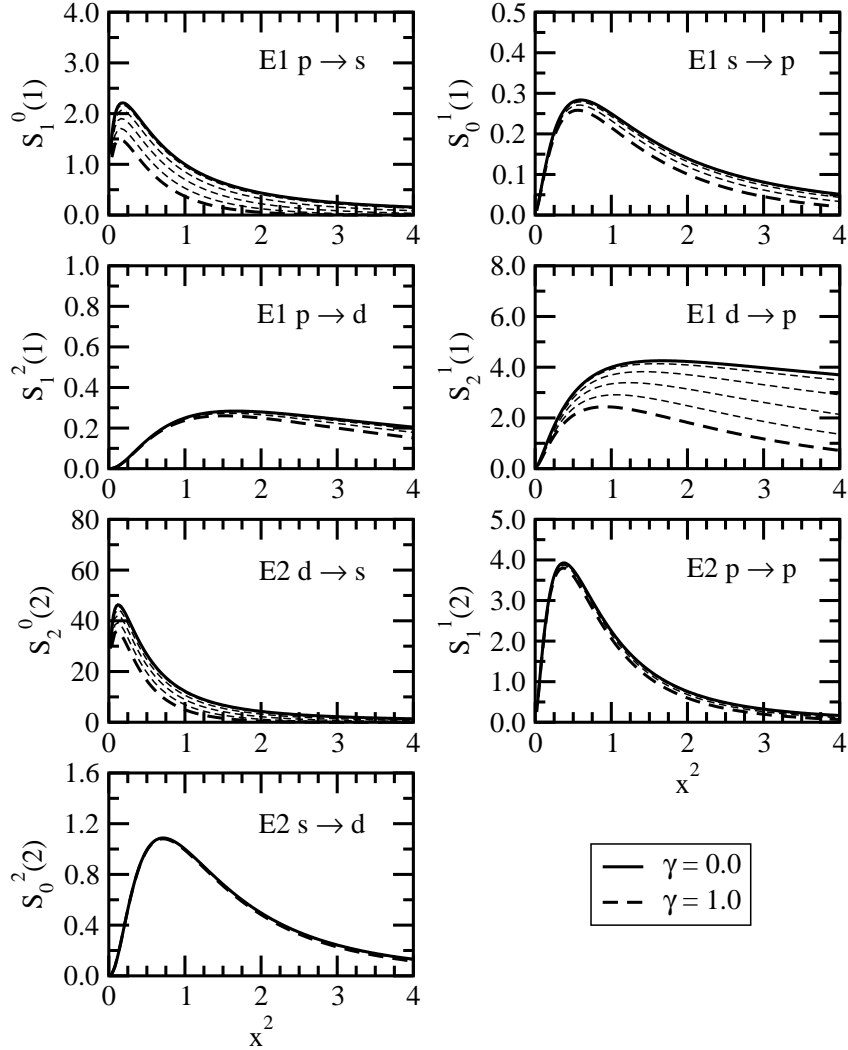}
\end{center}
\caption{\label{fig:1b} 
Shape functions ${\mathcal S}_{l_{i}}^{l_{f}}(\lambda)$
as a function of $x^{2}=E_{bc}/S_{n}$ for 
values of the parameter $\gamma$ between $0.0$ and $1.0$ in steps of
$0.2$.}
\end{figure}

In case of a finite cutoff radius $R$ in the radial integral
(\ref{eq:radint})
the shape function ${\mathcal S}_{l_{i}}^{l_{f}}(\lambda)$
depends on both $\gamma=qR$ and
$\kappa=kR$. Since $q$ is fixed for a given nucleus is it reasonable again
to use $\gamma$ and $x=\kappa/\gamma=k/q$ as independent parameters.
For halo nuclei with small nucleon separation energy $q$ will be
a small quantity. As long as $R$ does not 
become too large, e.g.\
for heavy nuclei, $\gamma$ will also be small. The variation of
the shape functions with $\gamma$ gives an estimate of the contribution
to the radial integral from the nuclear interior. In Fig.\ \ref{fig:1b}
the change of the shape functions with $\gamma$ is shown for the
transitions of Fig.\ \ref{fig:1}. There is a clear systematic trend.
The shape functions are less sensitive to a change in $\gamma$
for larger final state orbital angular momentum $l_{f}$ and
higher multipolarity $\lambda$ because of the suppression of
the integrand at small $r$ from the spherical Bessel functions
$j_{l_{f}}(kr)$ and $r^{\lambda}$ from the transition operator,
respectively. On the other hand, a larger orbital angular momentum
$l_{i}$ in the bound state increases the sensitivity since
the wave function introduces a $r^{-l_{i}}$ dependence at small radii.
This explains the strong $\gamma$-dependence of the shape function
${\mathcal S}_{2}^{1}(1)$.

\subsection{Shape functions in n+core systems with final-state interaction}
\label{subsec:fsinc}

For neutron scattering the finite-range expansion (\ref{eq:fre}) reduces to
\begin{equation} \label{eq:frex}
 k^{2l+1} \cot(\delta_{l}) = - \frac{1}{a_{l}} + \frac{r_{l}}{2} k^{2}
 + \dots
\end{equation}
since $\eta = 0$.
Taking only the contributions with $a_{l}$ and $r_{l}$ into account
the phase shift $\delta_{l}$ crosses the value $\pi/2$ at an energy
$E_{0}=\hbar^{2}/(\mu a_{l} r_{l})$.
This behaviour produces a resonance in the corresponding
partial wave if the phase shift is increasing that, however,
does not have to be physical.
In general, the actual energy-dependence of the phase shift
close to a resonance at energy $E_{R}$ is given by the
Breit-Wigner form
$ \tan \delta_{l} = \Gamma/(2(E_{R}-E))$
with positive width $\Gamma$
where the parameters $E_{R}$ and $\Gamma$ are not related to $a_{l}$
and $r_{l}$ in the effective-range expansion.
Particular values for these parameters that correctly reproduce
the phase shift at low momenta will not necessarily reproduce
the position and the width of an actual 
resonance with orbital angular momentum $l$.
However, the phase shift can be calculated with the help of the relation
\begin{equation} \label{eq:tanbl}
 \tan (\delta_{l}) = - (xb_{l})^{2l+1} 
\end{equation}
for general cases assuming an energy dependence of 
the function $b_{l}$.
It replaces the phase shift $\delta_{l}$
in order to take the FSI in the scattering wave function into account.
The dimensionless function $b_{l}$ is related by
\begin{equation}
 D_{l}(E) = - \left(\frac{b_{l}}{q}\right)^{2l+1}
\end{equation}
to the function $D_{l}(E)$ introduced in Ref.\ \cite{Bay04} for
a low-energy expansion of cross sections for radiative capture reactions.

The function $b_{l}$ is of the order of $\gamma=qR$ 
unless the logarithmic derivative $L$ of the scattering wave function
at radius $R$ is close to $-l$, see eq.\ (\ref{eqn:scalen_l}).
It is useful in expansions of the
shape functions ${\mathcal S}_{l_{i}}^{l_{f}}(\lambda)$ if
the limit $R \to 0$ is considered. 
On the other hand, the
scaled function 
\begin{equation} \label{eq:spf}
  c_{l} = \frac{b_{l}}{\gamma}
\end{equation} 
(usually of order one)
is the appropriate quantity
if the limit $\gamma \to 0$ is studied. In this case we have
\begin{equation} 
 a_{l} = \lim_{x\to 0} \left(c_{l}R\right)^{2l+1}  
\end{equation}
with constant radius $R$.

\begin{figure}
\begin{center}
\includegraphics[width=135mm]{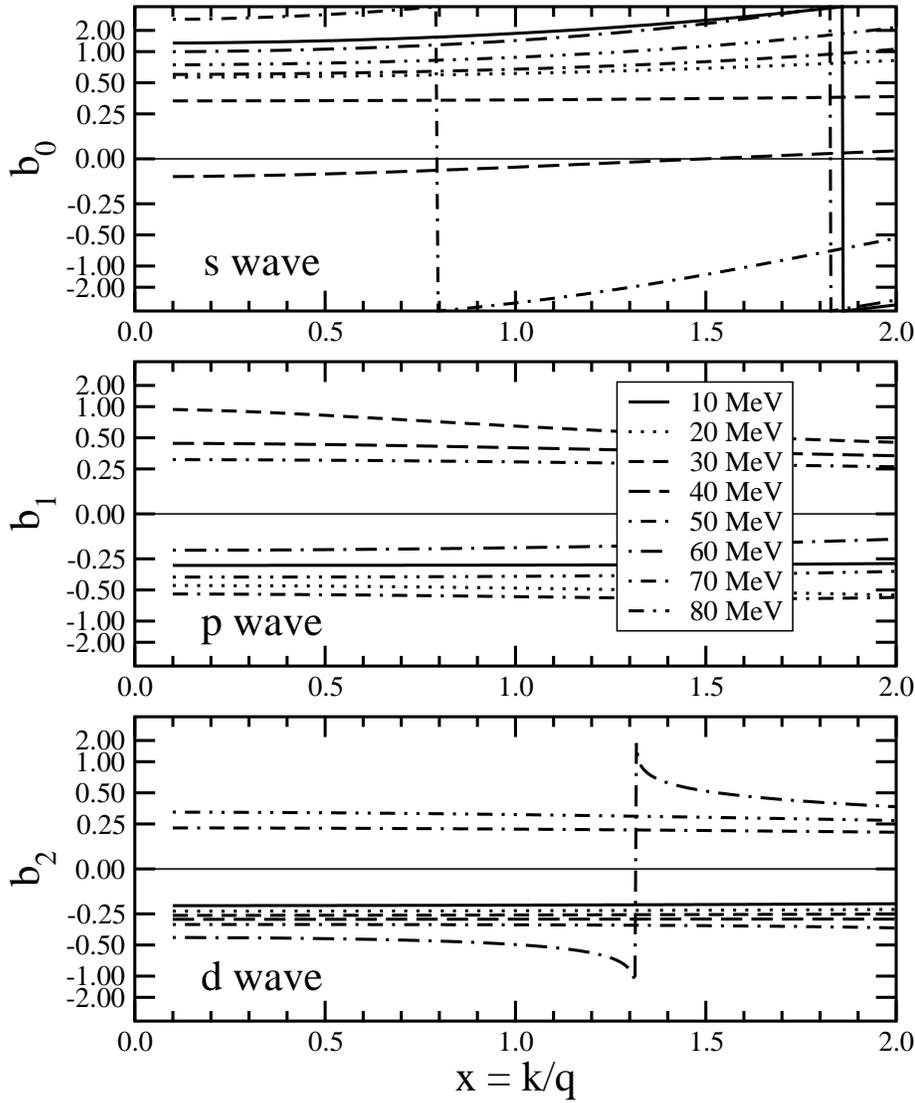}
\end{center}
\caption{\label{fig:2} 
Function $b_{l}$ 
in partial waves with orbital angular momentum $l=0,1,2$
as a function of $x=k/q$ for the
scattering of a neutron on ${}^{10}$Be in a single-particle model with
a Woods-Saxon potential of radius $R_{0}=2.78$~fm, diffuseness parameter
$a=0.65$~fm and various potential depths $V_{0}$.
(Note the nonlinear scale on the $y$ axis.)}
\end{figure}

Typical values of $b_{l}$ and $c_{l}$ can
be estimated from examples employing a single particle model
where a nuclear potential of Woods-Saxon form with typical
parameters is assumed.
In Figure~\ref{fig:2} the function $b_{l}$ 
for orbital angular momenta
$l=0,1,2$ is shown as a function of $x=k/q$ for the
scattering of a neutron on ${}^{10}$Be assuming different depths
$V_{0}$ of the potential. In most cases the function
$b_{l}$ is quite constant
as a function of $x$ except when there is a resonance 
(vertical lines in Fig.~\ref{fig:2}) in the continuum
for a certain fixed depth $V_{0}$. 
For $p$ and $d$ waves $b_{l}$ is
usually in the intervall $[-0.5,0.5]$; for $s$ waves $b_{0}$ covers
a larger range.
Assuming a zero-range potential for the neutron-core interaction
the scattering wave function
is given by $\psi^{(+)}(\vec{r}) = \exp(i\vec{k}\cdot \vec{r})
-\exp(ikr)/[(q+ik)r]$ with the bound state parameter $q$
determined by the binding energy.
The $s$ wave scattering length is just $a_{0} =1/q$ 
corresponding to $b_{0} = 1$ and $c_{0} = 1/\gamma$.
Since the ground state is
close to the threshold for a halo nucleus 
$c_{0}$ becomes unnaturally large and
there will be a large effect
on the continuum in the same partial wave. However, the partial waves in
the final state for an electric excitation have a different orbital angular
momentum 
and parity and usually there will be no resonance in the
energy range of interest.
In general it is reasonable to 
assume $|b_{l}|<0.5$ with a weak momentum dependence.

\begin{table}
\caption{\label{tab:3}Characteristic shape functions 
${\mathcal S}_{l_{i}}^{l_{f}}(\lambda)$ for $\gamma=0$ and $b_{l_{f}} \neq 0$.}
\begin{tabular}{ll}
 \hline 
 $\lambda = 0$ & $\lambda = 1$ \\
 \hline 
 $\displaystyle {\mathcal S}_{0}^{0}(0) = \frac{x}{(1+x^{2})^{2}} 
 \frac{(1-b_{0})^{2}}{1+(xb_{0})^{2}}$ &
 $\displaystyle {\mathcal S}_{1}^{0}(1)  = \frac{x}{(1+x^{2})^{4}} 
 \frac{(3-2b_{0}+x^{2})^{2}}{1+(xb_{0})^{2}}$  \\
 &
 $\displaystyle {\mathcal S}_{0}^{1}(1)  = \frac{x^{3}}{(1+x^{2})^{4}}
 \frac{[2-b_{1}^{3}(1+3x^{2})]^{2}}{1+(xb_{1})^{6}}$ \\
 \hline 
 $\lambda = 2$ \\
 \hline 
 $\displaystyle {\mathcal S}_{2}^{0}(2)  =  \frac{x}{(1+x^{2})^{6}} 
 \frac{(15-8b_{0}+10x^{2}+3x^{4})^{2}}{1+(xb_{0})^{2}}$ & \\
 $\displaystyle {\mathcal S}_{1}^{1}(2) = \frac{x^{3}}{(1+x^{2})^{6}} 
 \frac{4[5+x^{2}-b_{1}^{3}(1+5x^{2})]^{2}}{1+(xb_{1})^{6}}$ & \\
 $\displaystyle {\mathcal S}_{0}^{2}(2) = \frac{x^{5}}{(1+x^{2})^{6}} 
 \frac{[8-b_{2}^{5}(3+10x^{2}+15x^{4})]^{2}}{1+(xb_{2})^{10}}$ & \\
 \hline 
\end{tabular}
\end{table}

The phase shift in the general expression for the
reduced radial integral (\ref{eq:iredgen})
can be replaced by the function $b_{l}$ using the explicit relation
(\ref{eq:tanbl}).
For some of the transitions it is still possible
to take the limit $R\to 0$ in the reduced radial integrals
for $b_{l}\neq 0$. 
The corresponding shape functions are given in table \ref{tab:3}.
For the other transitions the reduced radial integral and the
shape function diverges.
It is obvious that the $x^{2l_{f}+1}$ dependence at small $x$ is
obtained again but the modulus is modified in the limit $x\to 0$
as compared to the generic shape functions in table \ref{tab:2}
because of the FSI.

\begin{figure}
\begin{center}
\includegraphics[width=135mm]{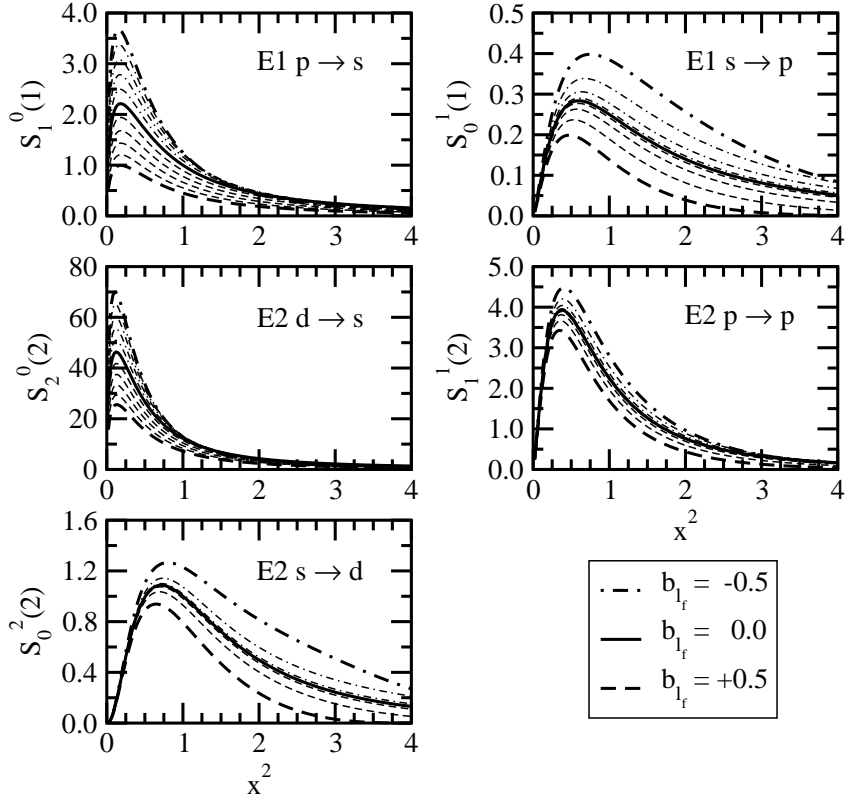}
\end{center}
\caption{\label{fig:3} 
Shape functions ${\mathcal S}_{l_{i}}^{l_{f}}(\lambda)$
for $\gamma=0$
as a function of $x^{2}=E/S_{n}$ for various 
values of the function $b_{l_{f}}$ in steps of
$0.1$.  The thick dashed, solid, and dot-dashed lines correspond to
$b_{l_{f}}=+0.5$, $0.0$, $-0.5$, respectively.}
\end{figure}

In Fig.~\ref{fig:3} the dependence of the shape function 
in the limit $R \to 0$ on $b_{l_{f}}$ is shown
for the cases of table \ref{tab:3}. 
(Transitions with $\lambda=0$
are not relevant.) The function $b_{l_{f}}$
varies in the interval $[-0.5,0.5]$. This corresponds
to reasonable values that are far from a resonance in the particular
continuum channel. Depending on the magnitude of $b_{l_{f}}$ 
the shape functions show a pronounced variation around
the generic form that reflects the effect
of the potential in the final state. Small changes in the reduced
scattering length lead to a smaller effect for higher orbital
angular momenta $l_{f}$ due to the $b_{l_{f}}^{2l_{f}+1}$ dependence
in the analytical expressions. For positive values of $b_{l_{f}}$
one finds an increase of 
${\mathcal S}_{l_{i}}^{l_{f}}(\lambda)$ whereas a negative $b_{l_{f}}$ leads
to a decrease in absolute value. There is also a change in the shape
observed and a shift of the maximum. This shift 
becomes noticable only for large values of $b_{l}$.
The position of the maximum moves to larger $x^{2}$ with
decreasing $b_{l}$ for $p$ and
$d$ waves in the final state, whereas the trend is opposite
for $s$-wave final states. There is also a clear hierachy observed.
The maximum of the shape function appears at higher $x^{2}$ with
larger orbital angular momentum $l_{f}$ in the continuum.

\begin{table}
\caption{\label{tab:4}Expansion of the characteristic shape functions 
${\mathcal S}_{l_{i}}^{l_{f}}(\lambda)$ in the parameter $x$
for finite $\gamma$ and $b_{l_{f}} \neq 0$.}
\begin{tabular}{l}
 \hline 
 $\lambda = 0$ \\
 \hline 
 $\displaystyle {\mathcal S}_{0}^{0}(0) =
 \exp(-2\gamma) \left( 1+\gamma-b_{0}\right)^{2} x$  \\
 $\displaystyle {\mathcal S}_{1}^{1}(0) =
 \frac{\exp(-2\gamma)}{9\gamma^{2}}
 \left[\gamma(3+3\gamma+\gamma^{2})-3b_{1}^{3}\right]^{2} x^{3}$ \\
 \hline 
 $\lambda = 1$ \\
 \hline 
 $\displaystyle {\mathcal S}_{1}^{0}(1) =
 \exp(-2\gamma) \left[ 3+3\gamma+\gamma^{2}-b_{0}(2+\gamma)\right]^{2} x$ \\
 $\displaystyle {\mathcal S}_{0}^{1}(1) =
 \frac{\exp(-2\gamma)}{9} 
 \left(6+6\gamma+3\gamma^{2}+\gamma^{3}-3b_{1}^{3} \right)^{2} x^{3}$ \\
 $\displaystyle {\mathcal S}_{2}^{1}(1) =
 \frac{\exp(-2\gamma)}{9\gamma^{2}} 
 \left[\gamma(15+15\gamma+6\gamma^{2}+\gamma^{3})
   -3b_{1}^{3}(3+\gamma)\right]^{2} x^{3}$ \\
 $\displaystyle {\mathcal S}_{1}^{2}(1) =
 \frac{\exp(-2\gamma)}{225\gamma^{2}} 
 \left[\gamma(30+30\gamma+45\gamma^{2}+5\gamma^{3}+\gamma^{4})
   -45b_{2}^{5} \right]^{2} x^{5}$\\
 \hline 
 $\lambda = 2$ \\
 \hline 
 $\displaystyle {\mathcal S}_{2}^{0}(2) =
 \exp(-2\gamma) \left[15+15\gamma+6\gamma^{2}+\gamma^{3}
   -b_{0}(8+5\gamma+\gamma^{2})\right]^{2} x$ \\
 $\displaystyle {\mathcal S}_{1}^{1}(2) =
 \frac{\exp(-2\gamma)}{9} 
 \left[30+30\gamma+15\gamma^{2}+5\gamma^{3}+\gamma^{4}
   -3b_{1}^{3}(2+\gamma) \right]^{2} x^{3}$ \\
 $\displaystyle {\mathcal S}_{0}^{2}(2) =
 \frac{\exp(-2\gamma)}{225} 
 \left(120+120\gamma+60\gamma^{2}+20\gamma^{3}+5\gamma^{4}+\gamma^{5}
   -45b_{2}^{5} \right)^{2} x^{5}$ \\
 \hline 
\end{tabular}
\end{table}

The analytic expressions of the shape functions in table \ref{tab:3}
were obtained by extending the asymptotic
form of the bound and scattering wave functions to zero radius
in the radial integral. The integrand with the regular scattering wave
function shows a $r^{\lambda-l_{i}+l_{f}+1}$ dependence; in contrast to that
one finds a $r^{\lambda-l_{i}-l_{f}}$ 
behaviour at small $r$ for the integrand with the
irregular scattering wave function.
One might expect that the irregular contribution 
is overestimated in this limit since it diverges for $r \to 0$. 
Expanding the general
shape functions in powers of $x$ 
with constant $\gamma = qR$
one finds the results as given in table \ref{tab:4}.
The functions again show the typical
$x^{2l_{f}+1}$ dependence at low $x=k/q$ but the slope depends
less strongly on $b_{l_{f}}$ 
for finite $\gamma$ than for the functions given in table \ref{tab:3}.
The shape functions of the transitions $d\to p$ and $p \to d$ 
($p\to p$) for $\lambda=1$ ($\lambda=0$) diverge in the limit $\gamma \to 0$
or $R \to 0$.

\begin{figure}
\begin{center}
\includegraphics[width=135mm]{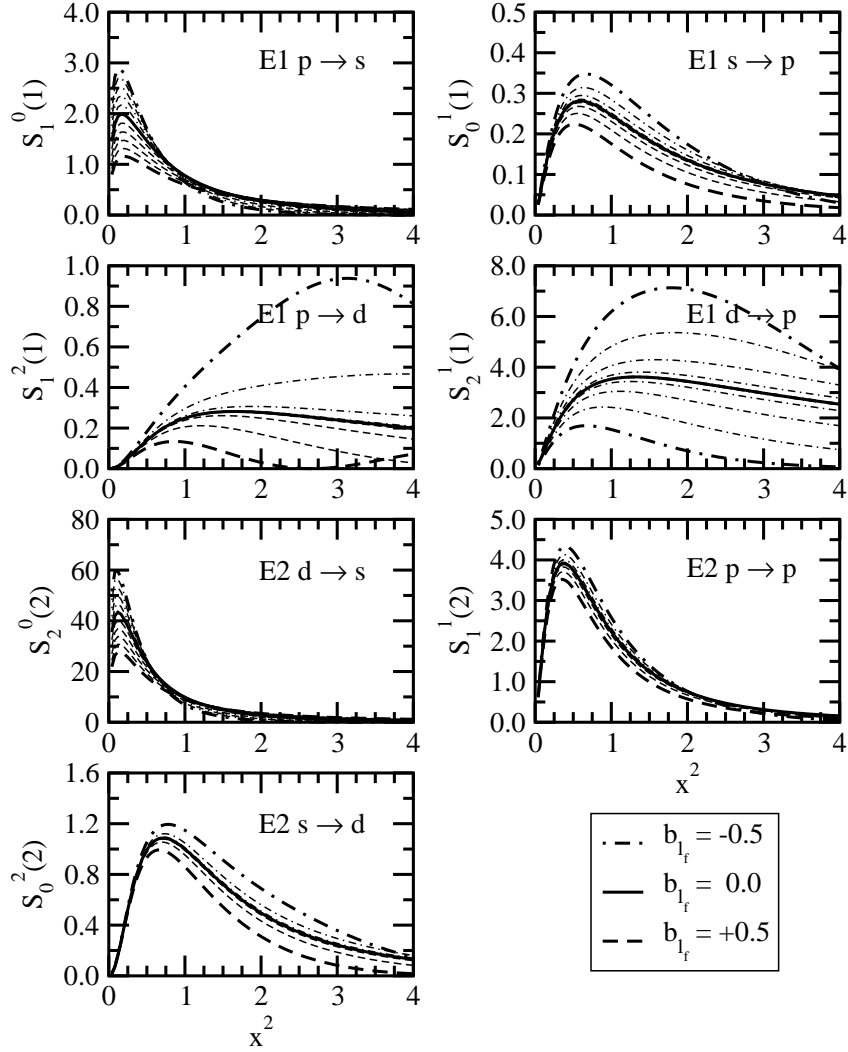}
\end{center}
\caption{\label{fig:4} 
Shape functions ${\mathcal S}_{l_{i}}^{l_{f}}(\lambda)$
for $\gamma=0.5$
as a function of $x^{2}=E/S_{n}$ for various 
values of $b_{l_{f}}$ in steps of
$0.1$.  The thick dashed, solid, and dot-dashed lines correspond to
$b_{l_{f}}=+0.5$, $0.0$, and $-0.5$, respectively.}
\end{figure}

\begin{figure}
\begin{center}
\includegraphics[width=135mm]{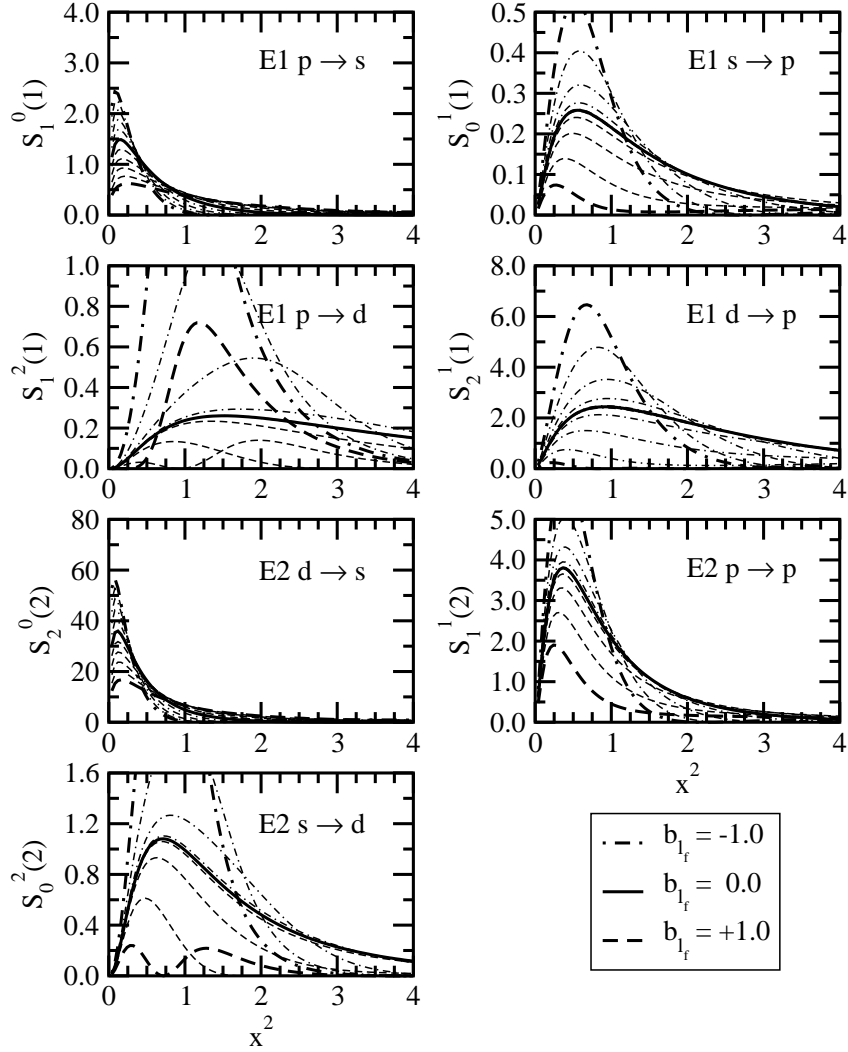}
\end{center}
\caption{\label{fig:5} 
Shape functions ${\mathcal S}_{l_{i}}^{l_{f}}(\lambda)$
for $\gamma=1.0$
as a function of $x^{2}=E/S_{n}$ for various 
values of $b_{l_{f}}$ in steps of
$0.2$.  The thick dashed, solid, and dot-dashed lines correspond to
$b_{l_{f}}= +1.0$, $0.0$, and $-1.0$, respectively. 
The scale of the $y$ axis is chosen to
be the same as in Fig.~\ref{fig:4} for a better comparison.}
\end{figure}

Fig.~\ref{fig:4} shows the dependence of the shape functions
on $b_{l_{f}}$ for constant $\gamma=0.5$, i.e.\ a constant 
cutoff radius $R$. 
This value of $\gamma$ 
corresponds to $R=3.36$~fm in case of neutron
scattering on ${}^{10}$Be with a bound state parameter 
$q=0.1487$~fm$^{-1}$ for a neutron separation energy of $S_{n}=0.504$~MeV.
The shape functions are very similar
to the case with $R\to 0$, cf.\ Fig.\ \ref{fig:3}, except for
two cases where ${\mathcal S}_{l_{i}}^{l_{f}}(\lambda)$ diverges
in the limit $R \to 0$ for finite $b_{l_{f}}$. In general the shape functions
for finite $R$ are slightly smaller than in the case $R=0$.
Final state effects
for a finite value of $\gamma$ are less pronounced than in the
case with $\gamma = 0$ but the differences is small.
Therefore the dependence of the shape functions 
in table \ref{tab:3}
on $b_{l_{f}}$ gives a reasonable impression about the importance
of the final state interaction.
For $E1$ transitions $p \to d$ or $d \to p$ a strong dependence
of the shape function on $b_{l_{f}}$
is found. The shape and magnitude of the reduced transition
probability depends sensitively on the choice of $b_{l_{f}}$ and
$R$.

The sensitivity of the shape functions to final-state effects
on the neutron separation
energy can be estimated from an expansion of 
${\mathcal S}_{l_{i}}^{l_{f}}(\lambda)$
in terms of $\gamma$ by replacing the function $b_{l}$
with the quantity $c_{l}$ as defined in equation (\ref{eq:spf}).
This approach was introduced in \cite{Typ04a}.
Analytical expressions for 
dipole transitions are given by
\begin{eqnarray}
 {\mathcal S}_{1}^{0}(1) & = & 
 \frac{x(3+x^{2})^{2}}{(1+x^{2})^{4}} 
 \left[ 1 - \frac{4c_{0}}{3+x^{2}} \gamma + \dots \right] \: ,
 \\
 {\mathcal S}_{0}^{1}(1) & = &
 \frac{4x^{3}}{(1+x^{2})^{4}}
 \left[1 - c_{1}^{3}(1+3x^{2})\gamma^{3} + \dots \right] \: ,
 \\
 {\mathcal S}_{2}^{1}(1) & = &
 \frac{x^{3}(5+3x^{2})^{2}}{(1+x^{2})^{4}} 
 \left[ 1 - \frac{(1+6c_{1}^{3})(1+x^{2})^{2}}{5+3x^{2}}\gamma^{2}
 + \dots \right] \: ,
 \\
 {\mathcal S}_{1}^{2}(1) & = &
 \frac{4x^{5}}{(1+x^{2})^{4}} 
 \left[ 1 - \frac{1+180c_{2}^{5}}{60}(1+x^{2})^{2}\gamma^{4} + \dots \right] 
 \: .
\end{eqnarray}
These expansions are consistent with the results 
in table \ref{tab:4}
in the limit $\gamma \to 0$ 
($c_{l} = b_{l}/\gamma$, see eq.\ (\ref{eq:spf})).
There are corrections to the $E1$ shape functions 
in table \ref{tab:3} 
linear in $\gamma$ for $p \to s$ transitions and
proportional to $\gamma^{3}$ for $s \to p$ transitions
for finite values of $c_{l_{f}}$. For $d \to p$ and
$p \to d$ transitions one finds corrections to the
forms in table \ref{tab:3} 
of order $\gamma^{2}$ and $\gamma^{4}$,
respectively.
They contain a contribution depending on the interaction
($c_{l_{f}}$) and a correction from the finite size of the system.
Transitions $l \to l-1$ are affected 
more strongly by the interaction
than transitions $l \to l+1$.

Another approach to estimate the sensitivity 
to the FSI is a comparison of the shape functions by scaling 
$\gamma$ and $b_{l}$ appropriately.
In Fig.~\ref{fig:4} the shape functions were shown for $\gamma=0.5$
and constant $b_{l_{f}} \in [-0.5,0.5]$. Multiplying $\gamma$ by a factor of
2 corresponds to an increase of $S_{n}$ by a factor of 4. In order
to cover the same range of scattering lengths $a_{l}$ as in Fig.~\ref{fig:4}
also the interval for the function $b_{l}$ 
has to be increased to $[-1.0,1.0]$ because the ratio $b_{l_{f}}/q$ has
to be kept constant, cf.\ Eq.~(\ref{eq:abrel}). 
The corresponding shape functions for $\gamma = 1.0$
are depicted in Fig.~\ref{fig:5}. It is obvious that a higher 
neutron separation energy leads to a much stronger dependence of
the shape functions on the strength of the final-state interaction.
This behaviour is easily understood from the radial dependence of the
bound state wave function. For larger separation energy 
the slope of the asymptotic wave function is much steeper since $q$ 
increases. The radial integral is more sensitive to a small shift
of the continuum wave function from a finite nuclear phase shift
because the main contribution to the integrals arises from a
smaller intervall in the radius on the surface of the nucleus.
The effect is less pronounced for $s$ waves in the final state
but increases dramatically for higher partial waves in the continuum.

\subsection{Shape functions in p+core systems}
\label{subsec:p+core}

\begin{figure}
\begin{center}
\includegraphics[width=135mm]{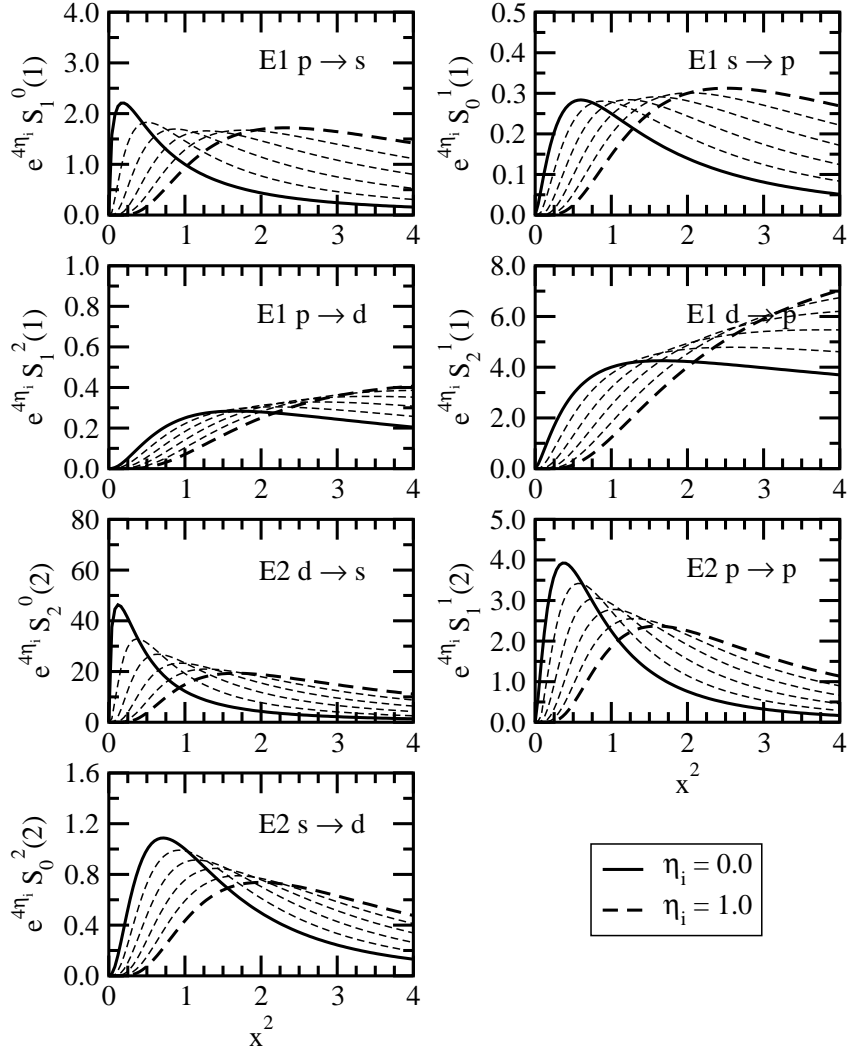}
\end{center}
\caption{\label{fig:1c} 
Scaled shape functions $e^{4\eta_{i}} {\mathcal S}_{l_{i}}^{l_{f}}(\lambda)$
as a function of $x^{2}=E_{bc}/S_{p}$ for $\gamma=0$ and
values of the parameter $\eta_{i}$ between $0.0$ and $1.0$ in steps of
$0.2$ without nuclear final-state interaction.}
\end{figure}

\begin{figure}
\begin{center}
\includegraphics[width=135mm]{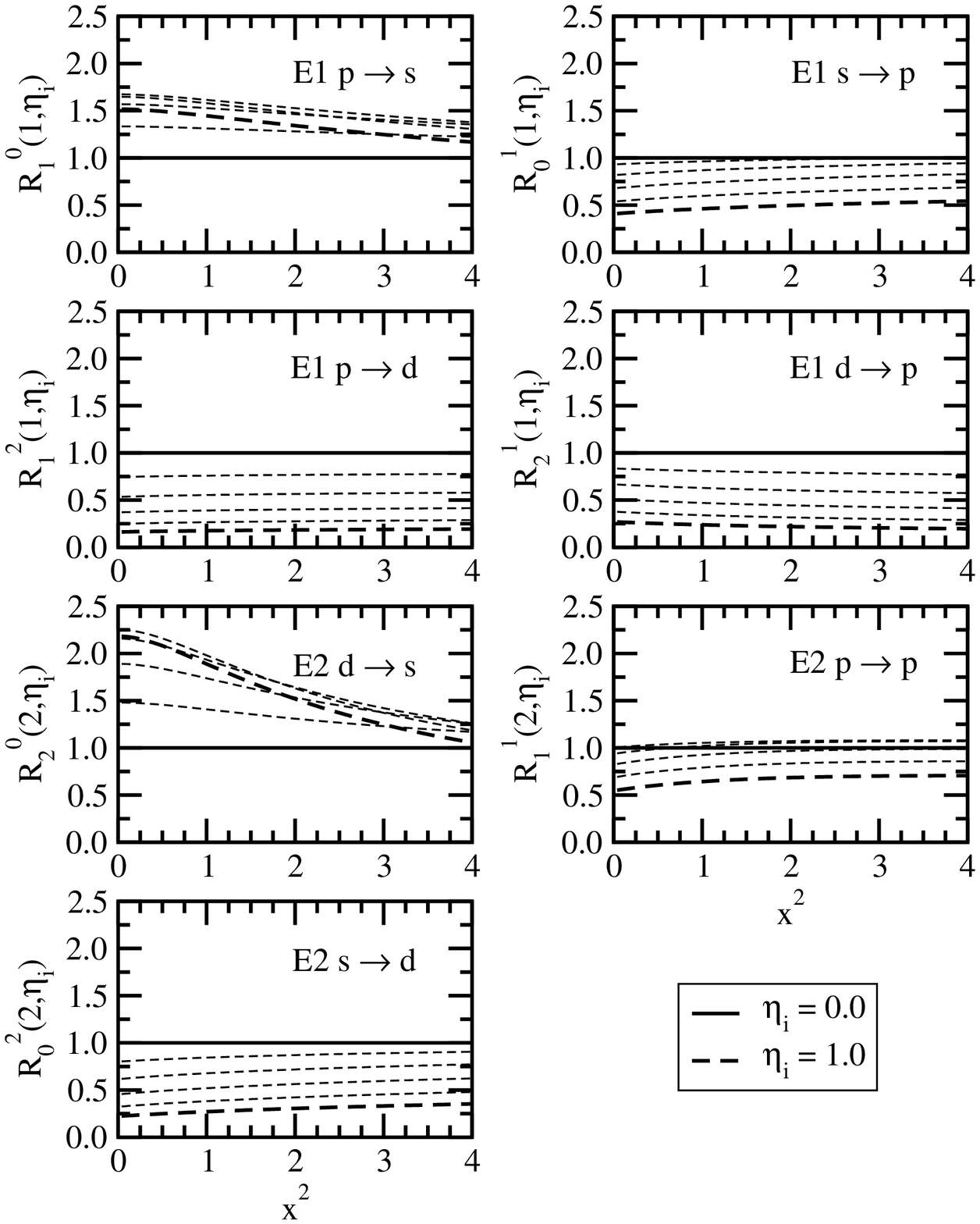}
\end{center}
\caption{\label{fig:1d} 
Ratios $R_{l_{i}}^{l_{f}}(\lambda,\eta_{i})$
as a function of $x^{2}=E_{bc}/S_{p}$ for $\gamma=0$ and
values of the parameter $\eta_{i}$ between $0.0$ and $1.0$ in steps of
$0.2$ without nuclear final-state interaction.}
\end{figure}

The Coulomb interaction in p+core systems leads to a systematic 
modification of the characteristic shape functions as compared to
n+core systems. Since analytical results are not available in general one
has to resort to a numerical integration of the radial integral
(\ref{eq:hdef})
if the dependence on arbitrary values of $x$ is studied.
In Ref. \cite{Muk02} an analytic result for the radial integral was 
obtained in the special case of scattering on a solid sphere 
with given radius.
For $x \ll 1$ it is possible to obtain analytical results
in an approach with an expansion for small energies as
presented in Refs.\ \cite{Jen98,Bay00}.
For small $x$ one finds a suppression of the shape function
approximately
proportional to $\exp(-2\pi \eta_{f})$ with the
Sommerfeld parameter $\eta_{f}$ of the scattering state.
This scaling is 
characteristic for a case with a Coulomb barrier.
However, at larger energies the absolute value of the shape functions
is determined by the Sommerfeld parameter $\eta_{i}$ of the 
bound state since it defines the range of radii with the largest
contribution to the radial integral. 
In figure~\ref{fig:1c} the variation of the 
shape functions with $\eta_{i}$ for $\gamma=0$
and for various transitions is shown. The functions are scaled 
with $\exp(4\eta_{i})$.
Effects from the nuclear
interaction in the final state are not taken into account.
For $\eta_{i}=0$ the results for neutrons are recovered.
With increasing $\eta_{i}$ the maximum of the shape function
shifts to higher $x^{2}$, i.e.\ to higher relative energies.
At the same time the width of the strength distribution increases
and the absolute value of ${\mathcal S}_{l_{i}}^{l_{f}}(\lambda)$
reduces considerably (note the scaling).

At low $x^{2}$ the suppression of the shape functions with
finite $\eta_{i}$ as compared to the case $\eta_{i}=0$
is quantitatively
given by the factor $C_{l_{f}}^{2}(\eta_{f})$ as defined in
equation (\ref{eq:coulfac}). For $x\to \infty$ corresponding
to $\eta_{f}=\eta_{i}/x \to 0$
we have $\lim_{\eta_{f} \to 0} C_{l_{f}}(\eta_{f})= 1$. 
This suggests to scale the shape functions
for the proton+core case with the penetrability factor
$C_{l_{f}}^{-2}(\eta_{f})$.
In figure \ref{fig:1d} the ratio
\begin{equation}
 R_{l_{i}}^{l_{f}}(\lambda,\eta_{i})
 = \left[ C_{l_{f}}(\eta_{f}) \right]^{-2} 
 \frac{{\mathcal S}_{l_{i}}^{l_{f}}(\lambda,\eta_{i})}{
 {\mathcal S}_{l_{i}}^{l_{f}}(\lambda,0)}
\end{equation}
is depicted for various transitions and parameters $\eta_{i}$.
The ratio only weakly depends on $x^{2}$. Therefore, the main
difference between the proton+core and neutron+core cases
is well described by the penetrability factor $C_{l_{f}}^{-2}(\eta_{f})$.

There is a major difference between n+core and p+core nuclei.
Due to the increasing Coulomb barrier the
observation of a large transition strength at low relative
energies will be lost for heavier p+core nuclei and the difference
between transitions of different orbital angular momenta in the
initial and final states will be become smaller. In contrast,
even for heavy n+core nuclei the strong transition strength 
at small energies will persist and the shape will still be
characteristic of the particular transition.
In principle, variations of the generic shape functions as shown
in figure~\ref{fig:1c} with finite parameters $\gamma=qR$ and
$b_{l}$ can be studied along the lines
as for the n+core case. 
Very similar trends are observed
and we omit a detailed discussion.

\section{Total transition strength and sum rules}
\label{sec:totsum}

The total strength for an $E\lambda$ transition
from a bound state with quantum numbers $J_{i}, s, j_{c}, l_{i}$
to all possible final states
is related by the non energy-weighted sum rule 
\begin{eqnarray} \label{eq:newsr}
 B(E\lambda, J_{i} s j_{c} l_{i}) & = &
 \sum_{J_{f} l_{f}} \int_{0}^{\infty} dE \: 
 \frac{dB}{dE}(E\lambda, J_{i} s j_{c} l_{i} \to 
  k J_{f} s j_{c} l_{f})
 \\ \nonumber & = &
 \left[ Z_{\rm eff}^{(\lambda)}e \right]^{2}
 \frac{2\lambda+1}{4\pi} \langle r^{2\lambda} \rangle_{l_{i}}
\end{eqnarray}
to the expectation value $\langle r^{2\lambda} \rangle_{l_{i}}$
of the initial bound state with orbital angular momentum $l_{i}$,
see, e.g., \cite{Jon04}.
Note that for the integration over the energy $E$ not only 
continuum states but also bound states
have to be taken into account if they can be reached by an $E\lambda$
transition from the initial bound state.
In the special case of dipole transitions 
the root-mean-square radius of the bound state wave function
determines the total $E1$ strength. 
From the scaling laws of $\langle r^{2} \rangle_{l_{i}}$
as discussed
in subsection \ref{subsec:prob} 
we expect a divergence of the total transition
strengths $B(E1, l_{i})\propto \gamma^{-2}$ and
$B(E1, l_{i})\propto \gamma^{-1}$ for $l_{i}=0$ and
$1$, respectively, in the limit $\gamma \to 0$. 
In contrast, the total $E1$ strength should remain
finite for larger orbital angular momenta of the bound  state
wave function.

The total reduced transition probability to the 
continuum is obtained in our approach with asymptotic wave functions
from
\begin{eqnarray} \label{eq:bel}
 \lefteqn{B_{\rm cont} (E\lambda, J_{i} s j_{c}l_{i}  
 \to J_{f} s j_{c} l_{f})=}
 \\ \nonumber & & 
   \left[Z_{\rm eff}^{(\lambda)}e\right]^{2} \sum_{J_{f} l_{f}}
 \frac{2J_{f}+1}{2J_{i}+1} 
 \left[  D_{J_{i}j_{i}l_{i}}^{J_{f}j_{f}l_{f}}(\lambda s j_{c}) \right]^{2}
  \frac{\left|C^{j_{c}}_{J_{i}j_{i}l_{i}}\right|^{2}}{q^{2\lambda+1}}
 {\mathcal T}_{J_{i}j_{i}l_{i}}^{J_{f}j_{f}l_{f}}(\lambda j_{c})
\end{eqnarray}
with the dimensionless integral
\begin{equation} \label{eq:tdef}
{\mathcal T}_{J_{i}j_{i}l_{i}}^{J_{f}j_{f}l_{f}}(\lambda j_{c})
 = \frac{2}{\pi q^{2}} \int_{0}^{\infty} dk \: k \:
{\mathcal S}_{J_{i}j_{i}l_{i}}^{J_{f}j_{f}l_{f}}(\lambda j_{c})
 = \frac{2}{\pi} \int_{0}^{\infty} dx \: \left|
{\mathcal I}_{J_{i}j_{i}l_{i}}^{J_{f}j_{f}l_{f}}(\lambda j_{c}) 
 \right|^{2} \: .
\end{equation}
Evidently, the continuum contribution is only a fraction of the total strength
if the FSI allows the occurrence of bound
states in the corresponding channels.
Since we use only the exterior contribution to the radial integral
it is not guaranteed that the high energy behaviour of the
reduced radial integrals (\ref{eq:iredgen}) is correct. Indeed,
generally we do not find the $x^{-4}$ dependence for $E1$ transitions as
expected from the general considerations in section \ref{subsec:xs}.
For large $x$ there is a sizable contribution from the interior
integral that is necessary to generate the correct $x$ dependence
at high energies. However, the main contribution to the integral
(\ref{eq:tdef}) arises at small $x$ close to the threshold
and as long as the integral converges it will give a reasonable
approximation of the total transition strength.

\begin{table}
\caption{\label{tab:5}Functions 
${\mathcal T}_{l_{i}}^{l_{f}}(\lambda)$ 
for finite $\gamma$ and $b_{l_{f}} = 0$.}
\begin{tabular}{l}
 \hline 
 $\lambda = 0$ \\
 \hline 
 $\displaystyle {\mathcal T}_{0}^{0}(0) = 
 \exp(-2\gamma)/2$ \\
 $\displaystyle {\mathcal T}_{1}^{1}(0) = 
 (2+\gamma)\exp(-2\gamma)/(2\gamma)$ \\
 \hline 
 $\lambda = 1$ \\
 \hline 
 $\displaystyle {\mathcal T}_{1}^{0}(1) = 
 (5+6\gamma+2\gamma^{2}) \exp(-2\gamma)/4$ \\
 $\displaystyle  {\mathcal T}_{0}^{1}(1) = 
 (1+2\gamma+\gamma^{2}) \exp(-2\gamma)/4 $\\
 $\displaystyle {\mathcal T}_{2}^{1}(1) = 
 (36+37\gamma+14\gamma^{2} +2\gamma^{3}) \exp(-2\gamma)/(4\gamma)$\\
 $\displaystyle {\mathcal T}_{1}^{2}(1) = 
 (5+6\gamma+2\gamma^{2}) \exp(-2\gamma)/4$\\
 \hline 
 $\lambda = 2$ \\
 \hline 
 $\displaystyle {\mathcal T}_{2}^{0}(2) = 
 (63+90\gamma+54\gamma^{2}+16\gamma^{3}+2\gamma^{4}) 
 \exp(-2\gamma)/4$ \\
 $\displaystyle {\mathcal T}_{1}^{1}(2) = 
 (7+14\gamma+14\gamma^{2}+8\gamma^{3}+2\gamma^{4}) \exp(-2\gamma)/4$ \\
 $\displaystyle {\mathcal T}_{0}^{2}(2) = 
 (3+6\gamma+6\gamma^{2}+4\gamma^{3}+2\gamma^{4}) \exp(-2\gamma)/4$ \\
 \hline
\end{tabular}
\end{table}

If one assumes that there is no nucleon-core interaction in the final states
(i.e.\ plane waves)
the full strength lies in the continuum.
In this case the functions 
${\mathcal T}_{l_{i}}^{l_{f}}(\lambda)$ 
can be calculated analytically for neutron+core systems in our approach.
These functions are given in table \ref{tab:5}
where we suppressed unimportant quantum numbers for simplification.
The integrals ${\mathcal T}_{1}^{1}(0)$ and 
${\mathcal T}_{2}^{1}(1)$ diverge in the limit  $\gamma\to 0$
but remain finite for $\gamma > 0$.  In the limit $\gamma \to \infty$, i.e.\
if the bound state does not develop a halo, the functions become very small.

The total transition strength (\ref{eq:bel})
is proportional to 
$\left| C_{l_{i}} \right|^{2} {\mathcal T}_{l_{i}}^{l_{f}}(\lambda)
/q^{2\lambda+1}$.
The scaling of the ANC $C_{l_{i}}$ in the limit $\gamma \to 0$ can be
obtained by considering the square-well model 
(see appendix \ref{app:A}).
We find 
\begin{equation} \label{eq:slanc}
 |C_{l_{i}}|^{2} \propto \left\{ 
\begin{array}{lll}
 q & \mbox{if} & l_{i} = 0 \\
 q \gamma^{2l_{i}-1} & \mbox{if} & l_{i} \geq 1
\end{array}
 \right.
\end{equation}
for $\gamma \to 0$.
Taking this dependence into consideration
we find the scaling of $B(E1,l_{i})$
consistent with the expectation from the non energy-weighted sum rule.

Now, let us consider the case with non-vanishing nuclear interaction
in the continuum states. The FSI is parametrized in our approach by the
function $b_{l_{f}}$ that in general depends on the energy $E$.
If we assume that $b_{l_{f}}$ is constant for all energies
the functions ${\mathcal T}_{l_{i}}^{l_{f}}(\lambda)$ can be
calculated analytically again.
One obtains, e.g.\, for the $E1$ $p \to s$ transition
\begin{eqnarray} \label{eq:t101}
 {\mathcal T}_{1}^{0}(1) & = &  e^{-2\gamma} 
 \left\{ \begin{array}{lll}
  \left[ \frac{5+6\gamma+2\gamma^{2}}{4} 
 - 2b_{0}\frac{[1+2b_{0}+\gamma(1+b_{0})]^{2}}{(1+b_{0})^{4}}
\exp\left( -\frac{2\gamma}{b_{0}}\right)\right]
 & \mbox{for} & b_{0} \geq 0 \\
 \frac{5+6\gamma+2\gamma^{2}}{4}  & \mbox{for} & b_{0} \leq 0 
 \end{array} \right. 
 \: .
\end{eqnarray}
For other transitions the expressions become quite complicated and
we abstain from giving them here.
Since the function
$b_{l_{f}}$ cannot be assumed constant over the whole range of
continuum energies one cannot expect that the calculated
${\mathcal T}_{l_{i}}^{l_{f}}(\lambda)$ corresponds to
a reasonable result in general.
However, it is sufficient to assume 
that the function $b_{l_{f}}$ does not vary too much
over the peak of the transition strength
that is described by the shape
function ${\mathcal S}_{l_{i}}^{l_{f}}(\lambda)$.
For large $|b_{0}|$ the excitation spectrum is distorted significantly
in the relevant low-energy region.

\begin{figure}
\begin{center}
\includegraphics[width=135mm]{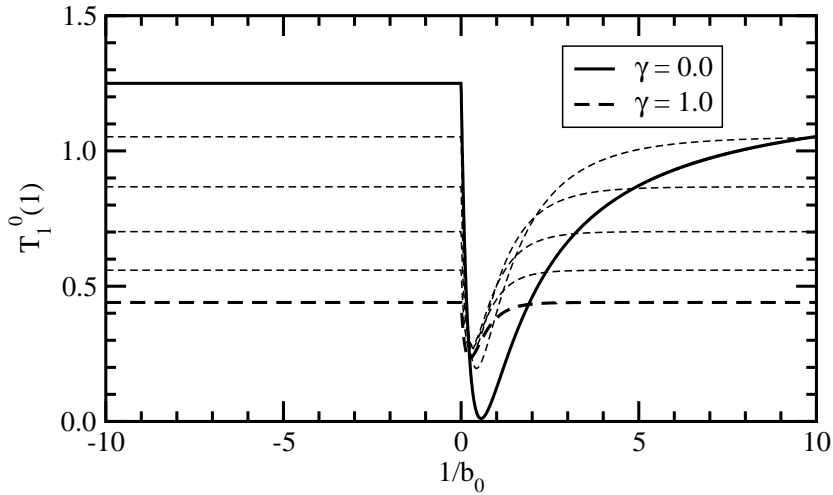}
\end{center}
\caption{\label{fig:3c} 
${\mathcal T}_{1}^{0}(1)$
as a function 
of  $1/b_{0}$ for values of
the parameter $\gamma$ between $0.0$ and $1.0$ in steps of $0.2$.}
\end{figure}

Equation (\ref{eq:t101}) is quite instructive.
It gives a good idea
how the $B(E1)$ strength is modified when the depth $V_{0}$
of the nuclear potential (with reasonable geometry) changes. 
For $b_{0} > 0$ there is a reduction of the total strength
in the continuum whereas in the case $b_{0} < 0$ no reduction
is found. This behaviour is related to the occurrence of bound states
in the $s$-wave continuum.
For $V_{0}=0$ the nuclear phase shift $\delta_{0}$
and the function $b_{0}$ is zero. 
Increasing the potential depth 
the phase shift at low energies becomes positive, corresponding to
a negative scattering length. Increasing the depth $V_{0}$ further
will lead to a more negative $b_{0}$ and a resonance will 
appear at low energies. Then there comes the point
where the potential will be able to support a bound state and
$b_{0}$ suddenly jumps to a large positive value. Increasing the
depth $V_{0}$ further will reduce the scattering length again.
Finally, $b_{0}$ becomes zero and the cycle starts again.
Considering this relation between strength of the potential
and the scattering length it is reasonable to plot
${\mathcal T}_{1}^{0}(1)$ as a function of $1/b_{0}$ as depicted 
in figure~\ref{fig:3c}. As long as the potential is not able to bind
a state the continuum $B(E1)$ strength does not change. As soon
as a resonance in the continuum becomes bound a sudden drop
of the transition strength is observed that recovers to its
initial value when $b_{0}$ approaches zero
from positive values. 
An analogous effect was found for the $s \to p$ transition
for the $E1$ excitation of ${}^{11}$Be in Ref.\ \cite{Typ04a}.
Similar observations are expected for other transitions.

This effect of the FSI has consequences for the extraction of the
ANC or the spectroscopic factor for an assumed ground-state
single-particle configuration from experimentally obtained
total reduced transition probabilities. When the value of $B(E1,l_{i})$ 
from the experiment is compared to the theoretical result assuming
a single-particle model with
plane waves in the continuum the extracted ANC or spectroscopic
factor will be underestimated if the actual nuclear potential
supports also bound states. 
In the latter case, a part of the total transition strength 
goes to these
bound states, even if they cannot be reached by a transition of the
halo nucleon because they are occupied by nucleons of the core.

Using the relation
\begin{equation}
 \sum_{f} (E_{f}-E_{i})\left| \langle f | r_{i} | i \rangle \right|^{2}
 = \frac{1}{2}
 \langle i | \left[ r_{i}, \left[ H, r_{i} \right] \right] | i \rangle
\end{equation}
the energy-weighted (or Thomas-Reiche-Kuhn) sum rule
\begin{eqnarray} \label{eq:ewsr}
 S(E1, J_{i} s j_{c} l_{i})
 & = &  \frac{2\mu}{\hbar^{2}}
 \sum_{J_{f} l_{f}} \int_{0}^{\infty} dE \: (E+S_{b}) \:
 \frac{dB}{dE}(E1, J_{i} s j_{c} l_{i} \to 
  k J_{f} s j_{c} l_{f})
 \\ \nonumber & = & \frac{9}{4\pi} \left[ Z_{\rm eff}^{(1)} e \right]^{2}
\end{eqnarray}
for dipole transitions is obtained. 
In constrast to the non energy-weighted sum rule (\ref{eq:newsr})
it gives a result independent of the orbital angular
momentum $l_{i}$ of the bound state. In principle, the sum contains
all bound and unbound states that can be reached 
from the initial state by an $E1$ transition.
Since the photon energy is $E_{\gamma} = E_{f}-E_{i} = E+S$
the sum rule (\ref{eq:ewsr}) can be expressed in terms of 
the total $E1$ photo absorption cross section
(\ref{eq:sigabs}) as
\begin{eqnarray}
 \lefteqn{\int_{0}^{\infty} dE_{\gamma} \: \sigma_{E1}(a+\gamma \to b+c)}
 \\ \nonumber 
 & = & \frac{2\pi^{2}\hbar}{\mu c} \left[ Z_{\rm eff}^{(1)} e \right]^{2} 
 =  \frac{2\pi^{2}\hbar e^{2}}{m c} 
\left( \frac{N_{a}Z_{a}}{A_{a}} - \frac{N_{b}Z_{b}}{A_{b}} 
 - \frac{N_{c}Z_{c}}{A_{c}} \right)
\end{eqnarray}
for the disintegration into core+nucleon.
It has the form of a cluster
sum rule since we only consider the excitation of the halo nucleon but not
of the core \cite{Alh82,Hen04}.
In our approach we can write
for the continuum contribution
\begin{eqnarray}  \label{eq:ewsr2}
 \lefteqn{S_{\rm cont} (E1, J_{i} s j_{c}l_{i})}
 \\ \nonumber & = & 
   \left[Z_{\rm eff}^{(1)}e\right]^{2} \sum_{J_{f} l_{f}}
 \frac{2J_{f}+1}{2J_{i}+1} 
 \left[  D_{J_{i}j_{i}l_{i}}^{J_{f}j_{f}l_{f}}(1 s j_{c}) \right]^{2}
  \frac{\left|C^{j_{c}}_{J_{i}j_{i}l_{i}}\right|^{2}}{q}
 {\mathcal U}_{J_{i}j_{i}l_{i}}^{J_{f}j_{f}l_{f}}(1 j_{c})
\end{eqnarray}
with the dimensionless integral
\begin{eqnarray} 
{\mathcal U}_{J_{i}j_{i}l_{i}}^{J_{f}j_{f}l_{f}}(1 j_{c})
 & = & \frac{2}{\pi q^{4}} \int_{0}^{\infty} dk \: k \: 
 (q^{2}+k^{2})\:
{\mathcal S}_{J_{i}j_{i}l_{i}}^{J_{f}j_{f}l_{f}}(1 j_{c})
 \\ \nonumber 
 & = & \frac{2}{\pi} \int_{0}^{\infty} dx \: (1+x^{2}) \: \left|
{\mathcal I}_{J_{i}j_{i}l_{i}}^{J_{f}j_{f}l_{f}}(1 j_{c}) 
 \right|^{2} \: .
\end{eqnarray}
If there are bound states in the final channel we expect
$S_{\rm cont} (E1, J_{i} s j_{c}l_{i}) < S (E1, J_{i} s j_{c}l_{i})$.
Here again, the remarks on the high energy behaviour of the integrand
apply as for the functions (\ref{eq:tdef}). For the most important
transition $s \to p$ 
without interaction in the continuum states we obtain the finite result
\begin{equation}
  {\mathcal U}_{0}^{1}(1) = 
  e^{-2\gamma} \frac{3+2\gamma}{2}
\end{equation}
where we suppressed irrelevant quantum numbers.
Considering the scaling law (\ref{eq:slanc}) of the ANC for $l_{i}=0$
we find the correct result (\ref{eq:ewsr})
for the  energy-weighted sum rule
in the limit $\gamma \to 0$. For the ground state with $l_{i}=1$ there
are two contributions from $l_{f}=0$ and $l_{f}=2$ in the final state.
In this case the integral ${\mathcal U}_{1}^{2}(1)$ diverges with
$\gamma^{-1}$ but this dependence is compensated by the scaling of the
ANC. However, since the radial integrals 
in the external approximation do not vanish fast enough with increasing $x$,
the correct value for the energy-weighted sum rule 
is not recovered in the limit $\gamma \to 0$.
 
\section{Examples for transition strengths of nucleon+core nuclei}
\label{sec:examples}

Let us now look at some specific 
examples of exotic nuclei with nucleon+core structure.
The dependence of the reduced transition probability and of the 
cross sections for electromagnetic transitions
on the strength of the fragment-fragment
interaction in the final state
can be studied in the simple but realistic single-particle model of
subsection~\ref{subsec:trans}. 
For the nuclear potential a
Woods-Saxon shape with
the representative parameters $a=0.65$~fm and 
$R_{0}=r_{0}A^{1/3}$ where $r_{0}=1.25$~fm is assumed.
The depth of the
potential in the bound state is adjusted to reproduce the experimental
neutron and proton separation energies, respectively. In order to
simulate a varying interaction strength in the final states 
the transition strength is studied for 
the depth of the Woods-Saxon potential in the continuum 
in the range from 0 to
80~MeV. In Tables \ref{tab:B1} and \ref{tab:B2} the separation energy $S_{b}$, 
potential radius $R_{0}$, potential depth $V_{0}$, and orbital angular
momentum $l_{i}$ 
for the ground state are given for the neutron+core nuclei 
and proton+core nuclei, respectively. For simplicity, no spin-orbit
potential is considered in the calculations.
Also the characteristic parameter $\gamma$ is given in Tables
\ref{tab:B1} and \ref{tab:B2}. Only nuclei with $\gamma <1$ can be
considered to be halo nuclei, however, the boundary to nuclei without
an extended nucleon distribution is smooth. For p+core nuclei
the parameter $\eta_{i}$ is given in addition. A larger value
indicates a stronger dominance of Coulomb distortions of the 
$dB(E1)/dE$ strength.
In the case of neutron+core nuclei $E1$ $s\to p$ and $d\to p$ transition
will be considered. In the case of proton+core nuclei we will study
especially
$E1$ $p \to s$ excitations.
For many observables, a square-well model would
also be a good approximation, especially for low
energy processes where the ``shape independence'' 
is valid to a large extent.  

\subsection{Neutron+core nuclei}

\begin{table}
\caption{\label{tab:B1}Neutron 
separation energy $S_{n}$, radius $R_{0}$, depth $V_{0}$
of the Woods-Saxon potential and parameter $\gamma=qR_{0}$
for the ground state with orbital angular momentum $l_{i}$
of the  neutron+core nuclei in the single-particle model.
The diffuseness parameter is $a=0.65$~fm.}
\begin{center}
\begin{tabular}{ccccccc}
  \hline
  &       & ${}^{11}$Be & ${}^{15}$C & ${}^{17}$O &
            ${}^{23}$O  & ${}^{23}$O \\
  \hline
  $S_{n}$ & [MeV] & 0.504   & 1.218   & 4.140   & 
                    2.740   & 2.740   \\
  $R_{0}$ & [fm]  & 2.78    & 3.08    & 3.21    & 
                    3.55    & 3.55    \\
  $V_{0}$ & [MeV] & 54.3765 & 48.562  & 55.3726 & 
                    42.416  & 43.1932 \\
  $l_{i}$ &       & 0       & 0       & 2       & 
                    0       & 2       \\
  $\gamma$ &      & 0.413   & 0.722   & 1.393   &
                    1.264   & 1.264   \\
  \hline 
\end{tabular}
\end{center}
\end{table}

\begin{figure}
\begin{center}
\includegraphics[width=120mm]{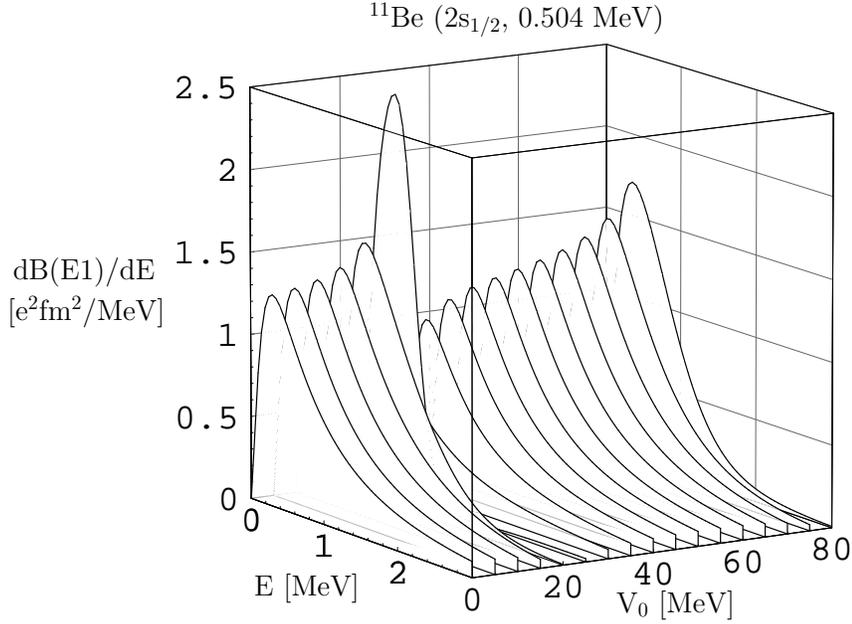}
\end{center}
\caption{\label{fig:6} 
Reduced transition probability
$dB(E1)/dE$ for the breakup of ${}^{11}$Be into a neutron and ${}^{10}$Be
as a function of the c.m.\ energy $E$ 
for various depths $V_{0}$ of the potential in the contiuum.
The quantum numbers for the single particle bound state and the
neutron separation energy in MeV are given in parenthesis.}
\end{figure}

The $\frac{1}{2}^{+}$ 
ground state of ${}^{11}$Be is considered as a neutron in a
$2s_{1/2}$ state coupled to the $0^{+}$ ground state of ${}^{10}$Be.
The experimental neutron separation energy of $S_{n} = 0.504$ 
MeV is
quite small so that ${}^{11}$Be is a typical example of a halo nucleus.
It was studied by Coulomb dissociation in the experiments 
\cite{Nak94,Ann94,Pal03,Fuk04}.
In Fig.~\ref{fig:6} the reduced transition probability 
$dB(E1)/dE$ for a transition into $p$ waves is shown for 
varying depths $V_{0}$ in the continuum between 0 and 80~MeV
in steps of 5~MeV. The E1 strength shows a large peak at relative
energies well below 1~MeV typical for a halo nucleus. The shape
changes smoothly with increasing potential depth except for $V_{0}$
close to 30~MeV where a $p$-wave resonance in the continuum becomes
bound. For a potential depth similar to the one required for
the correct separation energy in the ground state (cf. Tab.~\ref{tab:B1})
the shape of the E1 transition strength is very similar to the
result for a plane wave, i.e.\ $V_{0}=0$~MeV.
Experimental data for the $dB(E1)/dE$ distribution in ${}^{11}$Be 
were described in \cite{Typ04a} in our approach by fitting the
asymptotic normalization coefficient and the constants $c_{1}^{3/2}$
and $c_{1}^{1/2}$ in the effective-range expansion for the phase shifts
in the $p$ waves with total angular momentum $j=3/2$ and $j=1/2$.
An unnaturally large value for $c_{1}^{1/2}$ was found that is related 
to the low separation energy of the first bound excited state in ${}^{11}$Be.
The shape of the dipole strength function in the effective-range approach 
was found to agree very well to a calculation in the Woods-Saxon model.
A corresponding figure and more details can be found in Ref.\ \cite{Typ04a}.

\begin{figure}
\begin{center}
\includegraphics[width=120mm]{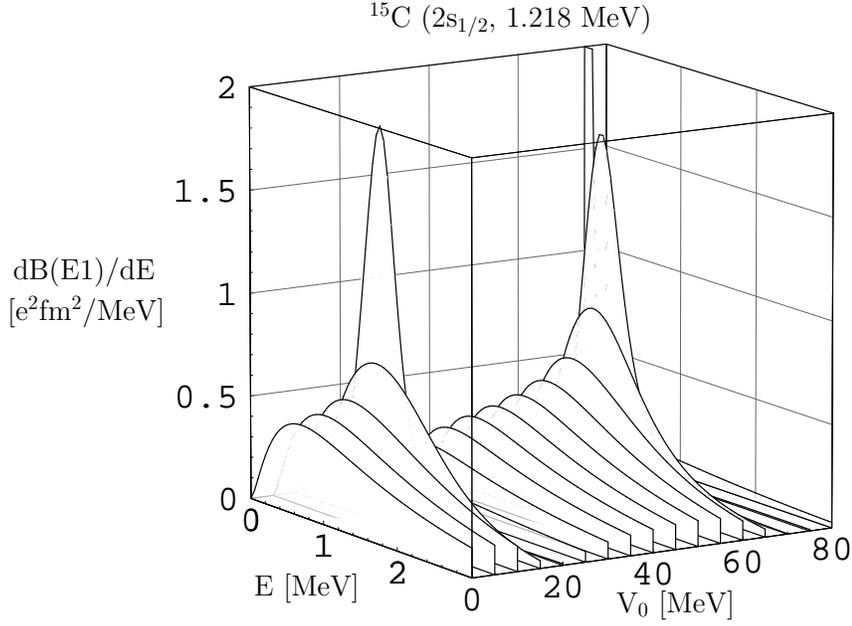}
\end{center}
\caption{\label{fig:7}
Same as Fig.~\ref{fig:6} but for the 
breakup of ${}^{15}$C into a neutron and ${}^{14}$C.}
\end{figure}

Another example of a neutron+core nucleus with the neutron in
a $s$ wave ground state is ${}^{15}$C but with a larger neutron separation
energy of $S_{n}=1.218$~MeV \cite{Dat03}. Comparing the dependence of 
the dipole transition strength $dB(E1)/dE$ on the potential depth $V_{0}$
as shown in Fig.~\ref{fig:7} with the case of ${}^{11}$Be one observes
a stronger variation of the shape and absolute value. 
This clearly shows the increased sensitivity of the transition
strength on the final-state interaction when the neutron separation
energy increases as expected from the analytical model in 
subsection~\ref{subsec:fsinc}. Again,
the reduced transition probability changes quite abruptly
when a $p$ wave resonance in the continuum becomes a bound state.

\begin{figure}
\begin{center}
\includegraphics[width=120mm]{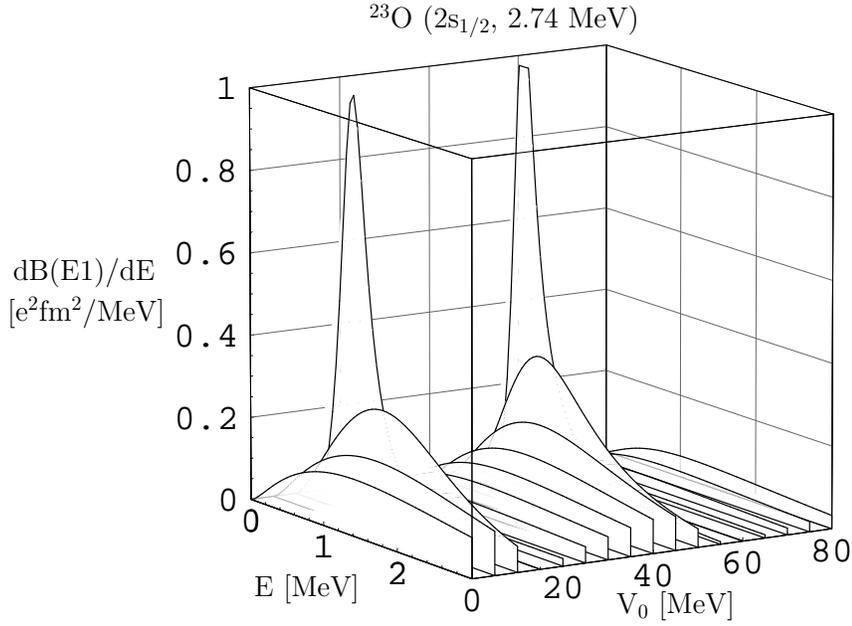}
\end{center}
\caption{\label{fig:8}
Same as Fig.~\ref{fig:6} but for the 
breakup of ${}^{23}$O in a $s$-wave ground state
into a neutron and ${}^{22}$O.}
\end{figure}

An even more drastic change of the transition strength is observed
in the case of ${}^{23}$O when the neutron is assumed to be in a
$s$-wave ground state, see Fig.~\ref{fig:8}. 
This nucleus with a neutron separation 
energy of $S_{n}=2.74$~MeV cannot be considered as a real halo nucleus.
Yet, it can be said that there is substantial low-energy 
$E1$ strength.

\begin{figure}
\begin{center}
\includegraphics[width=120mm]{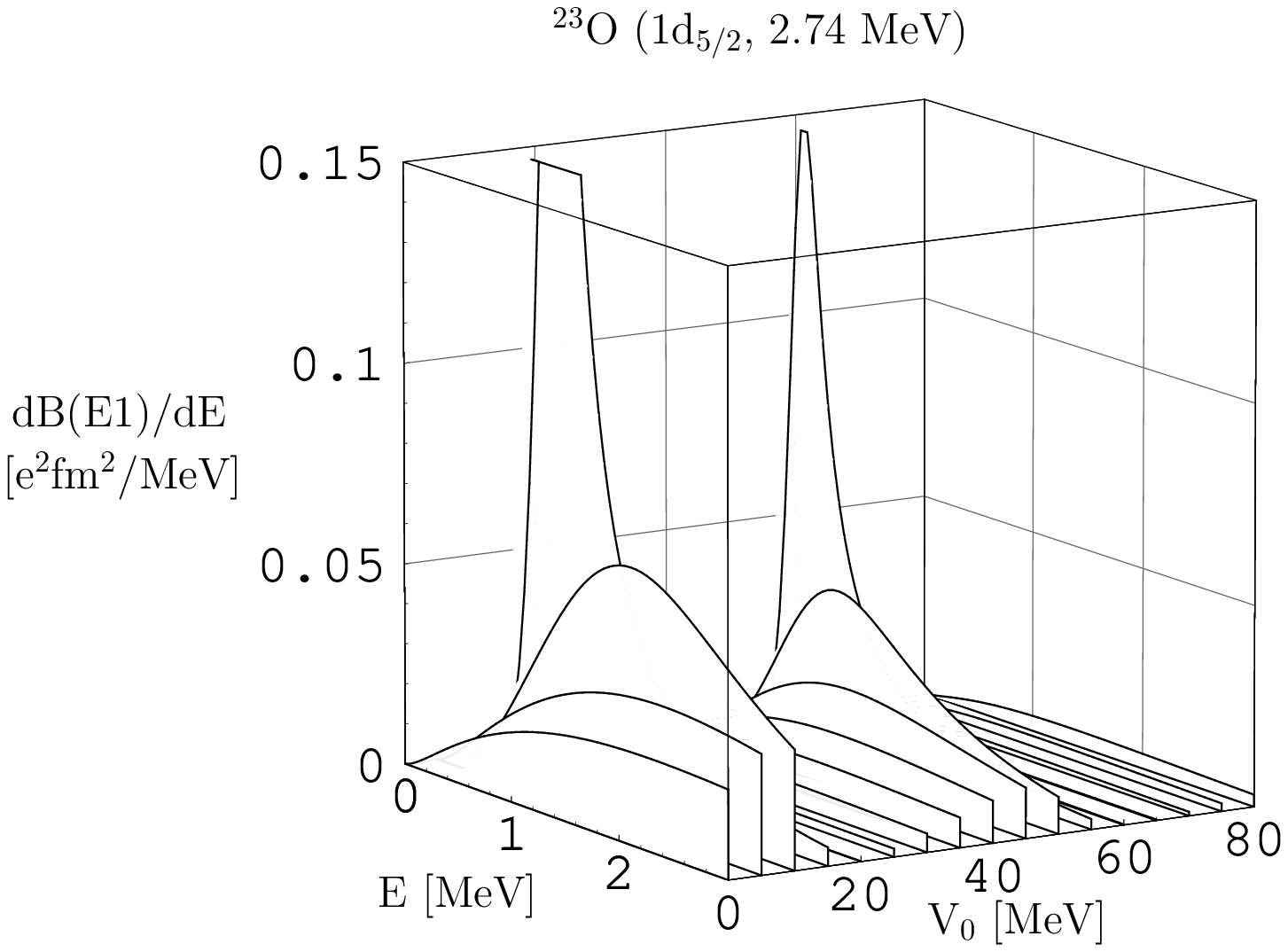}
\end{center}
\caption{\label{fig:9}
Same as Fig.~\ref{fig:6} but for the 
breakup of ${}^{23}$O in a $d$-wave ground state
into a neutron and ${}^{22}$O.}
\end{figure}

Comparing the three cases ${}^{11}$Be, ${}^{15}$C, and ${}^{23}$O
one observes a shift of the 
peak in the transition strength to higher relative energies in the
continuum. The position of the peak scales with the separation energy
as expected from the dependence of the shape functions on the
ratio $x^{2}=S_{n}/E$ in Section~\ref{sec:rrisf}. Additionally, one finds
a reduction of the overall strength as suggested from
Eqs.~(\ref{eq:dbelde}) and (\ref{eq:bel}) 
with an increasing value of the ground state 
parameter $q$.

Since the ground state angular momentum of ${}^{23}$O is not known
uniquely from experiments \cite{Noc04,Kan01,Kan02,Bro03,Tho03,Sau04,Cor04} 
it is also possible that the
neutron in the ground state occupies a $d_{5/2}$ single-particle state.
The strength function for an $E1$ transition to $p$-wave final
states under this assumption is shown in Fig.~\ref{fig:9}.
The absolute magnitude is about a factor 4 smaller than in the case
of a $s$-wave neutron in the ground state and the maximum is
shifted to larger relative energy. This is explained by 
the additional centrifugal barrier in the $d$ wave that reduces
the probability of finding the neutron at large distances from the
core in the classically forbidden region.

\begin{figure}
\begin{center}
\includegraphics[width=120mm]{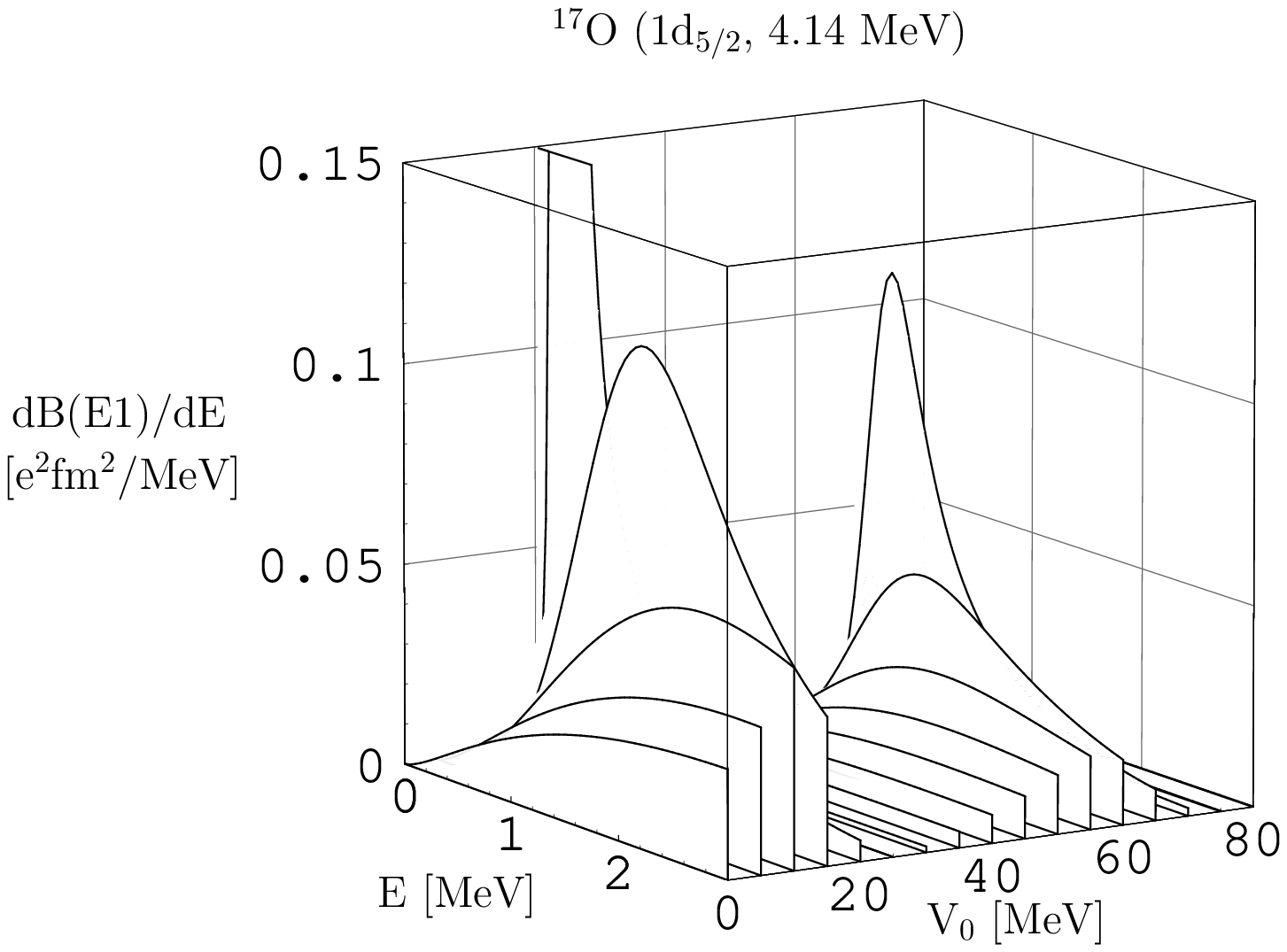}
\end{center}
\caption{\label{fig:10}
Same as Fig.~\ref{fig:6} but for the 
breakup of ${}^{17}$O into a neutron and ${}^{16}$O.}
\end{figure}

A further example for a neutron+core nucleus with a $d$-wave 
ground-state configuration is ${}^{17}$O, a nucleus close to the valley of
stability. Here the separation energy
of $S_{n}=4.14$~MeV is even larger than in the case of ${}^{23}$O
and the shape and magnitude of the reduced $E1$ transition probability
to $p$ waves in the final state vary drastically with an increasing
strength of the final-state interaction.

\begin{figure}
\begin{center}
\includegraphics[width=135mm]{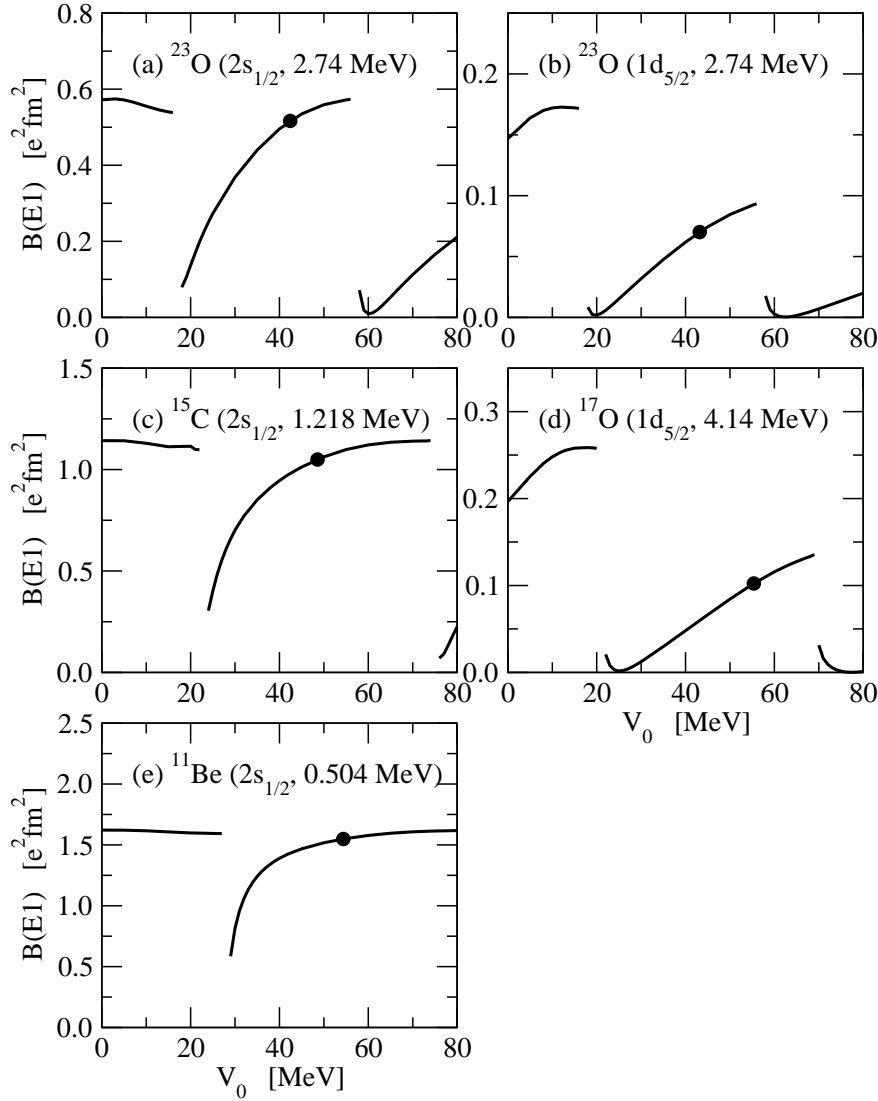}
\end{center}
\caption{\label{fig:11} Total reduced transition probability $B(E1)$
for the breakup into a neutron+core with c.m. energies $E$
between 0 and 10~MeV as a function of the potential
depth $V_{0}$ for the continuum states. The filled circle gives
the result for the same potential depth as for the single particle
bound state.
The quantum numbers for the single particle bound state and the
neutron separation energy for each system are given in parenthesis.}
\end{figure}

In Fig.~\ref{fig:11} the total $E1$ transition strength integrated
from 0 to 10~MeV is shown for the nuclei discussed above as a function
of the potential depth $V_{0}$ in the continuum. There is
a distinctive difference between nuclei with a $s$-wave and $d$-wave
neutron in the ground state. In the former case the integrated strength
stays almost constant when $V_{0}$ is increased starting at 0~MeV,
i.e.\ the plane-wave result. Then there is a sudden drop of the total
$E1$ strength when the $p$-wave resonance from the continuum becomes
a bound state. However, the total strength recovers to its value
for $V_{0}=0$ when the potential depth is increased further
until the next higher $p$-wave resonance becomes bound. 
The dependence of the total transition probability on the strength
of the FSI is
in qualitative agreement with the expectations from the analytical
neutron+core model as shown in figure \ref{fig:3c}.
The rise
of the total $E1$ strength after the continuum to bound state
crossing is faster when the neutron separation energy is smaller.
For halo nuclei there is only a small variation of the total
reduced transition probability if one is not too close to a potential
strength where the crossing appears. When the neutron is in the
$d$-wave ground state there is already a considerable
dependence of the total $E1$ strength on the depth of the 
potential close to the plane-wave limit. Beyond the sudden drop of the 
strength at the continuum-to-bound-state crossing the
absolute value increases again but it does not reach the plane-wave
limit again. Comparing the total $E1$ strength calculated
for a continuum potential that is identical to the bound state
potential (filled circle in Fig.~\ref{fig:11}) one immediately
finds that the integrated $E1$ strength is almost the same as
in the plane-wave case for a $s$-wave neutron in the ground state
but it is significantly smaller for a $d$-wave neutron.

\subsection{Proton+core nuclei}

\begin{table}
\caption{\label{tab:B2}Proton 
separation energy $S_{p}$, radius $R_{0}$, depth $V_{0}$
of the Woods-Saxon potential, parameters $\gamma = qR_{0}$
and $\eta_{i}$
for the ground state with orbital angular momentum $l_{i}$
of the  proton+core nuclei in the single-particle model.
The diffuseness parameter is $a=0.65$~fm.}
\begin{center}
\begin{tabular}{ccccc}
  \hline
  &       & ${}^{8}$B  & ${}^{9}$C   & ${}^{12}$N \\
  \hline
  $S_{p}$ & [MeV] & 0.1375  & 1.296   & 0.601  \\
  $R_{0}$ & [fm]  & 2.50    & 2.60    & 2.86   \\
  $V_{0}$ & [MeV] & 43.183  & 44.4456 & 36.474 \\
  $l_{i}$ &       & 1       & 1       & 1      \\
  $\gamma$ &      & 0.190   & 0.613   & 0.466  \\
  $\eta_{i}$ &    & 1.595   & 0.655   & 1.171  \\
  \hline 
\end{tabular}
\end{center}
\end{table}

Systems with proton+core structure show similar features as compared
to neutron+core systems, however, with modifications due to the
appearence of the 
Coulomb barrier. Since analytical results for the
shape functions are not available one has to resort to the numerical
calculations, e.g.\ in the simple single-particle model.

As examples we consider the three unstable nuclei 
${}^{8}$B \cite{Xu94,Tra01,Azh01,Oga03b,Tra03,Jun02,Sch03}, 
${}^{9}$C \cite{Bea01,Tra02,End03} and
${}^{12}$N \cite{Lef95,Tan03,Tim03}. 
Their ground state is well described by a proton in
a $p$-wave bound state. The proton wave function is calculated in
a single-particle model.  
The depth of the potentials was adjusted to reproduce the experimental 
binding energies as in the case of the neutron+core nuclei. 
The corresponding parameters of
the Woods-Saxon potential are given in table~\ref{tab:B2}.
As a consequence the ANC of the bound state 
wave functions is uniquely determined in this model.

\begin{figure}
\begin{center}
\includegraphics[width=120mm]{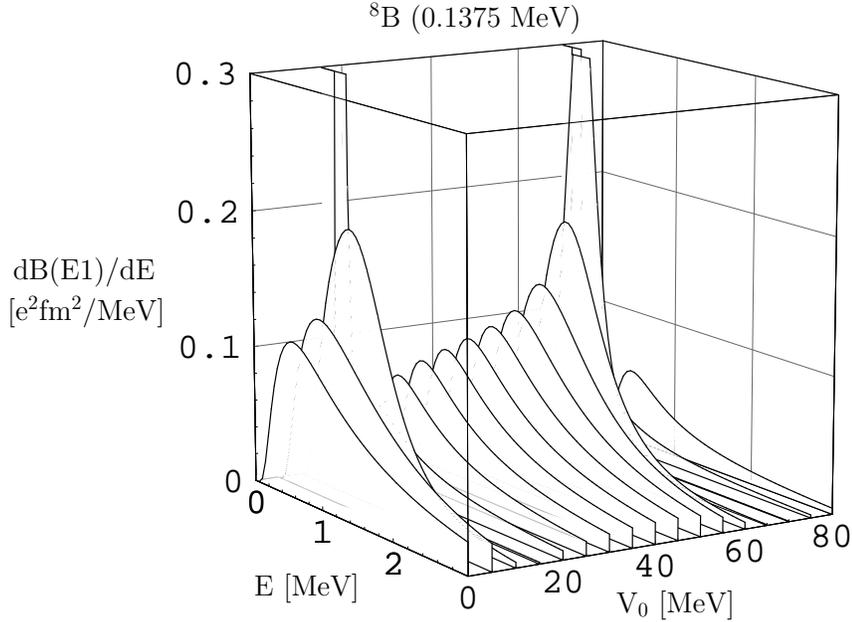}
\end{center}
\caption{\label{fig:13} 
Reduced transition probability
$dB(E1)/dE$ for the breakup of ${}^{8}$B into a proton and ${}^{7}$Be
into a $s$ wave continuum state
as a function of the c.m.\ energy $E$ 
for various depths $V_{0}$ of the potential in the contiuum.
The proton separation energy in MeV is given in parenthesis.}
\end{figure}

\begin{figure}
\begin{center}
\includegraphics[width=120mm]{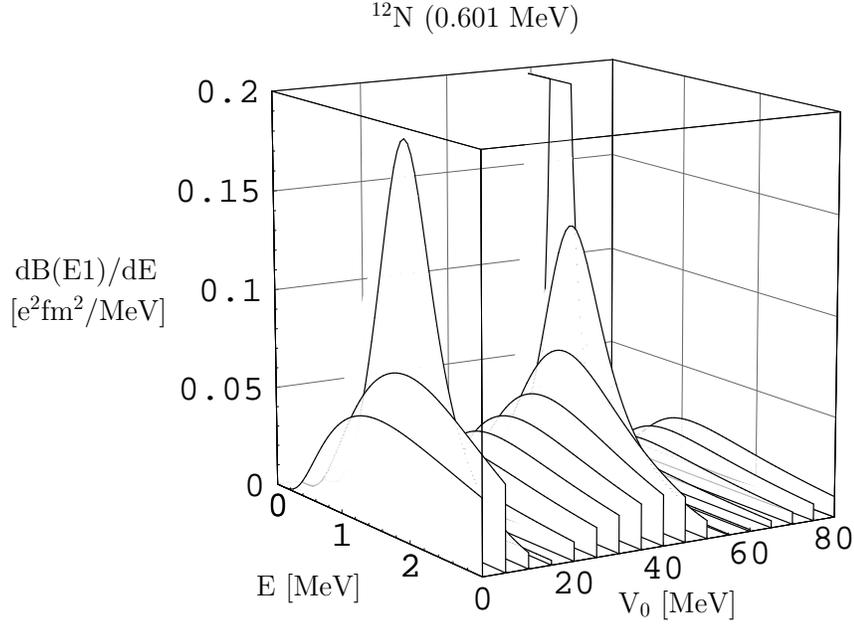}
\end{center}
\caption{\label{fig:14} 
Same as Fig.~\ref{fig:13} but for the 
breakup of ${}^{12}$N into a proton and ${}^{11}$C.}
\end{figure}

\begin{figure}
\begin{center}
\includegraphics[width=120mm]{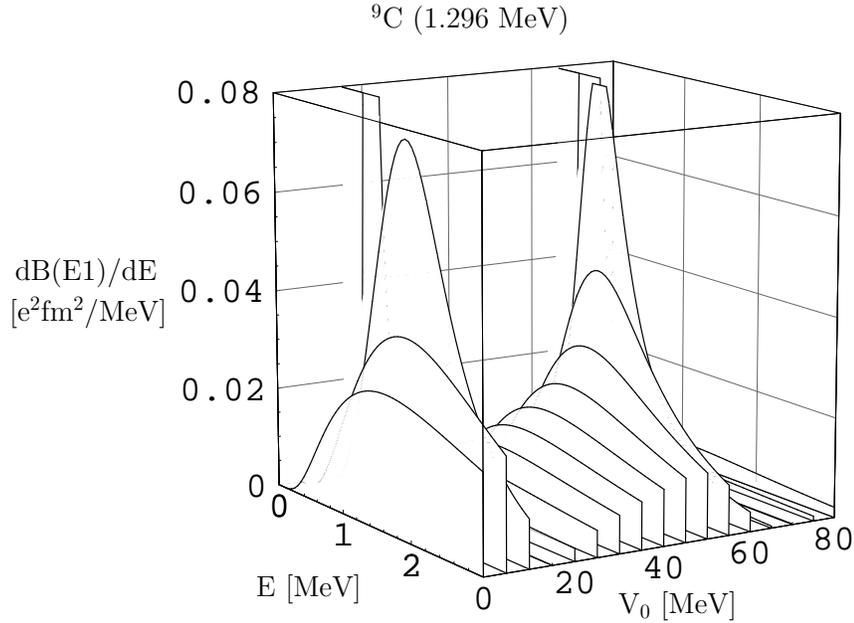}
\end{center}
\caption{\label{fig:15} 
Same as Fig.~\ref{fig:13} but for the 
breakup of ${}^{9}$C into a proton and ${}^{8}$B.}
\end{figure}

\begin{figure}
\begin{center}
\includegraphics[width=135mm]{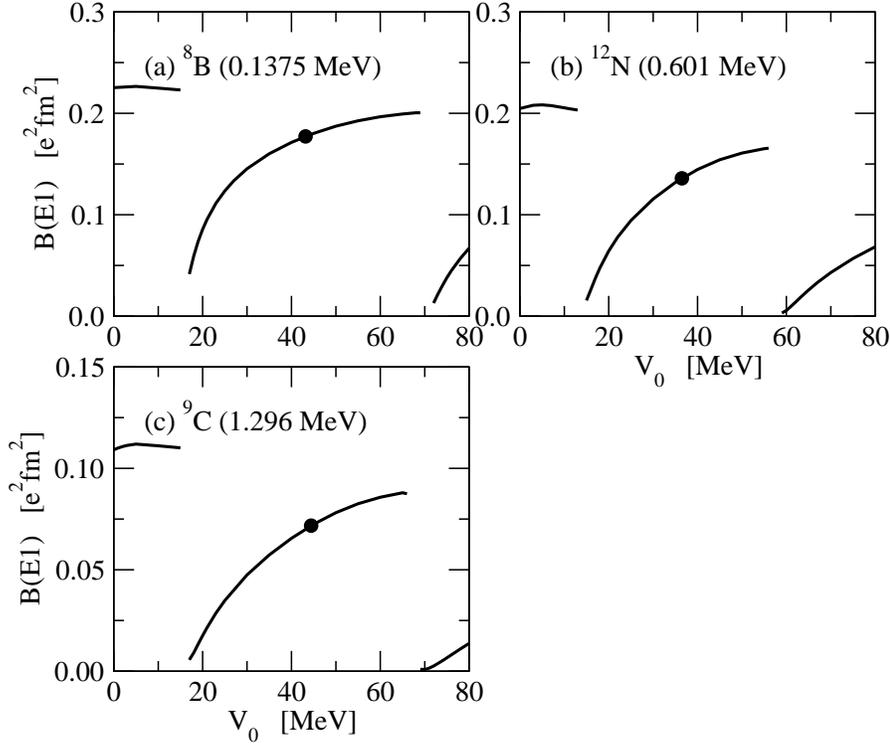}
\end{center}
\caption{\label{fig:16} Reduced transition probability $B(E1)$
for the breakup into a proton+core with c.m. energies $E$ between
0 and 10~MeV as a function of the potential
depth $V_{0}$ for the continuum states. The filled circle gives
the result for the same potential depth as for the single particle
bound state.
The proton separation energy for each system is given in parenthesis.}
\end{figure}

The reduced transition probability $dB(E1)/dE$ as a function of
the relative energy $E$ is shown
in figures \ref{fig:13}, \ref{fig:14}, and \ref{fig:15} for these
nuclei for various
depths of the continuum potential. In the calculation
it was assumed that the channel spin $S$ of the ground state is
given by $S=j_{c}+1/2$ with the core spin $j_{c}$.
The  additional Coulomb interaction between nucleon and core leads to a
substantial modification of the shape functions.
The transition strength still peaks at low relative energies,
however, the maximum is shifted to higher energies 
as compared to the neutron+core case.
For a pure Coulomb wave in the continuum state, i.e.\ $V_{0}=0$,
the maximum is reached at an energy of $E/S_{p}= 4.28$, $2.57$ and
$1.26$, for ${}^{8}$B, ${}^{12}$N, and ${}^{9}$C,
respectively. The larger shift of the maximum to higher energies 
corresponds to an increase of the parameter $\eta_{i}$,
cf.\ Table \ref{tab:B2}.
In the neutron+core case the maximum is expected
at a value of only $0.18$. 
The overall shape of the transition strength
changes similarly with the depth of the potential
as for neutron+core nuclei and the occurence of resonances in
the $p$-wave continuum is observed again. 
Also the integrated reduced transition probability, as depicted in figure
\ref{fig:16}, resembles in its dependence on the potential depth the 
results for neutron+core systems between the two cases of
$s \to p$ and $d \to p$ transitions. The $B(E1)$ value assuming
the same potential in the continuum as in the bound state is clearly
smaller than the value for pure Coulomb waves in the continuum.

\section{Low-energy behaviour and ANC method}
\label{sec:anc}

At small relative energies the reduced transition probability
$dB(E\lambda)/dE \propto {\mathcal S}_{l_{i}}^{l_{f}}(\lambda)$
is strongly suppressed 
due to penetration effects of the centrifugal barrier. It is useful to
change to a different quantity to study the low-energy
behaviour and the effects of the
nucleon-core interaction in the continuum. 
Considering the approximation of the shape functions
for small $x$ in table \ref{tab:4}
one can use
\begin{equation}
 \tilde{\mathcal S}_{l_{i}}^{l_{f}}(\lambda) 
 = x^{-2l_{f}-1} {\mathcal S}_{l_{i}}^{l_{f}}(\lambda)
\end{equation}
with a finite limit for $x \to 0$
in the case of neutron+core systems.
This quantitiy 
takes the angular momentum barrier into account. 
Correspondingly the quantities 
\begin{equation}
 \tilde{\mathcal S}_{l_{i}}^{l_{f}}({\rm abs},\lambda)
 =x^{-2l_{f}-1}
 {\mathcal S}_{l_{i}}^{l_{f}}({\rm abs},\lambda)
 = x^{-2l_{f}-1} (1+x^{2})^{2\lambda-1}{\mathcal S}_{l_{i}}^{l_{f}}(\lambda)
\end{equation}
and 
\begin{equation} \label{eq:scaptmod}
 \tilde{\mathcal S}_{l_{i}}^{l_{f}}({\rm capt},\lambda) 
 = x^{-2l_{i}+1}
 {\mathcal S}_{l_{i}}^{l_{f}}({\rm capt},\lambda)
 = x^{-2l_{i}-1}(1+x^{2})^{2\lambda+1}
  {\mathcal S}_{l_{f}}^{l_{i}}(\lambda)
\end{equation} 
are the relevant forms with a finite value for $x \to 0$
for photon absorption and radiative capture,
respectively,
cf.\ equations (\ref{eq:sabs}) and (\ref{eq:scapt}).
(Note that for the capture reaction the initial and final state 
are interchanged.)
The low-energy behaviour is clearly dominated by the centrifugal barrier
in the continuum state. From equation (\ref{eq:scaptmod})
one finds the famous $1/v \propto 1/x$ law of the cross section
for the neutron capture from a $s$ wave
in the continuum $l_{i}=0$.
The above quantities approach a finite value in the limit $E\to 0$
that depends on the function $b_{l_{f}}$.
Thus, the interaction in the continuum 
has an effect on the absolute normalization of the low-energy cross
sections.

Apart from the typical $x^{2l_{f}+1}$ dependence the expressions
for ${\mathcal S}_{l_{i}}^{l_{f}}(\lambda)$ contain a factor
$(1+x^{2})^{2\lambda+2}$ 
in the denominator that corresponds to a pole at 
$x^{2}=-1$. Considering equation (\ref{eq:scaptmod}) one sees
that the shape function for capture reactions 
has a simple pole at the separation energy $E=-S_{n}$.
This pole limits the range of convergence
of an expansion of the shape functions or cross sections
in terms of the parameter $x$ or the energy $E$ \cite{Bay04}.
Since the separation energy is very small for halo systems a
corresponding expansion has only limited applicability in this case.

A similar effect is observed for proton+core systems \cite{Jen98,Bay00}.
Here, the low-energy behaviour is dominated by the Coulomb barrier
in the continuum state. In nuclear astrophysics one needs, e.g., 
cross sections for radiative proton capture at very low energies. 
Instead of extrapolating the strongly energy dependent capture
cross section $\sigma_{\pi \lambda}(p+c \to a+\gamma)$ to low energies 
the astrophysical S factor
\begin{equation}
 S(E) = \sigma_{\pi \lambda}(p+c \to a+\gamma) E \exp(2\pi\eta_{i})
\end{equation}
is employed. It is weakly dependent on energy and approaches a finite
value for $E \to 0$.
The exponential term depending on the Sommerfeld
parameter $\eta_{i}$ in the continuum state cancels in leading order
the suppression of the capture cross section due to the Coulomb barrier.
The factor $E$ is proportional to the $k^{2}$ factor in the theorem
of detailed balance (\ref{eq:detbal}) from the phase space in the continuum.
Considering the scaling of the shape functions for
proton+core systems in section \ref{subsec:p+core}
the generalitzation of (\ref{eq:scaptmod}) is found.
The quantity
\begin{equation} 
 \tilde{\mathcal S}_{l_{i}}^{l_{f}}({\rm capt},\lambda) 
 = C_{l_{i}}^{2}(\eta_{i}) x^{2}
 {\mathcal S}_{l_{i}}^{l_{f}}({\rm capt},\lambda)
 =  C_{l_{i}}^{2}(\eta_{i}) (1+x^{2})^{2\lambda+1}
  {\mathcal S}_{l_{f}}^{l_{i}}(\lambda)
\end{equation} 
approaches a finite value for $x \to 0$ because
$C_{l_{i}}^{2}(\eta_{i}) \to \eta_{i}^{2l_{i}+1} \exp(-2\pi\eta_{i})$ 
for $\eta_{i} = \eta_{f}/x \to \infty$ with constant $\eta_{f}$.
It follows that
\begin{eqnarray}
 S(E) & \propto & \tilde{\mathcal S}_{l_{i}}^{l_{f}}({\rm capt},\lambda) 
 \propto \frac{{\rm const.}}{1+x^{2}}
\end{eqnarray}
for $x\to 0$.
For small separation energies of the proton in the bound state
a strong increase of the S factor at small energies is observed due to
the closeness of the pole at $E=-S_{p}$.
In general, the absolute value of $S(0)$ depends on the ANC of the ground state
and the scattering length, i.e.\ the strength of the continuum interaction.

An interesting case is the direct $E1$ radiative capture from the 
p + ${}^{16}$O
continuum to the first excited ($1/2^{+}$) state in ${}^{17}$F
that is relevant to nuclear astrophysics \cite{Bru96,Mor97}.
This state with $l=0$ and a small proton
separation energy of 105~keV can be considered as a typical halo state.
Assuming a radius of $R=3.15$~fm we find $\gamma=0.217$ and
$\eta_{i}=3.787$.
At low energies the $p \to s$ transition gives the main contribution 
to the astrophysical S factor (apart from the small contribution of the
$p \to d$ transition to the $5/2^{+}$ ground state with $S_{p} = 600$~keV).
Due to the closeness of the pole at $E=-S_{p}$ a strong rise of the
astrophysical S factor is observed for $E \to 0$ that is practically
independent of the potential in the continuum.

\begin{figure}
\begin{center}
\includegraphics[width=135mm]{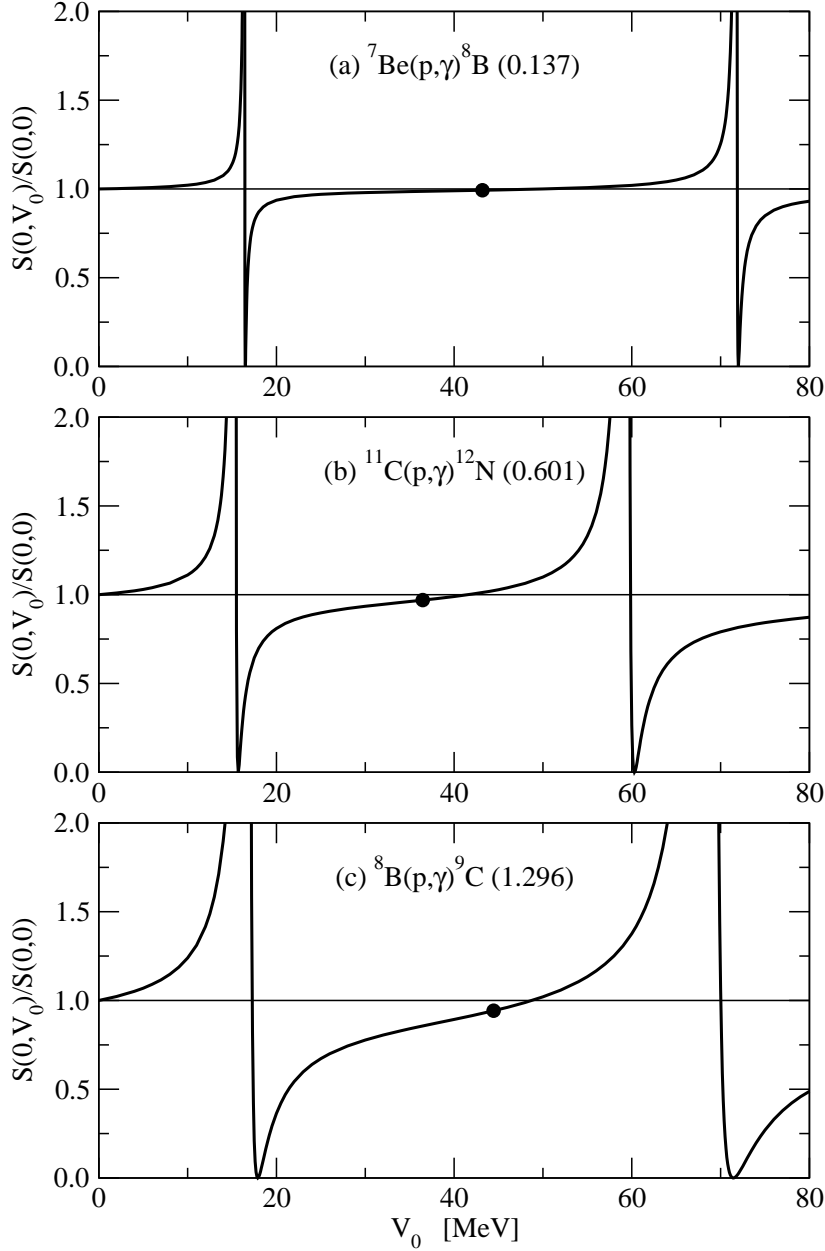}
\end{center}
\caption{\label{fig:12} S factor $S(E,V_{0})$
at energy $E=0$ as a function of the
depth $V_{0}$ of a Woods-Saxon potential in the continuum 
scaled to the S factor calculated with pure Coulomb waves for three
radiative capture reactions. The numbers in parenthesis are the 
separation energies of the proton. The filled circle shows
the result assuming the same potential depth in the continuum
as for the bound state.}
\end{figure}

The sensitivity of the low-energy S factor to the interaction in the
continuum can be
estimated by comparing $S(0)$ for various p+core systems
as a function of the interaction strength.
In figure~\ref{fig:12} the
dependence of the S factor at zero energy is shown for the
reactions (a) ${}^{7}$Be(p,$\gamma$)${}^{8}$B,
(b) ${}^{11}$C(p,$\gamma$)${}^{12}$N, and
(c) ${}^{8}$B(p,$\gamma$)${}^{9}$C.
In all cases we have an $E1$ $s \to p$ transition.
The S factor
is calculated with varying depths of the Woods-Saxon potential
in the continuum $s$ waves. Spin-orbit contribution to the potential
are neglected for simplicity. 
For an easier comparision all zero-energy S factors are divided by
the zero-energy S factor for pure Coulomb waves, i.e.\ $V_{0}=0$
for the nuclear potential. These are given by
(a) $22.07$~eV~b, (c) $129.4$~eV~b, and (b) $68.41$~eV~b
in the present model.
As is clearly seen in figure~\ref{fig:12}
the nuclear proton-core interaction
directly has an effect on the zero-energy S factor. There
is a considerable variation, especially if the depth $V_{0}$ approaches
a value where an $s$-wave continuum state becomes a bound state, e.g.\ at 
$V_{0}=16.3$~MeV and $71.6$~MeV in case of the
${}^{7}$Be(p,$\gamma$)${}^{8}$B reaction. Furthermore, we find that
the sensitivity to the potential in the continuum state increases
with the separation energy of the proton. Assuming the same depth
for the continuum potential as for the bound state a S factor is 
obtained that is somehow smaller, but close the value from a 
calculation with pure Coulomb waves.

Our results indicate that the zero-energy S factor
is not uniquely determined by the ground state ANC 
that can be extracted from
experiments.
In general it is necessary to
consider the nuclear interaction in the continuum that can be 
parametrized at small energies by the scattering lengths in the
relevant partial waves \cite{Bay04,Bay00}. 
One example of particular interest is
the ${}^{7}$Be(p,$\gamma$)${}^{8}$B reaction where experimental
data from direct and indirect experiments for the S factor
have to be extrapolated to zero energy assuming a certain
energy dependence from theoretical models \cite{Dav03}. 
For the ${}^{7}$Be+p system $s$-wave scattering
lengths $a_{S}$ for channel spin S=2 and S=1 (from coupling the spins of
the proton and ${}^{7}$Be) have been determined experimentally,
however with large uncertainties \cite{Ang03}.
More precise values are available
for the mirror system ${}^{7}$Li+n \cite{Koe83}. 
The scattering lengths are easily
reproduced in the single-particle model by adjusting the depth of the
Woods-Saxon potential in the continuum. As is seen clearly in figure
3 of Ref.\ \cite{Dav03} the energy dependence of the S factor shows
a considerable variation that affects the extrapolation of the
experimental data to zero energy.

\section{Conclusions and Outlook}

In this paper we study single-particle aspects
of neutron- and proton-halo nuclei,
i.e.\ the effects which appear when the separation energy 
of the least-bound nucleon tends to zero. 
These nuclear systems
display universal features, 
e.g.\ for the electromagnetic transition strength,
that can be described in simple models.

We use different approaches in our theoretical studies:
one is a square-well model, for which we obtain remarkably simple
analytical results. A more realistic and conventional 
approach is the model with Woods-Saxon potentials.
For our applications to halo nuclei
we find that it is, even quantitatively,
quite similar to the square-well model. 
The underlying reason is quite simple:
For the low energies relevant for the halo systems
there is shape independence: the properties of the 
potential are encoded in some low-energy parameters
which do not depend on the details of the chosen specific 
potential. Halo nuclei are low-energy phenomena, related to large 
wavelengths. It is well known that
a probe of a given wavelength is insensitive to details of 
structure at distances much smaller than this
wavelength \cite{Lep97}. This means that we can mimic the 
real short-distance structure (e.g.\ determined from
the nuclear many-body problem or a Woods-Saxon
single-particle model) by a simple short distant structure,
e.g.\ a square well model with its agreable analytical properties. 
The depth and radius parameters serve as adjustable 
parameters that, e.g., reproduce the binding
energy and the scattering length.

We mention also that there is a need to 
supplement full scale ab-initio microscopic 
approaches to nuclear structure by some empirical
fine tuning of parameters to reproduce the actual
position of loosely bound single particle structure,
see e.g.\ \cite{Sag01}. This is indispensible
in order to obtain a satisfactory description of
low-lying electromagnetic strength.

Following a study of some general single-particle properties 
we consider the radial matrix elements for electric excitations to the
continuum.
They essentially depend on the asymptotic wave functions of the
bound and scattering states.
We identify the important parameters and find 
rather simple analytical formulae for   
the most relevant cases for neutron
halo nuclei. The most important parameter is 
$\gamma=qR$ depending on the separation energy $S = \hbar^{2}q^{2}/(2\mu)$
and the size $R$ of the system.  It is shown that $\gamma$
serves as a convenient expansion parameter.
Proton halo nuclei are studied numerically.
We find simple scaling laws for the transition
strength. We also discuss typical
examples for light halo nuclei in the Woods-Saxon model.

We quantitatively analyze the effects of the 
final-state interaction in the continuum on the transition strength.
We find that it is determined mainly by the 
scattering length. In general, this nucleon-core
interaction has an influence on the radiative capture
process down to very small energies.
This can serve as a warning for the simple
application of the ANC method: the results are rather insensitive to the
potential in the continuum only for true halo systems.

The power of the present approach is 
demonstrated by recent applications to actual nuclei \cite{Typ04a,Noc04}.
It will serve also as a framework for further applications,
notably for the upcoming radioactive beam facilities RIKEN,
FAIR/GSI, RIA. One important conclusion is that we quantitatively 
understand halo effects. It is expected from our
studies that neutron-halo effects will show up
irrespective of the mass number $A$, whereas 
proton halo effects will tend to disappear
with increasing charge number $Z$.

We hope that methods similar to ones developed in this paper
will also be useful for the (much) more complicated
problems of two-, three, and more nucleon halo nuclei
\cite{Rii00,Jen04}.
One early discussion of that problem is 
\cite{Mig73}. A review on modern approaches in context of effective
field theories and references are found in \cite{Bra04}.
%


\section{Acknowledgments}

We greatfully acknowledge discussions with
T. Aumann, U. Datta Pramanik, D. Baye, H. Emling, G. Hansen, and C. Nociforo.

\section*{Note added in proof:}
In the meantime a paper ``Low-lying dipole strength for weakly bound
systems: A simple analytic estimate'' \cite{Nag05} appeared that contains 
results which are related to this paper.

\appendix 
\section{Neutron in a square-well potential}
\label{app:A}

The problem of a neutron in a square-well potential is well studied
for bound and scattering states
in the literature, see e.g.\ \cite{Mes61}, and many analytic results
have been obtained. However, often only results for 
a few selected cases are presented.
Here, formulas for arbitrary values of the orbital angular
momentum are given by generalizing earlier calculations
and simplifying the expressions. 
We also derive results that, to our knowledge, were not quoted in the
literature before.

\subsection{Wave functions and probabilities}
\label{app:A1}

The bound-state wave function
\begin{equation}
 \Phi_{nlm}^{i}(\vec{r}) = \frac{f_{nl}(r)}{r} Y_{lm} (\hat{r})
\end{equation}
for a neutron with orbital angular momentum $l$ and projection $m$
in a square-well potential of Radius $R$ and
depth $V_{n}$ is completely characterized by the separation
energy $S>0$ and the principal quantum number $n=1,2,\dots$ that counts the
number of nodes (including the node at $r=0$). 
The radial wave function
\begin{equation}
 f_{nl}(r) = \left\{ \begin{array}{ll}
 A_{nl} r j_{l}(\bar{q}_{nl} r) & \mbox{for} \quad r \leq R \\
 B_{nl} r i^{l} h^{(1)}_{l}(iq r) & \mbox{for} \quad r \geq R
 \end{array} \right.
\end{equation}
is expressed in terms of spherical Bessel and Hankel functions
$j_{l}$ and $h^{(1)}_{l}$ \cite{Abr65}. The quantities
$q = \sqrt{2\mu S}/\hbar$
and
$\bar{q}_{nl} = \sqrt{2\mu(V_{nl}-S)}/\hbar$
in the arguments are related by
\begin{equation}
 q^{2} + \bar{q}_{nl}^{2} = \frac{2\mu V_{nl}^{i}}{\hbar^{2}}
\end{equation}
where an increasing number of radial nodes $n$ requires
larger values of  $\bar{q}_{nl}$ and of the depth $V_{nl}^{i}$ 
of the potential in the ground state.
The relation between $q$ and $\bar{q}_{nl}$ for given $n$
is determined by the continuity of the logarithmic derivative
\begin{equation}
 L_{nl}^{i} = \left.
 \frac{r\frac{d}{dr}f_{nl}(r)}{f_{nl}(r)} \right|_{r=R}
\end{equation}
of the radial wave function $\phi_{nl}(r)$ at $r=R$. This condition
can be written as
\begin{equation}
 L_{nl}^{i} = l+1 - Y_{l}^{(+)} = Y_{l}^{(-)} - l
\end{equation}
with
\begin{equation}
 Y_{l}^{(\pm)} = 
 \bar{\gamma}_{nl} 
 \frac{j_{l\pm 1}(\bar{\gamma}_{nl})}{j_{l}(\bar{\gamma}_{nl})} = 
 i \gamma \frac{h^{(1)}_{l\pm 1}(i\gamma)}{h^{(1)}_{l}(i\gamma)} 
\end{equation}
where
\begin{equation} \label{eq:vi}
 \gamma = q R \: ,\quad \bar{\gamma}_{nl} = \bar{q}_{nl} R \quad
 \mbox{and} \quad
 \gamma^{2} + \bar{\gamma}_{nl}^{2} 
 = \frac{2\mu R^{2} V_{nl}^{i}}{\hbar^{2}}  = v_{i} \: .
\end{equation}
The (complex) constants $A_{nl}$ and $B_{nl}$ 
are determined by the continuity
condition for the wave function
\begin{equation} \label{eq:cont}
  A_{nl} j_{l}(\bar{\gamma}_{nl}) = B_{nl} i^{l} h^{(1)}_{l}(i\gamma)
\end{equation}
and the normalization condition
\begin{equation}
 1 = P^{<}_{nl} + P^{>}_{nl} 
\end{equation}
with
\begin{equation}
 P^{<}_{nl} = 
 \left|A_{nl}\right|^{2} 
 \int_{0}^{R} dr \: r^{2} \: \left[j_{l}(\bar{q}_{nl} r) \right]^{2}
\end{equation}
and
\begin{equation}
 P^{>}_{nl} = 
 \left|B_{nl}\right|^{2} \int_{R}^{\infty} 
 dr \: r^{2} \: \left|h^{(1)}_{l}(iq r) \right|^{2} \: .
\end{equation}
These radial integrals are easily evaluated (see Appendix \ref{app:E})
with the results
\begin{eqnarray}
 P^{<}_{nl} & = & 
  \frac{|B_{nl}|^{2}}{2}R^{3}  \left| h^{(1)}_{l}(i\gamma) \right|^{2}
   \left( 
 \frac{\gamma^{2}}{\bar{\gamma}_{nl}^{2}} X_{l} + 1 \right)
\end{eqnarray}
and
\begin{eqnarray}
 P^{>}_{nl} & = & 
   \frac{|B_{nl}|^{2}}{2}R^{3}  \left| h^{(1)}_{l}(i\gamma) \right|^{2}
   \left( X_{l} - 1  \right)
\end{eqnarray}
where the continuity equation (\ref{eq:cont}) was used.
The quantity
\begin{equation}
 X_{l} = -\frac{Y_{l}^{(-)}Y_{l}^{(+)}}{\gamma^{2}}
 = \frac{h^{(1)}_{l-1}(i\gamma)
 h^{(1)}_{l+1}(i\gamma)}{\left[ h^{(1)}_{l}(i\gamma) \right]^{2}} 
 = \frac{K_{l-1/2}(\gamma)
   K_{l+3/2}(\gamma)}{\left[K_{l+1/2}(\gamma)\right]^{2}}
\end{equation}
is a rational function in the variable $\gamma$.
From the ratio 
\begin{equation}
 \frac{P^{<}_{nl}}{P^{>}_{nl}} =  
 \frac{\gamma^{2}X_{l}+\bar{\gamma}_{nl}^{2}}{\bar{\gamma}_{nl}^{2}
 \left[X_{l}-1\right]} 
\end{equation}
the probability of finding the neutron at radii $r \leq R$ 
\begin{equation}
 P_{nl} = \frac{P^{<}_{nl}}{P^{<}_{nl}+P^{>}_{nl}} 
 =  \frac{\gamma^{2} X_{l} + \bar{\gamma}_{nl}^{2}}{\left(\bar{\gamma}_{nl}^{2}
+ \gamma^{2} \right) X_{l}}
\end{equation}
is obtained.
In the limit $\gamma \to 0$ we have
\begin{equation}
 \lim_{\gamma \to 0} X_{l} = \left\{ 
\begin{array}{ll}
 \infty & \mbox{for} \quad l=0 \\
 \frac{2l+1}{2l-1}& \mbox{for} \quad l>0
\end{array}\right.
\end{equation}
and 
\begin{equation}
 \lim_{\gamma \to 0} P_{nl} = \left\{ 
\begin{array}{ll}
 0 & \mbox{for} \quad l=0 \\
 \frac{2l-1}{2l+1}& \mbox{for} \quad l>0
\end{array}\right. 
\end{equation}
independent of the principal quantum number $n$.
Equivalent expressions for neutrons 
were given recently for $l=0,1,2$ 
in Ref.~\cite{Liu04}.
 
In case of a scattering state the wave function has the form
\begin{equation}
 \Phi_{lm}^{f}(\vec{r})
  = 4\pi i^{l} \frac{g_{l}(r)}{kr} Y_{lm}(\hat{r}) Y_{lm}^{\ast}(\hat{k})
\end{equation}
for energy $E=\hbar^{2}k^{2}/(2\mu)$
with the radial wave function  
\begin{equation}
 g_{l}(r) = \left\{ \begin{array}{ll} 
 \bar{A}_{nl} \bar{k}_{nl} r j_{l}(\bar{k}_{nl}r) & \mbox{for} \quad r\leq R
 \\
 \frac{1}{2i} 
 \left[ \exp(2i\delta_{l}) u_{l}^{(+)}(kr) - u_{l}^{(-)}(kr) \right]
 & \mbox{for} \quad r\geq R
 \end{array} \right.
\end{equation}
where 
\begin{equation}
 u_{l}^{(\pm)} (x) = x \left[-y_{l}(x) \pm i j_{l}(x) \right]
\end{equation}
and $j_{l}$ and $y_{l}$ are
spherical Bessel and Neumann functions, respectively \cite{Abr65}.
The phase shift is denoted by $\delta_{l}$ and 
$\bar{k}_{nl} = \sqrt{2\mu(E+V_{nl})}/\hbar$
where $V_{nl}$ is the depth of the square-well potential
that gives the correct separation energy $S$ of the neutron.
The continuity condition for the logarithmic derivative
\begin{eqnarray}
 L_{nl}^{f}  & = &
 \left. \frac{r \frac{d}{dr} g_{l}(r)}{g_{l}(r)} \right|_{r=R}
 = 1 + \frac{\bar{\kappa}_{nl}
 j_{l}^{\prime}(\bar{\kappa}_{nl})}{j_{l}(\bar{\kappa}_{nl})}
 \\ \nonumber & = & 
 \frac{\kappa\left[ \exp(2i\delta_{l}) u_{l}^{(+)\prime}(\kappa)
 - u_{l}^{(-)\prime}(\kappa)\right]}{\exp(2i\delta_{l})u_{l}^{(+)}(\kappa)
 - u_{l}^{(-)}(\kappa)}
\end{eqnarray}
at $r=R$ with 
\begin{equation}
 \kappa = kR \qquad \bar{\kappa}_{nl} = \bar{k}_{nl} R
\end{equation}
determines the phase shift $\delta_{l}$ from
\begin{equation}
 \exp(2i\delta_{l}) = 
 \frac{L_{nl}^{f}u_{l}^{(-)}(\kappa)-\kappa u_{l}^{(-)\prime}
 (\kappa)}{L_{nl}^{f}u_{l}^{(+)}(\kappa)-\kappa 
 u_{l}^{(+)\prime}(\kappa)} \: .
\end{equation}
The phase shift is a sum
\begin{equation}
 \delta_{l} = \tau_{l} + \rho_{l}
\end{equation}
of the hard-sphere phase shift $\tau_{l}$ with
\begin{equation}
 \exp(2i\tau_{l}) = \frac{u_{l}^{(-)}(\kappa)}{u_{l}^{(+)}(\kappa)}
\end{equation}
and the additional phase shift $\rho_{l}$ with
\begin{equation}
 \exp(2i\rho_{l}) = 
 \frac{L_{nl}^{f}-q_{l}^{(-)}(\kappa)}{L_{nl}^{f}
 -q_{l}^{(+)}(\kappa)}
\end{equation}
where
\begin{equation}
 q_{l}^{(\pm)}(\kappa) = 
 \frac{\kappa u_{l}^{(\pm)\prime}(\kappa)}{u_{l}^{(\pm)}(\kappa)}
 \: .
\end{equation}
The relation
\begin{equation} \label{eq:vf}
  \bar{\kappa}_{nl}^{2} - \kappa^{2} =
 \frac{2\mu V_{nl}^{f}R^{2}}{\hbar^{2}} = 
  v_{f}
\end{equation}
with the depth $V_{nl}^{f}$ of the potential in the
scattering state
determines the quantity $\bar{\kappa}_{nl}$. 
Generally, $V_{nl}^{i}$ in (\ref{eq:vi}) can be different from
$V_{nl}^{f}$.
In the limit $k \to 0$ we have
\begin{equation}
 \tan(\delta_{nl}) \to - a_{l} k^{2l+1}
\end{equation}
with the scattering length
\begin{equation} \label{eqn:scalen_l}
 a_{l} = a_{l}^{hs} 
 \left( 1 - \frac{2l+1}{L_{nl}^{f}(0)+l}\right)
\end{equation}
where
\begin{equation}
 a_{l}^{hs} = \left\{ \begin{array}{lll}
 R & \mbox{if} & l=0 \\
 \frac{R^{2l+1}}{(2l+1)!!(2l-1)!!} & \mbox{if} & l>0
\end{array} \right. 
\end{equation}
is the scattering length of a hard sphere of radius $R$, 
i.e.\ $g_{l}(R)=0$ corresponding to
$L_{nl}^{f} \to \infty$.

Defining the penetrability
\begin{equation}
 P_{l}(x) = \left| u_{l}^{(\pm)}(x) \right|^{-2}
 = x^{-2} \left[y_{l}^{2}(x) + j_{l}^{2}(x)\right]^{-1}
\end{equation}
we can write
\begin{equation}
 u_{l}^{(\pm)} (x) = P_{l}^{-\frac{1}{2}}(x) 
 \exp\left[ \mp i \tau_{l}(x)\right]
\end{equation}
with the hard-sphere phase shift $\tau_{l}$.
Then the continuity of the wave function at $r=R$
\begin{eqnarray} \label{eq:contf}
 \bar{A}_{nl} \bar{\kappa}_{nl} j_{l}(\bar{\kappa}_{nl}) & = &
\frac{1}{2i} 
 \left[ \exp(2i\delta_{l}) u_{l}^{(+)}(\kappa) - u_{l}^{(-)}(\kappa) \right] 
 \\ \nonumber & = & 
  P_{l}^{-\frac{1}{2}}(\kappa) \:
  \exp(i\delta_{l}) \:
  \sin \left( \delta_{l}-\tau_{l}\right)
\end{eqnarray}
fixes the constant $\bar{A}_{nl}$. 
The differential probability of finding the neutron
inside the square-well potential of radius $R$
is given by
\begin{eqnarray}
 \frac{dP_{nl}}{dk} & = & 
 k^{2} \int \frac{d\Omega_{k}}{(2\pi)^{3}} 
 \int d\Omega_{r} \int_{0}^{R} dr \: r^{2} \: \left| \Phi_{lm}^{f} \right|^{2}
 \\ \nonumber & = & 
 \frac{R}{\pi}
  \left| \bar{A}_{nl} \right|^{2} \bar{\kappa}_{nl}^{2}
 \left(\left[j_{l}(\bar{\kappa}) \right]^{2} 
 - j_{l-1}(\bar{\kappa}) j_{l+1}(\bar{\kappa})\right)
 \\ \nonumber & = & 
 R
 \frac{\sin^{2}\left( \delta_{l}
 -\tau_{l}\right)}{\pi P_{l}(\kappa)}
 \left( 1 - \frac{(L_{nl}^{f}+l)}{\bar{\kappa}_{nl}} 
 \frac{(l+1-L_{nl}^{f})}{\bar{\kappa}_{nl}} \right)
 \: .
\end{eqnarray}
Using the continuity relations for the logarithmic derivative
one obtains
\begin{eqnarray} \label{eq:pnlcont}
 \frac{dP_{nl}}{dx} & = & 
 \frac{\gamma}{4\pi}
 \left(  \left| \exp(i\delta_{l}) u_{l}^{(+)}(\kappa) 
- \exp(-i\delta_{l}) u_{l}^{(-)}(\kappa) \right|^{2} 
 \right. \\ \nonumber & & \left.
 - \frac{\kappa^{2}}{\bar{\kappa}_{nl}^{2}} 
 [\exp(i\delta_{l}) u_{l-1}^{(+)}(\kappa) 
- \exp(-i\delta_{l}) u_{l-1}^{(-)}(\kappa)]
 \right. \\ \nonumber & & \left.
 [\exp(i\delta_{l}) u_{l+1}^{(+)}(\kappa) 
- \exp(-i\delta_{l}) u_{l+1}^{(-)}(\kappa)]^{\ast}
 \right)
 \\ \nonumber & = & 
 \frac{\gamma}{\pi}
 \left( \frac{\sin^{2}\left( \delta_{l}
 -\tau_{l}\right)}{P_{l}(\kappa)}
 - \frac{\kappa^{2}}{\bar{\kappa}_{nl}^{2}} 
 \frac{\sin\left( \delta_{l}
 -\tau_{l-1}\right)\sin\left( \delta_{l}
 -\tau_{l+1}\right)}{\sqrt{P_{l-1}(\kappa)
 P_{l+1}(\kappa)}}
 \right)
 \: .
\end{eqnarray}
The probability is determined mainly by 
the penetrabilities $P_{l,l\pm 1}$ and the differences of the
phase shift $\delta_{l}$ from the hard-sphere values $\tau_{l,l\pm 1}$.

\subsection{Root-mean-square radius}
\label{app:A2}

The root-mean-square radius of the bound state wave function
is obtained from
\begin{equation}
 \langle r^{2} \rangle = \frac{R_{2}}{R_{0}}
\end{equation}
with
\begin{equation}
  R_{n} = \int_{0}^{\infty} dr \: r^{n+2} \: \left| \phi_{nl}(r) \right|^{2}
\end{equation}
With the help of the integrals in appendix \ref{app:E} and the continuity
equation for the wave function and the logarithmic derivative we find
\begin{eqnarray}
 R_{0} & = & 
 \left| A_{nl} \right|^{2} \int_{0}^{R} dr \: r^{2} \: 
 \left[ j_{l}(\bar{q}_{nl}r)\right]^{2} 
 + \left| B_{nl} \right|^{2} \int_{R}^{\infty} dr \: r^{2} \: 
 \left| h^{(1)}_{l}(iqr)\right|^{2} 
 \\ \nonumber & = & 
 \frac{\left| A_{nl} \right|^{2}}{\bar{q}_{nl}^{3}} 
\frac{\bar{\gamma}_{nl}}{2} \left[ j_{l}(\bar{\gamma}_{nl}) \right]^{2} 
 \left( \bar{\gamma}_{nl}^{2}
 -  Y_{l}^{(-)} Y_{l}^{(+)}
 \right)
 \\ \nonumber & & 
 - (-1)^{l} \frac{\left| B_{nl} \right|^{2}}{(iq)^{3}} 
\frac{i\gamma}{2} \left[ h^{(1)}_{l}(i\gamma) \right]^{2} 
 \left( [i\gamma]^{2}
   -   Y_{l}^{(-)} Y_{l}^{(+)}\right)
 \\ \nonumber & = & 
 - \frac{R^{3}}{2} \left| B_{nl} \right|^{2}
 \left| h^{(1)}_{l}(i\gamma) \right|^{2}\left(
 \frac{1}{\bar{\gamma}_{nl}^{2}} + \frac{1}{\gamma^{2}}
 \right) Y_{l}^{(-)} Y_{l}^{(+)} 
 \: .
\end{eqnarray}
Similarly
we obtain
\begin{eqnarray}
 R_{2} & = & 
 \left| A_{nl} \right|^{2} \int_{0}^{R} dr \: r^{4} \: 
 \left[ j_{l}(\bar{q}_{nl}r)\right]^{2} 
 + \left| B_{nl} \right|^{2} \int_{R}^{\infty} dr \: r^{4} \: 
 \left| h^{(1)}_{l}(iqr)\right|^{2} 
 \\ \nonumber & = & 
 \frac{\left| A_{nl} \right|^{2}}{\bar{q}_{nl}^{5}} 
\frac{\bar{\gamma}_{nl}}{12} [j_{l}(\bar{\gamma}_{nl})]^{2} 
 \left( 3 \bar{\gamma}_{nl}^{4}
 -2 \bar{\gamma}_{nl}^{2}
 Y_{l}^{(-)} Y_{l}^{(+)}
 \right.
 \\ \nonumber & & \left.
 -  \left\{(2l-1)Y_{l}^{(-)} 
 -\bar{\gamma}_{nl}^{2} \right\}
 \left\{(2l+3)Y_{l}^{(+)} 
 -\bar{\gamma}_{nl}^{2} \right\}
 \right) 
 \\ \nonumber & & 
 - (-1)^{l} \frac{\left| B_{nl} \right|^{2}}{(iq)^{5}} 
\frac{i\gamma}{12} [h^{(1)}_{l}(i\gamma)]^{2} \left( 3 (i\gamma)^{4}
 -2 (i\gamma)^{2} Y_{l}^{(-)} Y_{l}^{(+)}
 \right.
 \\ \nonumber & & \left.
 -  \left\{(2l-1)Y_{l}^{(-)} 
 -(i\gamma)^{2} \right\}
 \left\{(2l+3) Y_{l}^{(+)} 
 -(i\gamma)^{2} \right\}
 \right) 
 \\ \nonumber & = & 
 \left| B_{nl} \right|^{2} \left|h^{(1)}_{l}(i\gamma)\right|^{2} 
 \frac{R^{5}}{12}
 \left(\frac{1}{\bar{\gamma}_{nl}^{2}}+\frac{1}{\gamma^{2}} \right) 
 \left[-2 Y_{l}^{(-)} Y_{l}^{(+)}
\right. 
 \\ \nonumber & & 
 + \left\{ (2l+3) Y_{l}^{(+)} 
 + (2l-1) Y_{l}^{(-)}  
 \right\}
 \\ \nonumber & & \left.
 - \left(\frac{1}{\bar{\gamma}_{nl}^{2}} 
 - \frac{1}{\gamma^{2}} \right) \left\{
 (2l-1)(2l+3) Y_{l}^{(-)} Y_{l}^{(+)}
 \right\}  \right] \: .
\end{eqnarray}
The ratio $R_{2}/R_{0}$ gives
\begin{eqnarray}
 \langle r^{2} \rangle_{l} & = & 
    \frac{R^{2}}{6} \left[2
 -  \frac{2l+3}{Y_{l}^{(-)}}
 - \frac{2l-1}{Y_{l}^{(+)}} 
 + \left(\frac{1}{\bar{\gamma}_{nl}^{2}} - \frac{1}{\gamma^{2}} \right) 
 (2l-1)(2l+3) \right] \: .
\end{eqnarray}
We note
\begin{equation} \label{eq:recur}
 Y_{l}^{(-)} Y_{l-1}^{(+)} = - \gamma^{2}
 \quad \mbox{and} \quad
 Y_{l}^{(-)} + Y_{l}^{(+)} = 2l+1
\end{equation}
and find
\begin{eqnarray}
 \langle r^{2} \rangle_{l} & = & 
    \frac{R^{2}}{6} \left[2
 + \frac{2l+3}{Y_{l-2}^{(+)}}
 - \frac{2l-1}{Y_{l}^{(+)}} 
 + \frac{(2l-1)(2l+3)}{\bar{\gamma}_{nl}^{2}}
  \right]
\end{eqnarray}
Explicit expressions for $Y_{l}^{(\pm)}$ are found from the recursion
relations (\ref{eq:recur})
with 
\begin{equation}
 Y_{0}^{(+)} = 1+\gamma \qquad
 Y_{0}^{(-)} = - \gamma
\end{equation}
for $l=0$.
We have
\begin{equation}
 \langle r^{2} \rangle_{0} = 
 \frac{R^{2}}{6} \left[ \frac{3+2\gamma}{1+\gamma}
 + \frac{3}{\gamma} + \frac{3}{\gamma^{2}} 
 - \frac{3}{\bar{\gamma}_{nl}^{2}}\right]
\end{equation}
and
\begin{equation}
 \langle r^{2} \rangle_{1} = 
 \frac{R^{2}}{6} \left[ \frac{5+5\gamma+2\gamma^{2}}{3+3\gamma+\gamma^{2}}
 + \frac{5}{\gamma} 
 + \frac{5}{\bar{\gamma}_{nl}^{2}}\right]
\end{equation}
for the two lowest orbital angular momenta.
This leads to the divergences
\begin{equation}
 \langle r^{2} \rangle_{0} \to 
 \frac{R^{2}}{2\gamma^{2}} 
\quad \mbox{and} \quad
  \langle r^{2} \rangle_{1} \to
 \frac{5R^{2}}{6\gamma} 
\end{equation}
in the limit $\gamma \to 0$.
With the asymptotic behaviour
\begin{equation}
 Y_{l}^{(+)} \to 2l+1
 \quad \mbox{for} \quad l \geq 0
\end{equation}
for $\gamma \to 0$
we obtain
\begin{eqnarray}
  \langle r^{2} \rangle_{l} & \to & 
     \frac{(2l-1)(2l+3)}{6} R^{2} \left[  
 \frac{2}{(2l-3)(2l+1)} + \frac{1}{\bar{\gamma}_{nl}^{2}} 
  \right] 
\end{eqnarray}
for higher orbital angular momenta. There is no divergence
of the root-mean-square radius in the limit $\gamma \to 0$.
It approaches a finite value.

\subsection{Asymptotic normalization coefficient}
\label{app:A3}

The modulus of the quantity $B_{nl}$ 
is obtained from the normalization condition
\begin{equation}
 1 = P_{<}(nl) + P_{>}(nl) = 
 (-1)^{l} \frac{\left|B_{nl}\right|^{2}}{2}R^{3}  
   \left(1 + \frac{\gamma^{2}}{\bar{\gamma}_{nl}^{2}} \right)
 h^{(1)}_{l-1}(i\gamma)h^{(1)}_{l+1}(i\gamma) 
\end{equation}
It is related via $B_{nl}=q C_{nl} = \gamma C_{nl}/R$ to the ANC $C_{nl}$.
Thus we find for the ANC
\begin{equation}
 C_{nl} = 
 \sqrt{2q} \left[ (-1)^{l} \gamma^{3}
   \left(1 + \frac{\gamma^{2}}{\bar{\gamma}_{nl}^{2}} \right)
 h^{(1)}_{l-1}(i\gamma)h^{(1)}_{l+1}(i\gamma) \right]^{-\frac{1}{2}}
\end{equation}
In the case $l=0$ one has explicitly
\begin{equation}
 C_{n0} =
 \sqrt{2q} \exp(\gamma) \left[
   \left(1 + \frac{\gamma^{2}}{\bar{\gamma}_{n0}^{2}} \right)
 (1+\gamma) \right]^{-\frac{1}{2}}
\end{equation}
and
\begin{equation}
 \lim_{\gamma \to 0} C_{n0} = \sqrt{2q} \: .
\end{equation}
For $l=1$ we find
\begin{equation}
 C_{n1} \to \sqrt{\frac{2q}{3}} \: \gamma^{1/2} 
\end{equation}
and for $l>1$
\begin{equation}
 C_{nl} \to \sqrt{\frac{2q}{(2l-3)!!(2l+1)!!}}  \:
 \gamma^{l-1/2} 
\end{equation}
if $\gamma$ approaches zero.

\subsection{Transition integrals}
\label{app:A4}

Using the notation for the radial wave functions of appendix \ref{app:A1},
the interior and exterior contribution to the radial transition integral
(\ref{eq:radint}) are given by
\begin{eqnarray}
 I_{l_{i}}^{l_{f}}(\lambda,<) & = &
  A_{nl_{i}} \bar{A}_{nl_{f}}^{\ast}
  \bar{q}_{nl_{i}}^{-\lambda-2} 
 M_{l_{i}}^{l_{f}}(\lambda) 
\end{eqnarray}
and
\begin{eqnarray}
 I_{l_{i}}^{l_{f}}(\lambda,>) & = &
 B_{nl_{i}} q^{-\lambda-2}
 \left[ \frac{\exp(2i\delta_{l_{f}}) N_{l_{i}}^{(+)l_{f}}(\lambda) 
 - N_{l_{i}}^{(-)l_{f}}(\lambda) 
 }{2i} 
 \right]^{\ast}
\end{eqnarray}
with the dimensionless integrals
\begin{equation}
 M_{l_{i}}^{l_{f}}(\lambda) 
 =
   \bar{k} \bar{q}^{\lambda+2}  \int_{0}^{R} dr \: 
  j_{l_{i}}(\bar{q}r) \: r^{\lambda+2} \:   j_{l_{f}}(\bar{k}r)
\end{equation}
and
\begin{equation}
 N_{l_{i}}^{(\pm)l_{f}}(\lambda) 
 =
  q^{\lambda+2} \int_{R}^{\infty} dr \: i^{l_{i}}
 h^{(1)}_{l_{i}}(iq r) \: r^{\lambda+1} \:
 u_{l_{f}}^{(\pm)}(kr)
 \: .
\end{equation}
These integrals obey the recursion relations
\begin{eqnarray}
 \label{eq:recmf}
 M_{l_{i}}^{l_{f}+1}(\lambda+1) 
 & = & 
 \bar{q}  \left[ \frac{l_{f}+1}{\bar{k}} - \frac{d}{d\bar{k}} \right]
 M_{l_{i}}^{l_{f}}(\lambda) 
 \\
 \label{eq:recmi}
 M_{l_{i}+1}^{l_{f}}(\lambda+1) 
 & = & 
 \left[ l_{i}+\lambda+2 - \bar{q}\frac{d}{d\bar{q}} \right]
 M_{l_{i}}^{l_{f}}(\lambda) 
 \\
 \label{eq:recnf}
 N_{l_{i}}^{(\pm)l_{f}+1}(\lambda+1) 
 & = & 
 q  \left[ \frac{l_{f}+1}{k} - \frac{d}{dk} \right]
 N_{l_{i}}^{(\pm)l_{f}}(\lambda) 
 \\
 \label{eq:recni}
 N_{l_{i}+1}^{(\pm)l_{f}}(\lambda+1) 
 & = & 
  \left[ l_{i}+\lambda+2 - q\frac{d}{dq} \right]
 N_{l_{i}}^{(\pm)l_{f}}(\lambda) 
 \: .
\end{eqnarray}
Using the continuity equations (\ref{eq:cont}) and (\ref{eq:contf})
we obtain
\begin{eqnarray}
 I_{l_{i}}^{l_{f}}(\lambda,<) & = &
  A_{nl_{i}} \bar{A}_{nl_{f}}^{\ast} 
  \bar{q}_{nl_{i}}^{-\lambda-2} 
 M_{l_{i}}^{l_{f}}(\lambda) 
 \\ \nonumber & = & 
  B_{nl_{i}}  \bar{q}_{nl_{i}}^{-\lambda-2}
 \frac{i^{l_{i}} h^{(1)}_{l_{i}}(i\gamma)M_{l_{i}}^{l_{f}}(\lambda)}{
 j_{l_{i}}(\bar{\gamma}_{nl_{i}})\bar{\kappa}_{nl_{f}}  
 j_{l_{f}}(\bar{\kappa}_{nl_{f}})}
 \left[\frac{\exp(2i\delta_{l_{f}}) u_{l_{f}}^{(+)}(\kappa) 
 - u_{l_{f}}^{(-)}(\kappa)}{2i}  \right]^{\ast}
\end{eqnarray}
and the ratio
\begin{eqnarray}
 R_{l_{i}}^{l_{f}}(\lambda) & = & 
 \frac{I_{l_{i}}^{l_{f}}(\lambda,<)}{I_{l_{i}}^{l_{f}}(\lambda,>)}
 \\ \nonumber & = & 
 \left(\frac{\gamma}{\bar{\gamma}_{nl_{i}}}\right)^{\lambda+2}
 \frac{i^{l_{i}} h^{(1)}_{l_{i}}(i\gamma)
 M_{l_{i}}^{l_{f}}(\lambda)}{j_{l_{i}}(\bar{\gamma}_{nl_{i}})
 \bar{\kappa}_{nl_{f}}  
 j_{l_{f}}(\bar{\kappa}_{nl_{f}})}
\frac{\left[ \exp(2i\delta_{l_{f}}) u_{l_{f}}^{(+)}(\kappa) 
 - u_{l_{f}}^{(-)}(\kappa) \right]^{\ast}}{\left[ 
 \exp(2i\delta_{l_{f}}) N_{l_{i}}^{(+)l_{f}}(\lambda) 
 - N_{l_{i}}^{(-)l_{f}}(\lambda) 
 \right]^{\ast}} 
\end{eqnarray}
With the integral formula \cite{Abr65}
\begin{equation}
 \int dt \: f_{l}(at) \: t^{2} \: g_{l}(bt)
 = \frac{t^{2}}{a^{2}-b^{2}} 
 \left[ a f_{l+1}(at)g_{l}(bt)-bf_{l}(at)g_{l+1}(bt)\right]
\end{equation}
for spherical Bessel/Neumann/Hankel functions $f_{l}$ and $g_{l}$
we find the monopole functions
\begin{equation}
 M_{l}^{l}(0) =
  \frac{\bar{\kappa}\bar{\gamma}^{2}}{\bar{\gamma}^{2}-\bar{\kappa}^{2}}
 \left[\bar{\gamma} j_{l+1}(\bar{\gamma})j_{l}(\bar{\kappa})
 - \bar{\kappa} j_{l}(\bar{\gamma})j_{l+1}(\bar{\kappa})\right]
\end{equation}
and
\begin{equation}
 N_{l}^{(\pm)l}(0)   =
 \frac{\gamma^{2}}{\gamma^{2}+\kappa^{2}}
 \left[ i^{l+1}\gamma h_{l+1}^{(1)}(i\gamma) u_{l}^{(\pm)}(\kappa)
 -  i^{l} \kappa h_{l}^{(1)}(i\gamma)
 u_{l+1}^{(\pm)}(\kappa) \right]
\end{equation}
for general orbital angular momenta $l$.
The logarithmic derivative of the scattering wave function is given by
\begin{equation}
  L_{l_{f}}^{f}
 = \left. \frac{r\frac{d}{dr}g_{nl_{f}}}{g_{nl_{f}}} \right|_{r=R}
 = l_{f} + 1 - Z_{l_{f}}^{(+)} = Z_{l_{f}}^{(-)} - l_{f} 
\end{equation}
with
\begin{equation}
 Z_{l_{f}}^{(\pm)} =
  \bar{\kappa}_{nl_{f}} 
 \frac{j_{l_{f}\pm 1}(\bar{\kappa}_{nl_{f}})}{j_{l_{f}}
 (\bar{\kappa}_{nl_{f}})} 
 = \kappa \frac{\left[\exp(2i\delta_{l_{f}})u_{l_{f}\pm1}^{(+)}(\kappa)
 -u_{l_{f}\pm1}^{(-)}(\kappa)\right]}{\left[ 
 \exp(2i\delta_{l_{f}}) u_{l_{f}}^{(+)}(\kappa) 
     - u_{l_{f}}^{(-)}(\kappa) \right]} \: .
\end{equation}
We note
\begin{equation}
 Z_{l}^{(-)} Z_{l-1}^{(+)} = \kappa^{2}
 \quad \mbox{and} \quad
 Z_{l}^{(+)} + Z_{l}^{(-)} = 2l+1
\end{equation}
similar as in the case for the functions $Y_{l}^{(\pm)}$.
Introducing $Y_{l}^{(\pm)}$ and $Z_{l}^{(\pm)}$ we find
\begin{eqnarray}
 M_{l}^{l}(0)   
 & = & 
  \frac{\bar{\kappa}\bar{\gamma}^{2}}{\bar{\gamma}^{2}-\bar{\kappa}^{2}}
 j_{l}(\bar{\kappa}) j_{l}(\bar{\gamma})
 \left[ Y_{l}^{(+)} - Z_{l}^{(+)} \right]
\end{eqnarray}
and
\begin{eqnarray}
 \lefteqn{\exp(2i \delta_{l}) N_{l}^{(+)l}(0) - N_{l}^{(-)l}(0)}
 \\ \nonumber  & = &
 \frac{\gamma^{2}}{\gamma^{2}+\kappa^{2}}
 i^{l} h_{l}^{(1)}(i\gamma) 
 \left[ \exp(2i \delta_{l})  u_{l}^{(+)}(\kappa) 
 -  u_{l}^{(-)}(\kappa)\right] 
 \left[ Y_{l}^{(+)} - Z_{l}^{(+)} \right] \: .
\end{eqnarray}
The ratio of the monopole integrals is given by
\begin{equation}
 R_{l}^{l}(0) 
  = 
 \left(\frac{\gamma}{\bar{\gamma}_{nl}}\right)^{2}
  \frac{\bar{\kappa}_{nl}
 \bar{\gamma}_{nl}^{2}}{\bar{\gamma}_{nl}^{2}
 -\bar{\kappa}^{2}}
 \frac{\gamma^{2}+\kappa^{2}}{\gamma^{2}} \frac{1}{\bar{\kappa}_{nl}}
 = \frac{\gamma^{2}+\kappa^{2}}{v_{i}-v_{f}-\gamma^{2}-\kappa^{2}}
\end{equation}
where the relations (\ref{eq:vi}) and (\ref{eq:vf}) were used.
If the depths of the potential in the bound and scattering state are
identical, i.e.\ $v_{i} = v_{f}$, we have
\begin{equation}
 R_{l}^{l}(0) = -1
 \quad \mbox{i.e.} \quad
 I_{l}^{l}(0) = 0
\end{equation}
and the bound and scattering wave functions are orthogonal.

Applying the recursion relations (\ref{eq:recmf}) and
(\ref{eq:recnf}) the relevant integrals for
$E1$ transitions $l \to l+1$ are found to be
\begin{eqnarray}
 \lefteqn{M_{l}^{l+1}(1) =
 \frac{\bar{\kappa}\bar{\gamma}^{3}}{(\bar{\gamma}^{2}-\bar{\kappa}^{2})^{2}}
 j_{l}(\bar{\gamma}) j_{l+1}(\bar{\kappa})}
 \\ \nonumber & & \times
 \left[ (\bar{\gamma}^{2} - \bar{\kappa}^{2})
 \left( Y_{l}^{(+)} +Z_{l+1}^{(-)} \right) -2 Y_{l}^{(+)} Z_{l+1}^{(-)}
 + (2l+3) \bar{\kappa}^{2} - (2l+1) \bar{\gamma}^{2} \right]
\end{eqnarray}
and
\begin{eqnarray}
 \lefteqn{\exp(2i\delta_{l+1})N_{l}^{(+)l+1}(1) - N_{l}^{(-)l+1}(1) =}
 \\ \nonumber &  & 
 \frac{\gamma^{3}}{(\gamma^{2}+\kappa^{2})^{2}}
  i^{l} h_{l}^{(1)}(i\gamma)
 \left[ \exp(2i\delta_{l+1})u_{l+1}^{(+)}(\kappa)-u_{l+1}^{(-)}(\kappa)\right]
 \\ \nonumber & &  \times
 \left[ (\gamma^{2}+\kappa^{2}) 
 \left(  Y_{l}^{(+)} + Z_{l+1}^{(-)}\right)
 +  2  Y_{l}^{(+)} Z_{l+1}^{(-)}
 - (2l+1) \gamma^{2} - (2l+3) \kappa^{2} \right] \: .
\end{eqnarray}
Then the radial integrals are
\begin{eqnarray} \label{eq:intfk}
 \lefteqn{I_{l}^{l+1}(1,<) =
  \frac{R^{2}}{(\bar{\gamma}^{2}-\bar{\kappa}^{2})^{2}}
 f_{nl}(R) g_{nl+1}^{\ast}(R)}
 \\ \nonumber & & \times
 \left[ (\bar{\gamma}^{2} - \bar{\kappa}^{2})
 \left( Y_{l}^{(+)} +Z_{l+1}^{(-)} \right) -2 Y_{l}^{(+)} Z_{l+1}^{(-)}
 + (2l+3) \bar{\kappa}^{2} - (2l+1) \bar{\gamma}^{2} \right]
\end{eqnarray}
and
\begin{eqnarray} \label{eq:intfg}
 \lefteqn{I_{l}^{l+1}(1,>) =
  \frac{R^{2}}{(\gamma^{2}+\kappa^{2})^{2}}
  f_{nl}(R) g_{nl+1}^{\ast}(R)}
 \\ \nonumber & & 
 \times \left[ (\gamma^{2}+\kappa^{2}) 
 \left(  Y_{l}^{(+)} + Z_{l+1}^{(-)}\right)
 +  2  Y_{l}^{(+)} Z_{l+1}^{(-)}
 - (2l+1) \gamma^{2} - (2l+3) \kappa^{2} \right] \: .
\end{eqnarray}
Similarly, the relevant integrals for $E1$ transitions $l+1 \to l$
\begin{eqnarray} 
 \lefteqn{M_{l+1}^{l}(1) =
 \frac{\bar{\kappa}\bar{\gamma}^{3}}{(\bar{\gamma}^{2}-\bar{\kappa}^{2})^{2}}
  j_{l+1}(\bar{\gamma}) j_{l}(\bar{\kappa})}
 \\ \nonumber & & 
 \times \left[ (2l+3) \bar{\gamma}^{2} - (2l+1) \bar{\kappa}^{2}
 - \left( \bar{\gamma}^{2}-\bar{\kappa}^{2}\right)
 \left( Y_{l+1}^{(-)} + Z_{l}^{(+)} \right)
 - 2  Y_{l+1}^{(-)} Z_{l}^{(+)} \right]
\end{eqnarray}
and
\begin{eqnarray} 
 \lefteqn{ \exp(2i\delta_{l}) N_{l+1}^{(+)l}(1)
 - N_{l+1}^{(-)l}(1) = }
 \\ \nonumber &  & 
 \frac{\gamma^{3}}{(\gamma^{2}+\kappa^{2})^{2}}
 i^{l+1} h_{l+1}^{(1)}(i\gamma)
 \left[ \exp(2i\delta_{l}) u_{l}^{(+)}(\kappa) - u_{l}^{(-)}(\kappa) \right]
 \\ \nonumber & & \times
 \left[ (2l+3) \gamma^{2} + (2l+1) \kappa^{2}
 + 2 Y_{l+1}^{(-)} Z_{l}^{(+)} - (\gamma^{2}+\kappa^{2}) 
 \left( Y_{l+1}^{(-)}+ Z_{l}^{(+)}\right) \right]
\end{eqnarray}
are found with the recursion relations
(\ref{eq:recmi}) and (\ref{eq:recni}).
Correspondingly, the radial integrals
\begin{eqnarray} \label{eq:intik}
 \lefteqn{I_{l+1}^{l}(1,<) =
 \frac{R^{2}}{(\bar{\gamma}^{2}-\bar{\kappa}^{2})^{2}} 
 f_{nl+1}(R) g_{nl}^{\ast}(R)}
 \\ \nonumber & & \times
 \left[ (2l+3) \bar{\gamma}^{2} - (2l+1) \bar{\kappa}^{2}
 - 2  Y_{l+1}^{(-)} Z_{l}^{(+)}
 - \left( \bar{\gamma}^{2}-\bar{\kappa}^{2}\right)
 \left( Y_{l+1}^{(-)} + Z_{l}^{(+)} \right)
  \right]
\end{eqnarray}
and
\begin{eqnarray} \label{eq:intig}
 \lefteqn{I_{l+1}^{l}(1,>) =
 \frac{R^{2}}{(\gamma^{2}+\kappa^{2})^{2}}
 f_{nl+1}(R) g_{nl}^{\ast}(R)}
 \\ \nonumber & & \times
 \left[ (2l+3) \gamma^{2} + (2l+1) \kappa^{2}
 + 2 Y_{l+1}^{(-)} Z_{l}^{(+)} - (\gamma^{2}+\kappa^{2}) 
 \left( Y_{l+1}^{(-)}+ Z_{l}^{(+)}\right) \right]
\end{eqnarray}
are obtained. 

Assuming that the depths of the potential are the same in
the bound and the scattering state, i.e.
\begin{equation}
 v = \bar{\gamma}^{2}+\gamma^{2} = v_{i} = 
 v_{f} = \bar{\kappa}^{2}-\kappa^{2}
\end{equation}
we find for the total integrals
\begin{equation}
 I_{l}^{l+1}(1) =
  \frac{2R^{2}v}{(\gamma^{2}+\kappa^{2})^{2}}
 f_{nl}(R) g_{nl+1}^{\ast}(R)
\end{equation}
and
\begin{equation}
 I_{l+1}^{l}(1) =
 \frac{2R^{2}v}{(\gamma^{2}+\kappa^{2})^{2}}
 f_{nl+1}(R) g_{nl}^{\ast}(R)
\end{equation}
consistent with the result (\ref{eq:radintgf}) of the commutator relation.
The ratios of the interior to the exterior dipole integral are given by
\begin{eqnarray}
 \lefteqn{R_{l}^{l+1}(1) = \frac{I_{l}^{l+1}(1,<)}{I_{l}^{l+1}(1,>)}}
 \\ \nonumber & = & 
  \frac{(\bar{\gamma}^{2} - \bar{\kappa}^{2})
 \left( Y_{l}^{(+)} +Z_{l+1}^{(-)} \right) -2 Y_{l}^{(+)} Z_{l+1}^{(-)}
 + (2l+3) \bar{\kappa}^{2} - (2l+1) \bar{\gamma}^{2}}{
 (\gamma^{2}+\kappa^{2}) 
 \left(  Y_{l}^{(+)} + Z_{l+1}^{(-)}\right)
 +  2  Y_{l}^{(+)} Z_{l+1}^{(-)}
 - (2l+1) \gamma^{2} - (2l+3) \kappa^{2}}
\end{eqnarray}
and
\begin{eqnarray}
 \lefteqn{R_{l+1}^{l}(1) = \frac{I_{l+1}^{l}(1,<)}{I_{l+1}^{l}(1,>)}}
 \\ \nonumber & = & 
 \frac{(2l+3) \bar{\gamma}^{2} - (2l+1) \bar{\kappa}^{2}
 - 2  Y_{l+1}^{(-)} Z_{l}^{(+)}
 - \left( \bar{\gamma}^{2}-\bar{\kappa}^{2}\right)
 \left( Y_{l+1}^{(-)} + Z_{l}^{(+)} \right)}{
 (2l+3) \gamma^{2} + (2l+1) \kappa^{2}
 + 2 Y_{l+1}^{(-)} Z_{l}^{(+)} - (\gamma^{2}+\kappa^{2}) 
 \left( Y_{l+1}^{(-)}+ Z_{l}^{(+)}\right)}
\end{eqnarray}
for the transitions $l \to l+1$ and $l+1 \to l$, respectively.
The scaling behaviour of these ratios is obtained by an
expansion
in terms of small
$\gamma$ with $\kappa \to 0$. 
In the case with a $s$ wave bound state we have the relation
\begin{equation}
 Y_{0}^{(-)} = -\gamma = \bar{\gamma} \cot \bar{\gamma} \: .
\end{equation}
Then we find the expansion
\begin{equation}
 \bar{\gamma}_{n} = 
 s_{n} \left( 1 + \frac{\gamma}{s_{n}^{2}} 
 - \frac{\gamma^{2}}{s_{n}^{4}}
 + \left(2-\frac{s_{n}^{2}}{3} \right) \frac{\gamma^{3}}{s_{n}^{6}}
 - \left(5-\frac{4s_{n}^{2}}{3} \right) \frac{\gamma^{4}}{s_{n}^{8}} + \dots
 \right)
\end{equation}
with $s_{n} = (2n-1)\pi/2$
for $\gamma \to 0$
depending on the principal quantum number $n=1,2,\dots $.
Similarly, for the case of a $p$ wave bound state the relation
\begin{equation}
 Y_{1}^{(-)} = -\frac{\gamma^{2}}{1+\gamma} 
 = \frac{\bar{\gamma}^{2}}{1- \bar{\gamma} \cot \bar{\gamma}}
\end{equation}
leads to the expansion
\begin{equation}
 \bar{\gamma}_{n} = 
 s_{n} \left( 1 + \frac{\gamma^{2}}{s_{n}^{2}} 
 - \frac{\gamma^{3}}{s_{n}^{2}}
 + \left(s_{n}^{2}-2 \right) \frac{\gamma^{4}}{s_{n}^{4}}
 + \dots
 \right)
\end{equation}
with $s_{n} = n\pi$.
Using
\begin{equation}
 Z_{1}^{(-)} = \frac{\bar{\kappa}^{2}}{1-\bar{\kappa} \cot \bar{\kappa}}
 \quad \mbox{and} \quad
 Z_{0}^{(+)} = 1-\bar{\kappa} \cot \bar{\kappa}
\end{equation}
we obtain
\begin{equation}
 \lim_{\kappa \to 0} R_{0}^{1}(1) =  \frac{\gamma^{4}}{8s_{n}^{2}}
 + \dots
 \quad \mbox{\and} \quad
 \lim_{\kappa \to 0} R_{1}^{0}(1)  =  - \frac{\gamma^{2}}{4s_{n}^{2}}
 + \dots
\end{equation}
for small $\gamma$.

\section{Integrals of spherical cylinder functions}
\label{app:E}

With the recursion relations \cite{Abr65}
\begin{eqnarray}
 & & f_{l}^{\prime}(z) = f_{l-1}(z) - \frac{l+1}{z} f_{l}(z)
 = \frac{l}{z} f_{l}(z)-f_{l+1}(z)
 \\
 & & 
 f_{l+1}(z) + f_{l-1}(z) = \frac{2l+1}{z}f_{l}(z)
\end{eqnarray}
for any  spherical Bessel, Neumann
or Hankel function $f_{l}(z)$  the well-known integral formula
\begin{equation}
 \int dz \: z^{2} \left[ f_{l}(z) \right]^{2}
 = \frac{z^{3}}{2} \left( \left[ f_{l}(z) \right]^{2}
 -  f_{l-1}(z)f_{l+1}(z) \right)
\end{equation}
is easily proven. In a similar fashion the new relation
\begin{equation}
 \int dz \: z^{4} \: \left[ f_{l}(z)\right]^{2}
 = \frac{z^{5}}{12} \left( 3 [f_{l}(z)]^{2}
 -2 f_{l-1}(z) f_{l+1}(z) - f_{l-2}(z) f_{l+2}(z)
 \right) 
\end{equation}
is obtained. Similarly, integrals with higher powers in $z$ can be treated.
Introducing the logarithmic derivative
\begin{equation}
 L_{l}(z) = z\frac{f_{l}^{\prime}(z)}{f_{l}(z)} 
 = l-Y_{l}^{(+)}(z)
 =  Y_{l}^{(-)}(z)-l-1 
\end{equation}
with
\begin{equation}
 Y_{l}^{(\pm)}(z) = z\frac{f_{l\pm1}(z)}{f_{l}(z)} \: .
\end{equation}
one finds
\begin{eqnarray}
 f_{l-1}(z)f_{l+1}(z) & = & 
\frac{[f_{l}(z)]^{2}}{z^{2}} 
\left[ l+1+L_{l}(z)\right]\left[ l - L_{l}(z)\right] 
\end{eqnarray}
and
\begin{eqnarray}
 f_{l-2}(z)f_{l+2}(z) & = & 
 \frac{[f_{l}(z)]^{2}}{z^{4}} 
 \left\{ (2l-1)[L_{l}(z)+l+1]-z^{2}\right\}
 \\ \nonumber &  &
 \left\{ (2l+3)[l-L_{l}(z)]-z^{2}\right\}
\end{eqnarray}
With these relations we obtain
\begin{eqnarray}
 \int dz \: z^{2} \left[ f_{l}(z) \right]^{2}
 & = & 
\frac{z}{2} \left[ f_{l}(z) \right]^{2} \left( z^{2}
- \left[ l+1+L_{l}(z)\right]\left[ l - L_{l}(z)\right] \right)
\end{eqnarray}
and
\begin{eqnarray}
 \lefteqn{\int dz \: z^{4} \: \left[ f_{l}(z)\right]^{2}}
 \\ \nonumber 
 & = & \frac{z}{12} [f_{l}(z)]^{2} \left( 3 z^{4}
 -2 z^{2}
\left[ l+1+L_{l}(z)\right]\left[ l - L_{l}(z)\right]  \right.
 \\ \nonumber & & \left.
 -  \left\{(2l-1)\left[ l+1+L_{l}(z)\right] - z^{2}\right\}
 \left\{(2l+3)\left[ l - L_{l}(z)\right] - z^{2} \right\}
 \right) \: .
\end{eqnarray}

\section{Radial transition integrals for neutron+core systems}
\label{app:B}

In the case when $b$ is a neutron analytical expressions of the functions
${\mathcal H}_{l_{i}}^{l_{f}}(\lambda)$
are found. With $\eta = 0$
the Coulomb wave functions 
\begin{eqnarray}
 F_{l}(\eta;z) 
 \to z j_{l}(z)
 \qquad
 G_{l}(\eta;z) 
 \to - z y_{l}(z) 
\end{eqnarray}
in the integral (\ref{eq:hdef})
reduce to spherical Bessel and Neumann functions \cite{Abr65}. 
Similarly, the Whittaker function 
\begin{equation}
 W_{-\eta,l+\frac{1}{2}}(2qr) \to -qr \: i^{l} \: h_{l}^{(1)}(iqr)
\end{equation}
reduces to a spherical Hankel function.
This leads to
\begin{eqnarray}
 {\mathcal H}_{l_{i}}^{l_{f}}(\lambda) & = &
 - i^{l_{i}} \kappa \gamma^{\lambda+2}  \int_{1}^{\infty} dt  \: 
 t^{\lambda+2} \: 
  h^{(2)}_{l_{f}}(\kappa t) \:
   \:  h^{(1)}_{l_{i}}(i\gamma t)
\end{eqnarray}
with the quantities $\kappa = kR$ and $\gamma = qR$.
From  the recursion relation 
\begin{eqnarray}
 g_{l+1}(z) & = & \left[ \frac{l}{z}  - \frac{d}{dz} \right] g_{l}(z)
\end{eqnarray}
for any spherical Bessel function $g_{l}$
the recursion relations
\begin{eqnarray} \label{eq:recf}
 {\mathcal H}_{l_{i}}^{l_{f}+1}(\lambda+1) & = &
 \gamma \left[ \frac{l_{f}+1}{\kappa} 
 - \frac{d}{d\kappa} \right] {\mathcal H}_{l_{i}}^{l_{f}}(\lambda)
 \\ \label{eq:reci}
 {\mathcal H}_{l_{i}+1}^{l_{f}}(\lambda+1) & = &
 \left[ l_{i}+\lambda+2 
 - \gamma \frac{d}{d\gamma} \right] {\mathcal H}_{l_{i}}^{l_{f}}(\lambda)
\end{eqnarray}
for the functions ${\mathcal H}_{l_{i}}^{l_{f}}(\lambda)$ are obtained
that allow the calculation of all relevant functions for increasing 
values of $l_{i}$, $l_{f}$, and $\lambda$ from basic integrals with
$\lambda = 0$.
Explicitly one finds
\begin{eqnarray}
 {\mathcal H}_{0}^{0}(0) & = &
 i \gamma \int_{1}^{\infty} dt \: \exp[-(\gamma+i \kappa)t]
 = \frac{i\gamma}{\gamma+i\kappa} \exp(-\gamma-i\kappa)
\end{eqnarray}
and
\begin{eqnarray}
 {\mathcal H}_{1}^{1}(0) & = &
  -   \gamma \int_{1}^{\infty} dt  \:   \exp[-(\gamma+i \kappa)t]
  \left( 1 + \frac{1}{i\kappa t} 
  + \frac{1}{\gamma t} + \frac{1}{i\kappa\gamma t^{2}}\right)
 \\ \nonumber & = &
   \left( \frac{i}{\kappa}  - \frac{\gamma}{\gamma+i\kappa}
 \right) \exp(-\gamma-i\kappa)
\end{eqnarray}
with the help of the exponential integral and its recursion relations
\cite{Abr65}.
Separating real and imaginary parts one obtains the reduced
monopole integrals 
\begin{eqnarray} \label{eq:i000}
 {\mathcal I}_{0}^{0}(0) & = & 
 \frac{\gamma \exp(-\gamma)}{\gamma^{2}+\kappa^{2}} 
 \left[  \kappa \cos (\kappa+\delta_{0}) 
 + \gamma \sin(\kappa+\delta_{0})\right]
 \\ \label{eq:i110}
 {\mathcal I}_{1}^{1}(0) & = & 
 \frac{\gamma \exp(-\gamma)}{\gamma^{2}+\kappa^{2}} 
 \left[ -\gamma \cos (\kappa+\delta_{1}) 
 + \left( \kappa +\frac{\gamma^{2}+\kappa^{2}}{\gamma \kappa}\right) 
 \sin(\kappa+\delta_{1})\right]
\end{eqnarray}
where the addition theorems of the sine and cosine
functions have been used. 
The reduced integrals ${\mathcal I}_{l_{i}}^{l_{f}}(\lambda)$
for fixed phase shift $\delta$
obey the same recursion relations (\ref{eq:recf},\ref{eq:reci}) 
as the functions ${\mathcal H}_{l_{i}}^{l_{f}}(\lambda)$.

Considering the general form (\ref{eq:iredgen}) 
for the reduced radial integrals
the monopole functions 
\begin{equation}
 {\mathcal R}_{0}^{(+)0}(0) = \kappa  \qquad 
 {\mathcal R}_{0}^{(-)0}(0) = \gamma 
\end{equation}
and
\begin{equation}
 {\mathcal R}_{1}^{(+)1}(0) = - \kappa  \qquad 
 {\mathcal R}_{1}^{(-)1}(0) = 
 [\kappa^{2}(1+\gamma)+\gamma^{2}]/\gamma^{2}
\end{equation}
are extracted from Eqs.\ (\ref{eq:i000}) and (\ref{eq:i110}).
General expression for  ${\mathcal R}_{l_{i}}^{(\pm)l_{f}}(\lambda)$
in Eq.~(\ref{eq:iredgen}) 
for larger values of $l_{f}$, $l_{i}$, and
$\lambda$ can be obtained 
by applying the recursion relations
\begin{eqnarray}
 {\mathcal R}_{l_{i}}^{(\pm)l_{f}+1}(\lambda+1)
 & = & \left[ 2\kappa^{2}(\lambda+1)+
 (\gamma^{2}+\kappa^{2})(2l_{f}+1)\right]
 {\mathcal R}_{l_{i}}^{(\pm)l_{f}}(\lambda)
 \\ \nonumber & & 
 - \kappa (\gamma^{2}+\kappa^{2}) \left[
 \frac{d}{d\kappa} {\mathcal R}_{l_{i}}^{(\pm)l_{f}}(\lambda)
 \pm {\mathcal R}_{l_{i}}^{(\mp)l_{f}}(\lambda)\right]
\end{eqnarray}
\begin{eqnarray}
 {\mathcal R}_{l_{i}+1}^{(\pm)l_{f}}(\lambda+1)
 & = & \left[ 2\gamma^{2}(\lambda+1)+
 (\gamma^{2}+\kappa^{2})(l_{i}+\lambda+1-l_{f}+\gamma)\right]
 \\ \nonumber & & \times
 {\mathcal R}_{l_{i}}^{(\pm)l_{f}}(\lambda)
 - \gamma (\gamma^{2}+\kappa^{2}) 
 \frac{d}{d\gamma} {\mathcal R}_{l_{i}}^{(\pm)l_{f}}(\lambda)
\end{eqnarray}
Explicitly one finds
\begin{eqnarray}
 {\mathcal R}_{0}^{(+)1}(1) & = & -\kappa[\kappa^{2}(\gamma-2)+\gamma^{3}]
 \\
 {\mathcal R}_{0}^{(-)1}(1) & = & 
 \kappa^{4}+\kappa^{2}\gamma(3+\gamma)+\gamma^{3}
 \\
 {\mathcal R}_{1}^{(+)0}(1) & = & 
 \kappa[\kappa^{2}(1+\gamma)+\gamma^{2}(3+\gamma)]
 \\
 {\mathcal R}_{1}^{(-)0}(1) & = & \gamma^{2}[\kappa^{2}+\gamma(2+\gamma)]
 \\
 {\mathcal R}_{1}^{(+)2}(1) & = & -
 \kappa[\kappa^{4}(1+\gamma)+\kappa^{2}\gamma^{2}(6+\gamma)
 +3\gamma^{4}]/\gamma^{2}
 \\
 {\mathcal R}_{1}^{(-)2}(1) & = & 
 [\kappa^{4}(3+3\gamma-\gamma^{2})+\kappa^{2}\gamma^{2}(6+\gamma-\gamma^{2})
 +3\gamma^{4}]/\gamma^{2}
 \\
 {\mathcal R}_{2}^{(+)1}(1) & = & -
 \kappa[\kappa^{2}(1+\gamma)+\gamma^{2}(3+\gamma)]
 \\
 {\mathcal R}_{2}^{(-)1}(1) & = & 
 [\kappa^{4}(3+3\gamma+\gamma^{2})+\kappa^{2}\gamma^{2}(6+6\gamma+\gamma^{2})
 +\gamma^{4}(3+\gamma)]/\gamma^{2}
\end{eqnarray}
for $\lambda=1$ and
\begin{eqnarray}
 {\mathcal R}_{0}^{(+)2}(2) & = &
 -\kappa [\kappa^{6}+\kappa^{4}(-8+7\gamma+2\gamma^{2})
 +\kappa^{2}\gamma^{3}(10+\gamma)+3\gamma^{5}]
 \\
 {\mathcal R}_{0}^{(-)2}(2) & = & 
 \kappa^{6} (5-\gamma) +\kappa^{4}\gamma(15+6\gamma-2\gamma^{2}) 
 + \kappa^{2}\gamma^{3}(10+\gamma-\gamma^{2}) 
 \\ \nonumber & &
 + 3\gamma^{5}
 \\
 {\mathcal R}_{1}^{(+)1}(2) & = &
 \kappa[\kappa^{4}(2+2\gamma-\gamma^{2})+2\kappa^{2}\gamma^{2}(5-\gamma^{2})
 -\gamma^{5}(2+\gamma)]
 \\
 {\mathcal R}_{1}^{(-)1}(2) & = & 
 \kappa^{6}(1+\gamma) +\kappa^{4}\gamma^{2}(7+2\gamma) 
 +\kappa^{2}\gamma^{3}(10+7\gamma+\gamma^{2})
 \\ \nonumber & & 
 +\gamma^{5}(2+\gamma)
 \\
 {\mathcal R}_{2}^{(+)0}(2) & = &
 \kappa [\kappa^{4}(3+3\gamma+\gamma^{2})
 +2\kappa^{2}\gamma^{2}(5+5\gamma+\gamma^{2})
 \\ \nonumber & & 
 +\gamma^{4}(15+7\gamma+\gamma^{2})]
 \\
 {\mathcal R}_{2}^{(-)0}(2) & = & 
 \gamma^{2}[\kappa^{4}(1+\gamma)+2\kappa^{2}\gamma^{2}(3+\gamma)
 +\gamma^{3}(8+5\gamma+\gamma^{2})]
\end{eqnarray}
for $\lambda=2$. They cover most of the relevant $E1$ and $E2$ 
transitions between $s$, $p$, and $d$ waves in the initial and final
states.


\end{document}